\documentclass[]{aa}
\usepackage{array}
\usepackage{graphicx}
\usepackage{float}
\usepackage{fixltx2e}
\usepackage{hyperref}
\usepackage{multirow}

\usepackage{txfonts}
\begin{document}

   \title{Relationship between solar energetic particle intensities and coronal mass ejection kinematics using STEREO/SECCHI field of view}

  % \subtitle{I. Overviewing the $\kappa$-mechanism}

   \author{Anitha Ravishankar
          %\inst{1}
          \and
          Grzegorz Micha$\l$ek
          %\inst{1}%\fnmsep\thanks{Just to show the usage
         % of the elements in the author field}
          }

 \institute{Astronomical Observatory of Jagiellonian University, Krakow, Poland\\
 \email{anitha@oa.uj.edu.pl}\\
 \email{grzegorz.michalek@uj.edu.pl}
 }

%   \date{Received September 15, 1996; accepted March 16, 1997}

% \abstract{}{}{}{}{}
% 5 {} token are mandatory

  \abstract
   { Solar energetic particles (SEPs) accelerated from shocks driven by coronal mass ejections (CMEs) are one of the major causes of geomagnetic storms on Earth. Therefore, it is necessary to predict the occurrence and intensity of such disturbances. For this purpose we analyzed in detail 38 non-interacting halo and partial halo CMEs, as seen by the \emph{Solar and Heliospheric Observatory/Large Angle and Spectrometric Coronagraph} (SOHO/LASCO), generating SEPs (in >10 MeV, >50 MeV, and >100 MeV energy channels) during the quadrature configuration of the  \emph{Solar TErrestrial RElations Observatory} (STEREO) twin spacecrafts with respect to the Earth, which marks the ascending phase of solar cycle 24 (i.e., 2009–2013). The main criteria for this selection period is to obtain height--time measurements of the CMEs without significant projection effects and in a very large field of view. Using the data from STEREO/\emph{Sun Earth Connection Coronal and Heliospheric Investigation} (STEREO/SECCHI) images we determined several kinematic parameters and instantaneous speeds of the  CMEs. First, we compare instantaneous CME speed and Mach number versus SEP fluxes for events originating at the western and eastern limb;  we observe high correlation for the western events and anticorrelation for the eastern events. Of the two parameters, the Mach number offers higher correlation. Next we investigated instantaneous CME kinematic parameters such as maximum speed, maximum Mach number, and the CME speed and Mach number at SEP peak flux versus SEP peak fluxes. Highly positive correlation is observed for Mach number at SEP peak flux for all events. The obtained instantaneous Mach number parameters from the emperical models was verified with the start and end time of type II radio bursts, which are signatures of CME-driven shock in the interplanetary medium. Furthermore, we conducted estimates of delay in time and distance between CME, SEP, and shock parameters. We observe an increase in the delay in time and distance when SEPs reach peak flux with respect to CME onset as we move from the western to the eastern limb. Western limb events (longitude 60$^{\circ}$) have the best connectivity and this decreases as we move towards the eastern limb.  This variation is due to the magnetic connectivity from the Sun to the Earth, called  the Parker spiral interplanetary magnetic field (IMF). Comparative studies of the considered energy channels of the SEPs also throw light on the reacceleration of suprathermal seed ions by CME-driven shocks that are pre-accelerated in the magnetic reconnection. }

\maketitle



%
%-------------------------------------------------------------------

\section{Introduction}

Solar energetic particles (SEPs) are  one of the major causes of geomagnetic disturbances on Earth (\cite{cane1987}; \citealt{gosling1993}; \citealt{reames1999};  \citealt{kahler2001}; \citealt{Aschwanden12}). The timescales, spectra, composition and charge states, and the associated radio bursts observed at 1 AU of these particles categorize them into impulsive SEP events accelerated at coronal flare reconnection sites (\citealt{cane1986}), and gradual SEP events accelerated by coronal mass ejection (CME) shocks or interplanetary shocks (\citealt{gosling1993}; \citealt{reames1995}; \citealt{reames1999}). In the aspect of potential space weather impacts, the gradual SEP events with high proton fluxes are the prime threats that cause disturbances of Earth’s magnetosphere and upper atmosphere. 

The diffusive shock acceleration theory has been well studied, and it is a widely accepted mechanism for energizing the ions in gradual SEP events (e.g., \citealt{Jokipii1982}; \citealt{Lee1983}, \citealt{Lee2000}; \citealt{Lee2012}; \citealt{DesaiGiacalone2016}). Charged particles can be accelerated by collisionless shocks, provided the spatial diffusion allows some particles to traverse the shock many times. They gain energy because the scattering centers are embedded in converging plasma flows across the shock. Apart from the CME driver speed, other characteristic speeds (such as the Alfvén speed) of the ambient medium determine the strength of the shock (\citealt{Krogulec1994}; \citealt{Mann1999}; \citealt{gopalswamy01}; \citealt{Mann2003}; \citealt{gopalswamy08a}; \citealt{gopalswamy08b}; \citealt{gopalswamy2010}). As explained by \citealt{makela2011}, the formation of a fast-mode shock occurs in front of the CME when the CME speed relative to the ambient medium exceeds the local Alfvén speed. Thus, the particle acceleration in the CME-driven shocks in the corona and IP space can be affected by the variations in the CME speed due to evolution of the  propelling Lorentz and aerodynamic drag forces (\citealt{gopalswamy2000}; \citealt{yashiro2004}; \citealt{gopalswamy2006}) and in the Alfvén speed (see, e.g., \citealt{gopalswamy01}; \citealt{Mann2003}).

The Mach number is an important parameter used to determine the strength of shock fronts. Of the several methods used to calculate the Mach number (e.g., \citealt{Vinas1986}; \citealt{gopalswamy2010}; J. C. Kasper's database\footnote{http://www.cfa.harvard.edu/shocks/}), we use the standard approach of considering the Alfvén speed, solar wind speed, and the CME speed. These three parameters change with heliocentric distance (mostly decreasing), which influences the Mach number. Here, the shock forms when the CME speed exceeds the sum of the Alfvén speed and the solar wind speed (\citealt{gopalswamy2010}).

A good indicator of SEPs accelerated due to coronal and interplanetary shocks are type II bursts (\citealt{cliver}). These bursts are caused by electrons accelerated by shocks  (see, e.g., \citealt{Kahler1982}; \citealt{Kahler2000}; \citealt{Cane2002}; \citealt{Cliver2004}; \citealt{gopalswamy2005}; \citealt{cho2008}). Gradual SEPs are often associated with metric type II bursts (150 to 15 MHz) and are generated close to the Sun $\leq$ 3 R$_{sun}$ (\citealt{gopalswamy2009b}). Other methods of direct detection of shocks are the in situ measurements of the discontinuous jump in density, temperature, flow speed, and magnetic field in the solar wind data \citealt{gopalswamy2010}.

The average speed of CMEs is the widely used parameter for correlation studies with the associated SEP peak flux (see \citealt{kahler2001};  \citealt{Vourlidas10}; \citealt{Richardson14}; \citet{Richardson15}; \citealt{Pande};; \citealt{Xie19}).  However, \citealt{Liou2011} presented the approach of correlating fast-forward shock Mach numbers with the intensity of solar energetic oxygen (O) and helium-4 ($^{4}$He) particles at (E >$\approx$10 MeV n$^{-1}$), and obtained a good linear correlation for two SEP events that occurred on 28--31 October 2003. The results suggest that the Mach number of IP shocks is one of the primary parameters controlling the intensity of SEPs measured in the vicinity of the Earth. \citealt{Anitha2020a} further investigated this approach on a sample of 25 non-interacting CMEs and their associated SEPs that occurred during the period 2009–2013, using multiple spacecrafts: the \emph{Solar TErrestrial RElations Observatory} (STEREO)-A and -B and the \emph{SOlar and Heliospheric Observatory} (SOHO) for CMEs, and the \emph{Geostationary Operational Environmental Satellite} (GOES-13) for SEPs. Instantaneous speeds such as the CME maximum speed, Mach number at CME maximum speed, and   CME speed and Mach number at SEP peak flux were investigated, and better correlations were obtained compared to the average speed.

The main limitation of our work in \citealt{Anitha2020a} was the small population of the considered CMEs (25 events). In the present work we significantly expanded our statistical research with a sample of 38 non-interacting CMEs and their associated SEP events near the quadrature configuration of STEREO. The CME kinematics were determined using the data from STEREO/\emph{Sun Earth Connection Coronal and Heliospheric Investigation} (SECCHI) (\citealt{Brueckner95}; \citealt{Howard08}). STEREO/SECCHI data for CMEs was used instead of SOHO as STEREO offered a larger field of view, and the possibility to determine velocities at instances of SEP onset and peak flux, which occur at varied distances  from the Sun, with minimal projection effects. The GOES-13/\emph{Energetic Particle Sensor} (EPS), part of the \emph{Space Environment Monitor} (SEM), was used to study the associated SEP intensities at three energy bands (>10 MeV, >50 MeV, and >100 MeV).  We also considered different models of the solar magnetic field to obtain  accurate Alfv\'{e}n and solar wind speeds. We verified the obtained the Mach number with the start and end time of type II radio bursts, which are signatures of CME-driven shocks in the interplanetary medium. The start and end time of type II radio bursts should be consistent with instances when the CMEs reach a speed of Mach 1. We carefully consider correlations between different speeds of CMEs and SEP peak fluxes in the three energy channels. We also investigated these coefficients for different subsamples of events based on their longitudes (disk, disk-west, disk-east). The important result of this paper is that the Mach number at SEP peak flux can be very good indicator of peak intensities of SEPs. This good correlation is observed even for eastern limb events where the magnetic connectivity of Sun and the Earth are poor.

This article is organized as follows. The data and method used for the study are described in Section 2. In Section 3 we present results of our study. The conclusions and discussions are presented in Section 4.

%1
\begin{figure}[h!]
\includegraphics[width=9cm,height=6cm]{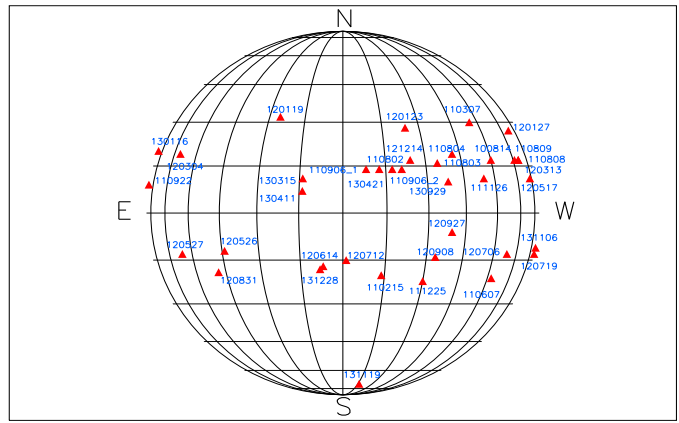}
   \caption{Heliographic locations of the solar flares associated with 38 non-interacting CMEs generating SEPs.}
              \label{FigGam}%
    \end{figure}

%--------------------------------------------------------------------
\section{Data and method}
In our study we used observations from the STEREO/SECCHI telescopes, and employed the technique to determine the instantanous speed of CMEs \citep{Anitha2020a}. In the following subsection we describe the method we used for our study.

%2
 
     \begin{figure*}[h!] 
   \centering
   \includegraphics[width=9.1cm,height=6cm]{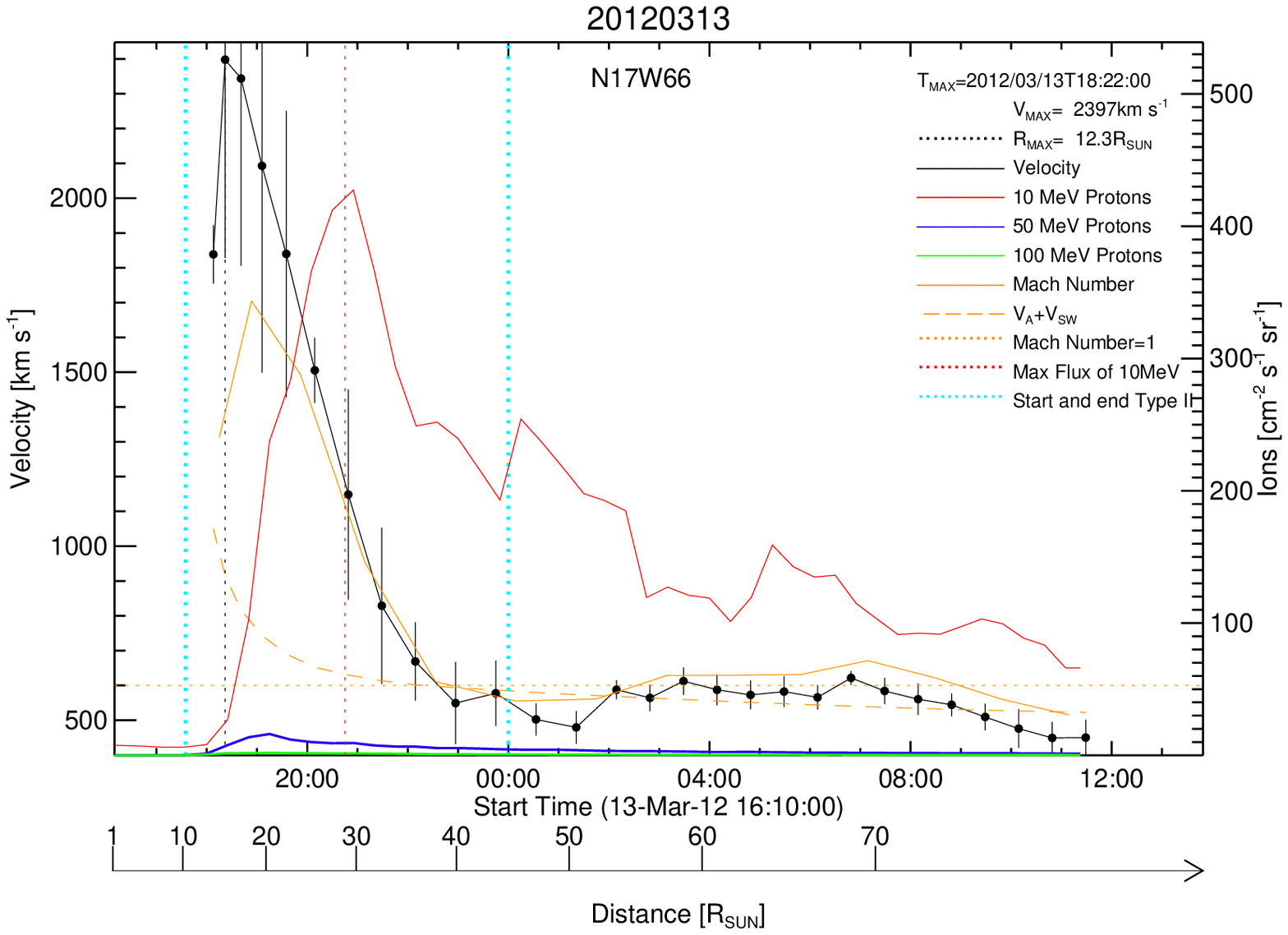}
   \includegraphics[width=9.1cm,height=6cm]{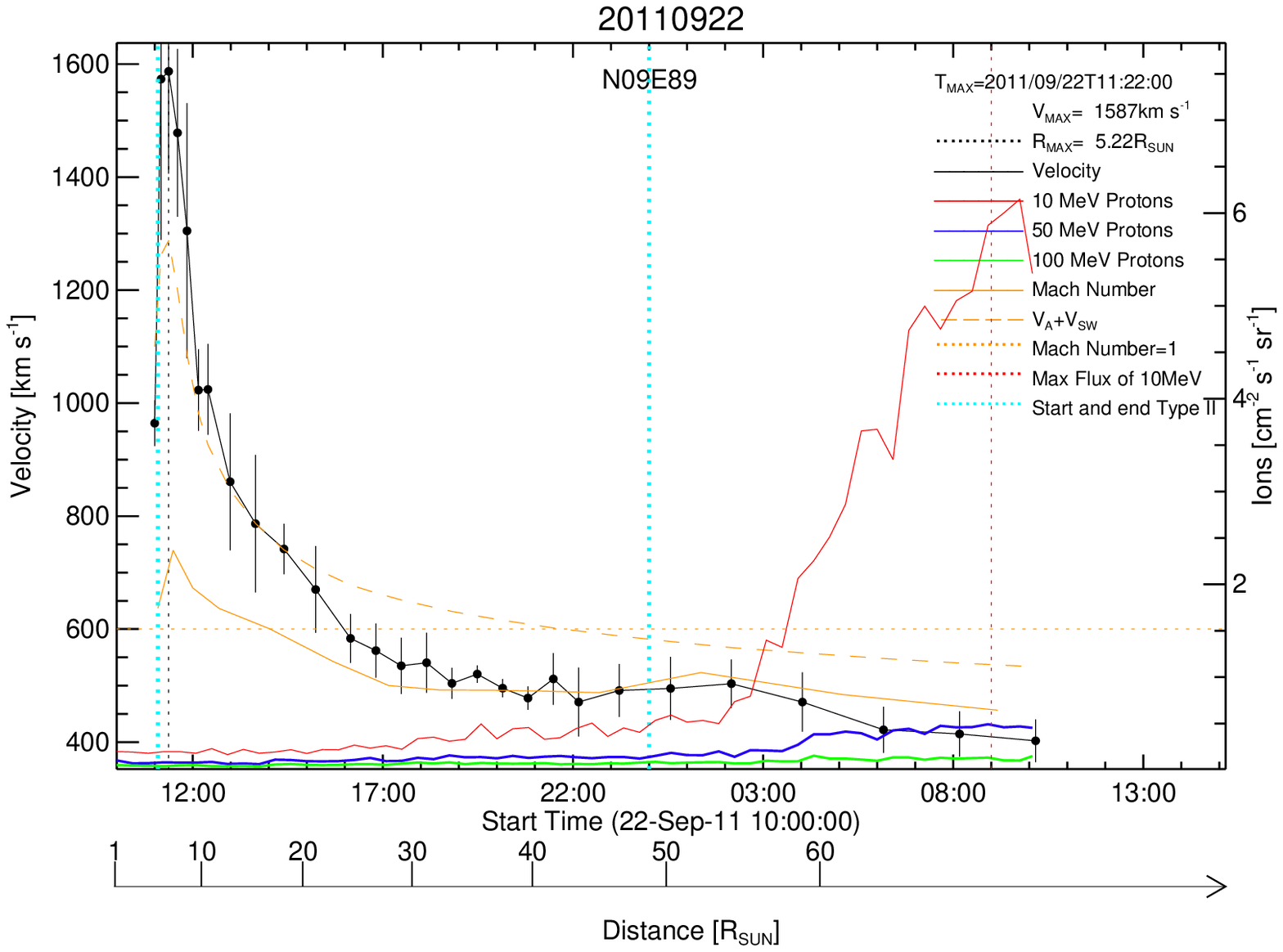}
   \caption{13 March 2012 event located at the west limb (left panel) and 22 September 2011 event located at the east limb (right panel). The plot shows the CME speed from STEREO (black line) with error bars,   SEP flux in the >10 MeV energy band (red line),   >50 MeV energy band (blue line), and   >100 MeV energy band (green line).  The sum of Alfv\'{e}n and solar wind speed [V$_{A}$ + V$_{SW}$] (dashed orange line) and the scaled Mach number (orange line) are shown. The start and end times of the associated type II burst are added (dotted cyan line). The CME maximum velocity [V$_{MAX}$] and   time [T$_{MAX}$] and the distance at CME peak velocity [R$_{MAX}$] (dotted black line) at V$_{MAX}$, SEP peak flux in the >10 MeV energy band (dotted red line), and the scaled Mach number=1 (dotted orange line) are shown in the figure.}
              \label{FigGam}%
    \end{figure*}

\subsection{Event selection}
In our study we concentrate only on CMEs generating SEP events. On further investigation the events were classified into halo (width=360$^{\circ}$) and partial-halo (width  >120$^{\circ}$) CMEs by the SOHO/LASCO CME catalog\footnote{cdaw.gsfc.nasa.gov/CME$\_$list} (\citealt{yashiro2004}, \citealt{gopalswamy2009a}).  These events are better observed from STEREO instruments during their quadrature configuration (\citealt{Bronarska2018}). We constricted the event sample to the period of ascending phase of solar cycle 24 (i.e., 2009–2013) as it marks the approximate quadrature configuration of STEREO. The twin spacecrafts STEREO-A and -B are at $\approx$90$^{\circ}$ separation with respect to the Earth. This position was chosen as it offered advantages in the accurate determination of plane-of-sky speeds, which are close to the true radial speed of halo CMEs with insignificant projection effects. The data from the STEREO/SECCHI coronagraphs COR1 and COR2, and the heliospheric imagers HI1 and HI2  can be obtained from the \emph{UK Solar System Data Centre (UKSSDC)\footnote{https://www.ukssdc.ac.uk/solar/stereo/data.html}} database;  they were used to perform manual measurements of height--time data points to determine the speed of CMEs. We focused our study on non-interacting CMEs as the velocities of interacting CMEs could be changed unpredictably along its propagation in the interplanetary medium. Each event in our sample was checked in the images from both of the twin satellites so that the measurements were done from the images that showed better quality. 

The SEPs associated with the CMEs were selected in the >10 MeV, >50 MeV, and >100 MeV energy bands. The events with flux value $\geq$1 pfu in the >10 MeV, $\geq$0.1 pfu in >50 MeV, and $\geq$0.05 pfu in >100 MeV energy bands were considered for the study as their proton flux is higher than the average background flux. The threshold value for >100 MeV is changed to $\geq$0.05 pfu from $\geq$0.1 pfu,   compared to the criteria taken in \citealt{Anitha2020a} paper, as we observe in our analysis differences in generation of SEP fluxes with respect to the location of event eruption. Hence, we  investigate these differences further in this article. The data from the SEM instrument on board the GOES-13 geostationary satellite recorded in the \emph{National Oceanic and Atmospheric Administration} (NOAA) database\footnote{https://satdat.ngdc.noaa.gov/sem/goes/data/avg/} was used to analyze the SEP fluxes. Source locations of CMEs were obtained from associated X-ray flares using the Hinode Flare Catalogue\footnote{https://hinode.isee.nagoya-u.ac.jp/flare$\_$catalogue/} (\citealt{watanabe}), and are shown in figure 1. The properties of the DH type II bursts that are signatures of these CME-driven shocks can be obtained from the WIND/Waves and STEREO database\footnote{https://cdaw.gsfc.nasa.gov/CME$\_$list/radio/waves$\_$type2.html} of  (\citealt{bougeret}).

During the quadrature configuration of STEREO (2009 - 2013) we found 61 SEP events with the above-mentioned criteria for their fluxes, but we could only analyze 38 among them due to limitations with their associated CMEs. We observed 15 interacting CMEs, 5 CMEs erupting on the backside of the Sun, and 3 CMEs that are too weak from which we could not obtain sufficient height--time data points for the analysis. Of the 38 events in the sample, 19 events (50\%) originate at the disk center (-20$^\circ$ < longitude < 45$^\circ$), 12 events (31\%)  at the west limb (longitude > 45$^\circ$), and 7 events (18\%)  at the east limb (longitude < -20$^\circ$). The majority of the observed events originate at the west limb and disk center, and only a few on the east limb. This is due to the magnetic connectivity of Sun with the Earth, which is explained in detail in the following subsections. It is worth noting that the presented sample of events are the complete list of non-interacting halo or partial halo CMEs that generate SEPs with the above-mentioned flux values in the three energy bands during the period 2009 - 2013, which also marks the ascending phase of solar cycle 24. A summary of these 38 events is given  in Table 6. The data presented in the table and explained in figure 2 are the basis of our study. They are explained in detail in the following sections.

The SEPs generated by the backside CMEs are of particular interest. \citealt{gopalswamy2020} demonstrated that backside CMEs can produce significant fluxes of energetic protons. In our research we found five such events. A thorough analysis of their associated active region on the Sun allowed us to determine the source location of the backside   CME responsible for the production of energetic  particles. Two of them (28 January 2011 and 23 July 2012) were located just over the edge of the solar disk, but the other three (21 March 2011, 04 June 2011, and 08 November 2012) were located as far as $\approx$30$^\circ$ behind the west limb of the Sun. This means that SEPs can be produced from sources located not only in the visible part of the Sun's disk, but even very far ($\approx$30$^\circ$) beyond the east (\citealt{gopalswamy2020}) and west (our study) limb of the disk. These results demonstrate the ability of backsided events to cause space weather effects at Earth, and therefore accurate predictions of SEPs will need to include such events.

\subsection{Method}

  The average speed determined using the linear fit method in the STEREO field of view is not a good indicator of CME kinematics as the speed varies significantly during its propagation in the interplanetary medium (\citealt{ravishankar}). Therefore, it is important to study other parameters that do not offer approximate relations. \citealt{Anitha2020a} presented the first set of results on comparative studies on average speeds and instantaneous speed to determine which of these parameters offers an accurate correlation with the peak fluxes of the associated SEPs. The results  show that instantaneous speeds such as the maximum speed and speed at SEP peak flux offer better correlation. In addition, the Mach number at CME maximum speed and the Mach number at SEP peak flux showed promising results. The comparative study of SOHO/\emph{Large Angle and Spectrometric Coronagraphs} (LASCO) and STEREO/SECCHI for CMEs   shows that the correlation obtained for STEREO is much higher as its quadrature configuraturion point of view helped with accurate measurements of true radial speed of halo events, whereas these halo events are subject to significant projection effects by SOHO/LASCO (\citealt{Bronarska2018}). In this paper we follow the same method, but unlike \citealt{Anitha2020a} the study   uses the full STEREO/SECCHI suite of instruments with completely new measurements of height--time data points. To obtain the instantaneous velocities we applied linear fits to five successive height--time points. By shifting the linear fits point by point through all of the  height--time points we acquired the instantaneous CME speeds. Practically, two successive  height--time points are sufficient to determine the speed, but as manual measurements are subject to unpredictable random errors, we used five successive points to obtain the most reasonable results. Details of this method are described by \citealt{Bronarska2018} in their article. If we have the instantaneous velocities of CMEs, we could determine the instantaneous Mach numbers and other interesting kinematic parameters of CMEs.

In figure 2 two CME events (13 March 2012 and 22 September 2011) and their associated parameters varying with time and distance are presented separately. These panels demonstrate the parameters considered in our study in relation to SEP fluxes. For comparison we present diagrams for a western disk event (left panel) and an eastern limb event (right panel). The figure shows the instantaneous CME speeds (with error bars) obtained from manual measurements of STEREO data (errors were obtained using the  bootstrap method (\citealt{Michalek17})). The values of the maximum velocity (V$_{MAX}$) and the time (T$_{MAX}$) and distance at (R$_{MAX}$) are shown on the right corner of the figures. The peak fluxes of the SEPs in the respective energy channels are determined. The start and end time of the type II burst represents the CME-driven shock, obtained from WIND/Waves and STEREO database, are shown as the two vertical dotted cyan lines, respectively.

%3
     \begin{figure*}[h!] 
   \centering
   \includegraphics[width=9.1cm,height=6cm]{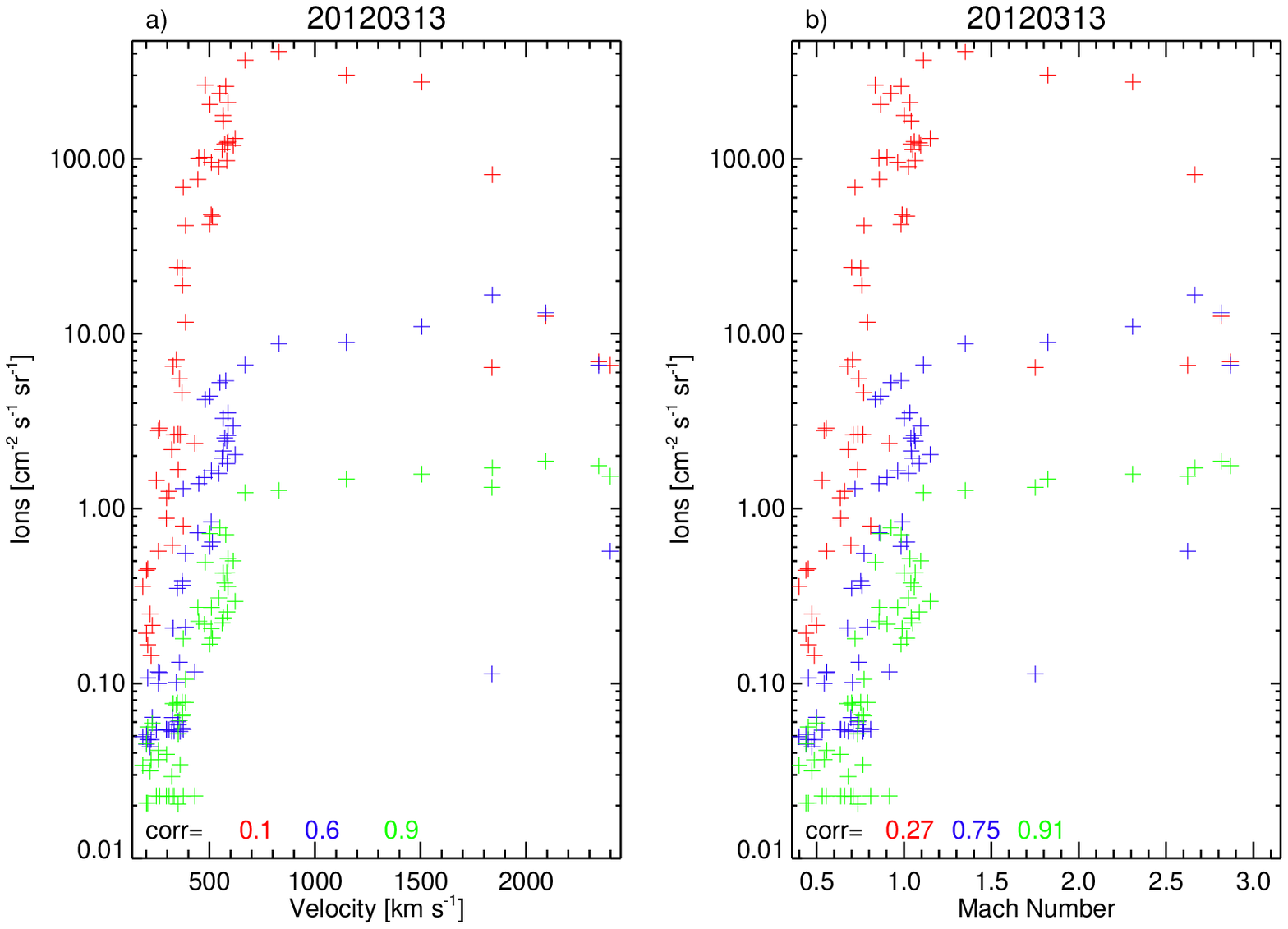}
   \includegraphics[width=9.1cm,height=6cm]{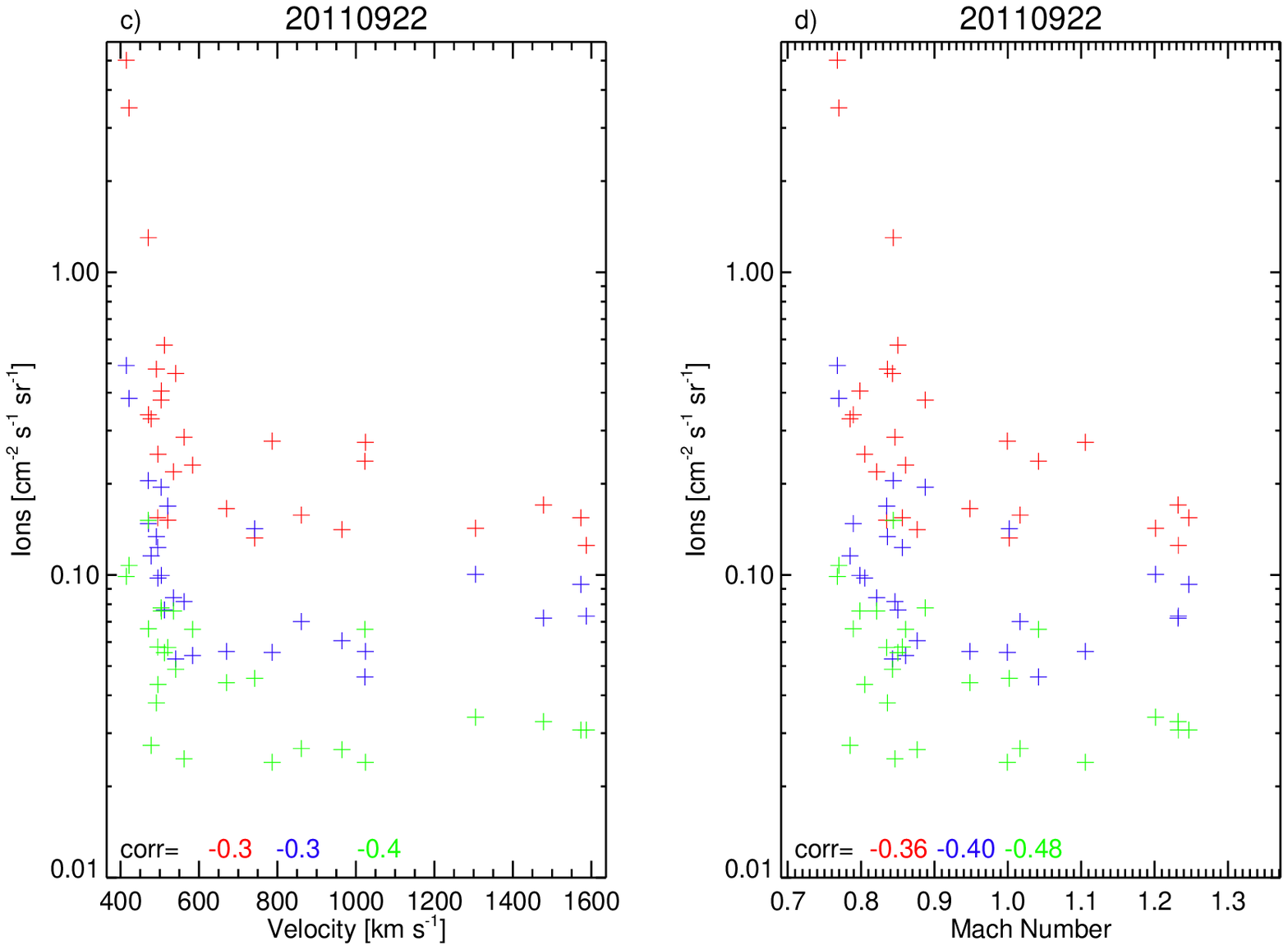}
   \caption{Plots showing correlation between instantaneous velocity and Mach number vs SEP flux for the 13 March 2012 event (panels a and b) and the 22 September 2011 event (panels c and d).}
              \label{FigGam}%
    \end{figure*}

The propagation times for accelerated protons to reach the Earth vary in the considered energy range. SEPs take 69 (>10 MeV), 31 (>50 MeV), and 22 (>100 MeV) minutes to reach the Earth. Their detection is formally delayed by about an hour compared to the observations carried out by coronagraphs as this means that the slowest protons arrive one hour later than light. The delay is about 10 minutes less because the peaks of SEPs are reached when the CMEs are at some distance from the Sun. This delay has been taken into account in figure 2 and in our considerations. However, for the consideration of the relationship between SEP flux peak and the maximum CME speed, this problem is completely negligible. This effect can only be relevant to the correct determination of the speed of CME at SEP peak. As can be seen in figure~2 (dotted red line), this speed is determined at some distance from the Sun, where its change is very slow. The CME, after reaching  maximum velocity, propagates at almost constant speed. In one hour the CME velocity can change by not more than 5$\%$. On the other hand, the error in determining the speed using a linear fit is about 15$\%$ (\citealt{Michalek17}). Therefore, we could neglect this effect in our study.

Mach number, which is the most important parameter considered in our study, mostly depends on the magnetic field and density of the plasma. We investigated two methods of calculating the magnetic field needed to determine the Alfv\'{e}n speed. The    \citet{dulk} method  determines the coronal magnetic field above active regions when the CMEs erupt, which is more common during the ascending phase of the solar cycle. The  \citet{leblanc}, \citet{Mann1999}, \citet{gopalswamy01}, and \citet{eselevich} method is applicable to determining the magnetic field for quiet regions, mainly related to prominence eruptions, usually during the descending phase of the solar cycle. Using these two methods and the plasma density model by \citet{leblanc}, we determined the Alfv\'{e}n speed for each considered event. Our analysis  shows that a much better prediction was obtained with the \citet{dulk} model as it matches the time frame chosen for study (i.e., ascending phase of solar cycle 24), thus in our further considerations we employed the Alfv\'{e}n speed obtained from this model. The solar wind speed was determined using the model presented by \citet{sheeley}. Having determined the Alfv\'{e}n speed (V$_{A}$) and solar wind speed (V$_{SW}$), along with the measured instantaneous CME speed (V$_{CME}$), we can simply determine the instantaneous Mach number (M$_{A}$):   M$_{A}$ = V$_{CME}$/(V$_{A}$ + V$_{SW}$). The estimated sum of V$_{A}$ + V$_{SW}$ is shown in both panels of figure 2 as dashed orange line and the Mach number (scaled by 600 for better visualization) is represented by the continuous orange line. The horizontal dotted orange line (at 600 km~s{$^{-1}$}) reflects the value of Mach number equal to 1. The significance of Mach number is explained in detail in section 3.5.

    The important factors that determine the peak intensities of these accelerated particles are the CME ejection speed and their magnetic connectivity with the Earth. West limb events (longitude=60$^{\circ}$) have the best connectivity, and this decreases as we move towards east and to the farther western limb (longitude >60$^{\circ}$). Due to this variation in connectivity along the solar disk, we observe delays in the time at which the SEPs reach maximum intensity with respect to the onset of the associated CMEs. For well-connected events, the SEPs reach peak fluxes quickly after the CME onset and maximum velocity from the Sun (see  left panel of figure 2) and the delay increases in the case of eastern events as the ejections must expand enough so that their fronts are well connected magnetically to the Earth (right panel of figure 2). Similar delays in both distance and time are observed between the maximum Mach number and SEP peak flux. Consequently, when the CME expands, its speed decreases. This means that when we observe the maximum intensities of energetic particles, especially for eastern events, the ejection speed may be much lower than their maximum value. The V$_A$ and V$_{SW}$ are much lower at these points as they decrease slowly with distance (r), but as V$_{CME}$ decreases signifcantly the observed Mach numbers are consequently lower as well. To be precise, we observe a Mach number <1 for all seven eastern limb events. This occurs due to projection effects. According to \citealt{Bronarska2018}, the real or space velocity should be V$_{INS}$ + 0.5V$_{INS}$, where V$_{INS}$ is the measured instantaneous velocity. As we observe the projected speeds, the Mach number obtained due to these speeds for limb events is less than 1. It could be also be a result of determination of V$_A$, and depends on the model of the magnetic field, which is not perfect.
    
     Interesting observations are made concerning the start and end times of the associated type II bursts shown by the vertical cyan lines in figure 2. In the left panel that shows the western event, we observe the onset of type II burst, CME velocity, Mach number, and SEPs approximately at the same time. However for the eastern event shown in right panel, the SEP peak in the >10 MeV channel is delayed by about 22 hours with respect to the CME onset. Among the seven eastern events in our sample, three   are not associated with type II bursts as seen in the database. The remaining four eastern events also display the delays   with respect to the CME onset. The type II bursts, which are signatures of CME-driven shocks, do not depend on the magnetic connectivity of the Sun and Earth; instead,    the SEP propagation is affected by the connectivity. For the eastern events the determined CME speed and Mach number are lower by 50\% (\citealt{Bronarska2018}) and we observe significant delays in the SEPs. The SEP fluxes for eastern events are among the lowest in our sample, and probably much higher fluxes of energetic particles were produced earlier that could not be detected due to poor connectivity. We observe that the type II bursts start when the Mach number reaches 1 and end when the Mach number goes back to 1 for all events. Therefore, type II bursts are good indicators of production of large SEP fluxes. These delays are discussed in detail in section 3.4. In addition, we observe the decline of the >10 MeV protons at the end of the type II bursts in the left panel, but in the right panel we see that the onset of the SEPs is well beyond the end of the radio bursts. This may be due to the threshold of observations of the instrument measuring the DH type II bursts. The bursts may have prolonged for a longer time, but in the reduced intensity, which the instrument was unable to measure.

%4
 \begin{figure*}[h!]
   \centering
   \includegraphics[width=9.1cm,height=6cm]{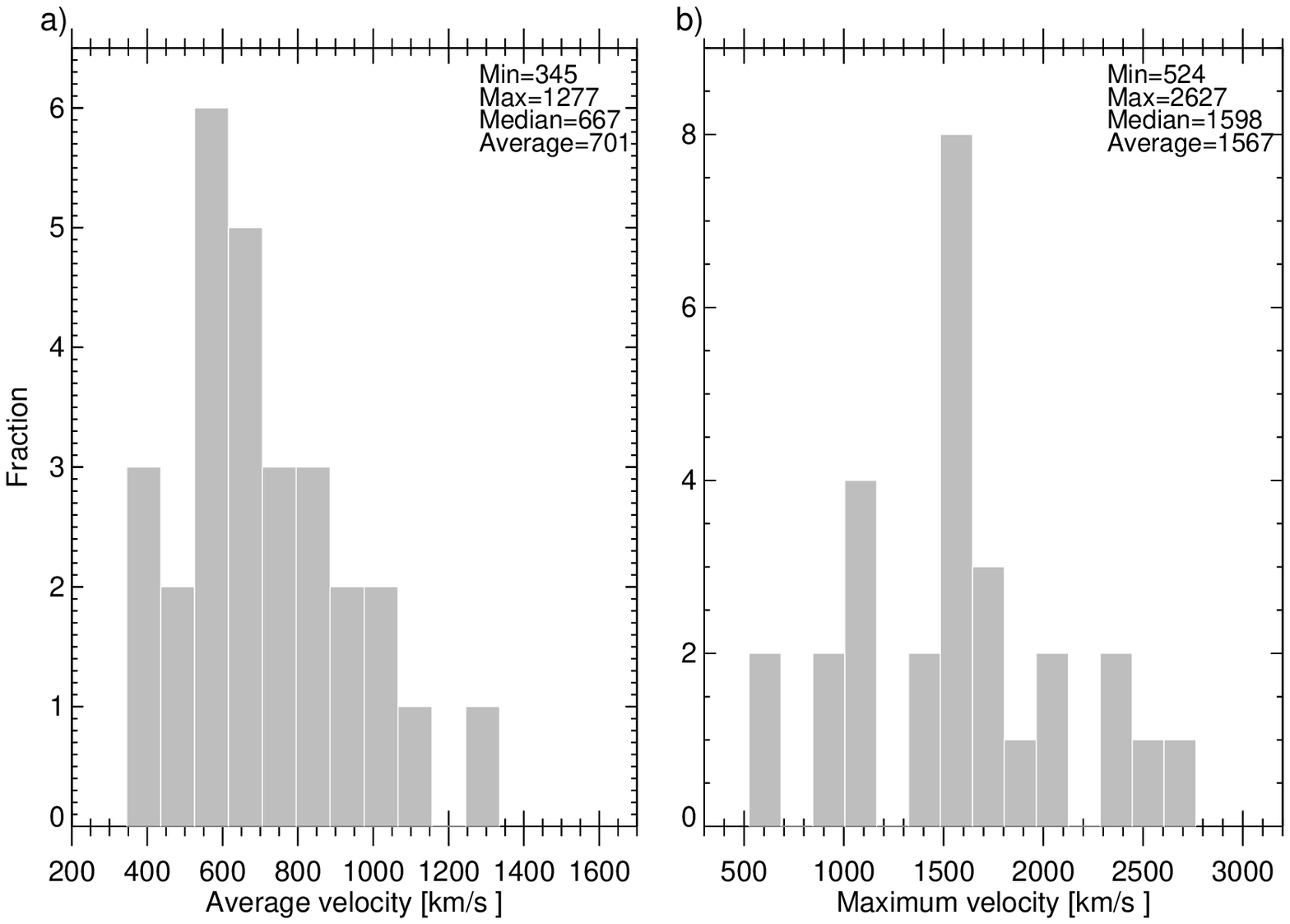}
   \includegraphics[width=9.1cm,height=6cm]{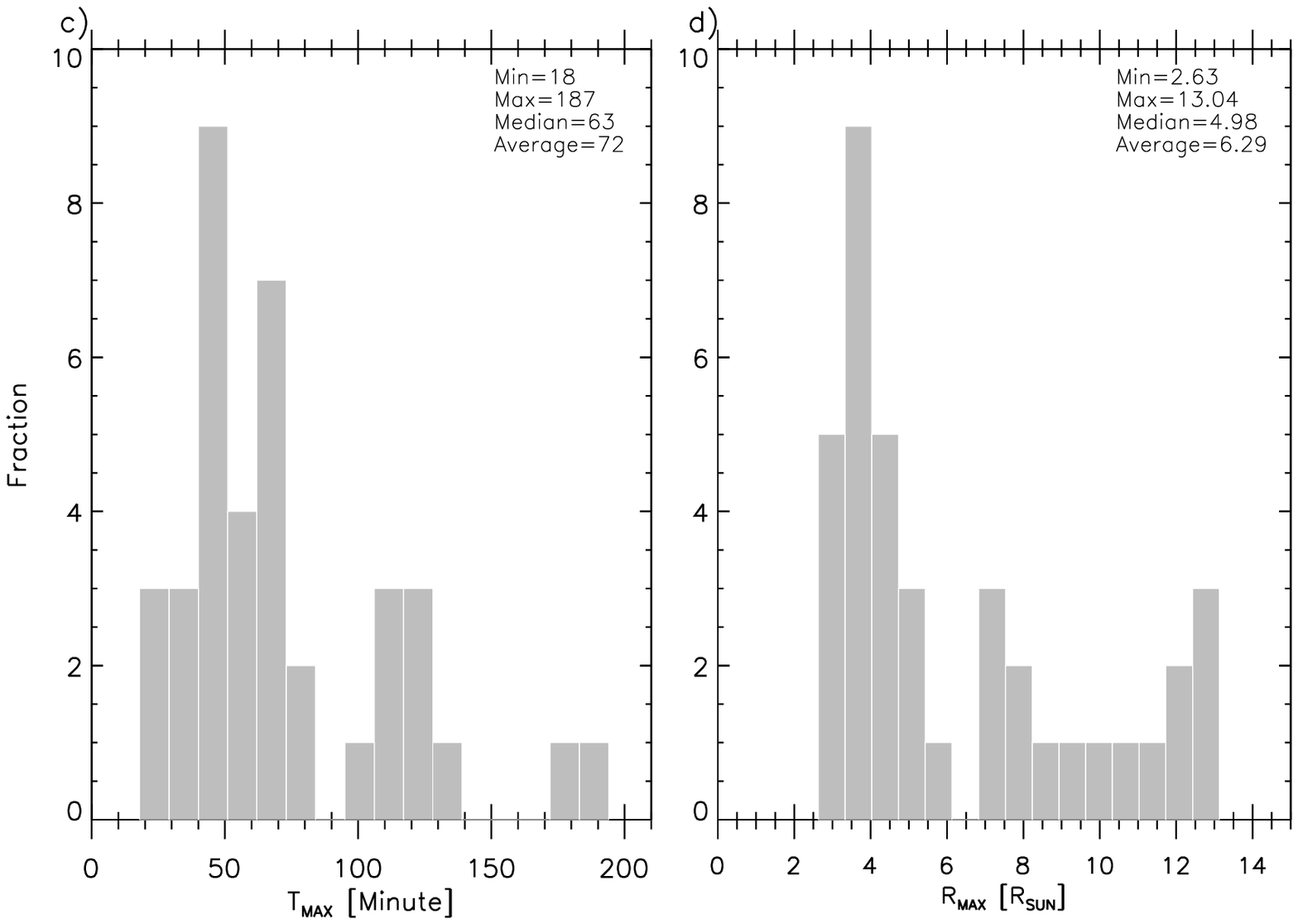}   
   \caption{Distributions of average (panel a) and maximum (panel b) velocity and the time (T$_{MAX}$, panel c) and distance (R$_{MAX}$, panel d) when CMEs reach maximum velocity.}
              \label{FigGam}%
    \end{figure*}

Figure 3 shows the correlation between instantaneous velocity and Mach number versus instantaneous fluxes of SEP in the three energy bands, >10 MeV, >50 MeV, and >100 MeV in red, blue, and green, respectively. Panels a and b represent the western disk event on 13 March 2012, and panels c and d represent the eastern limb event on 22 September 2011;   their correlations are shown in the bottom left of the figure for the respective energy bands. We observe perfect positive correlations in panels a and b, and anticorrelation in panels c and d. Furthermore, the correlation and anticorrelation tends to 1 and -1, respectively, for the higher energy bands   compared to the lower energy band for both the velocity and Mach number parameter with the SEP fluxes. These differences in correlation for the two events are observed due to varying delays caused by magnetic connectivity of the Sun and Earth with respect to the longitude, as explained earlier. These delays can be clearly seen in figure 2, and is observed for SEPs in all   three considered energy bands.  For  this reason we observe the western events showing positive correlation (panels a and b) and the eastern events showing anticorrelation (panels c and d).  Additionally, the correlation of the instantaneous Mach number with SEP fluxes, shown in panels b and d, offers improved correlation   compared to the instantaneous velocities. This shows that for studying the acceleration of particles driven by the CME the Mach number is the best parameter to consider. A detailed analysis of this is dicussed in section 3.5. It is important to note that significant correlations are observed for >100 MeV compared to >10 MeV SEP fluxes (panels a and b). This is due to the larger delay between the SEP and maximum velocity or maximum Mach number for >10 MeV compared to higher energetic particles. Therefore, few points from the initial phase of propagation decreases this correlation compared to higher energies. 

%5
\begin{figure*}[h!] 
   \centering
   \includegraphics[width=9.1cm,height=6cm]{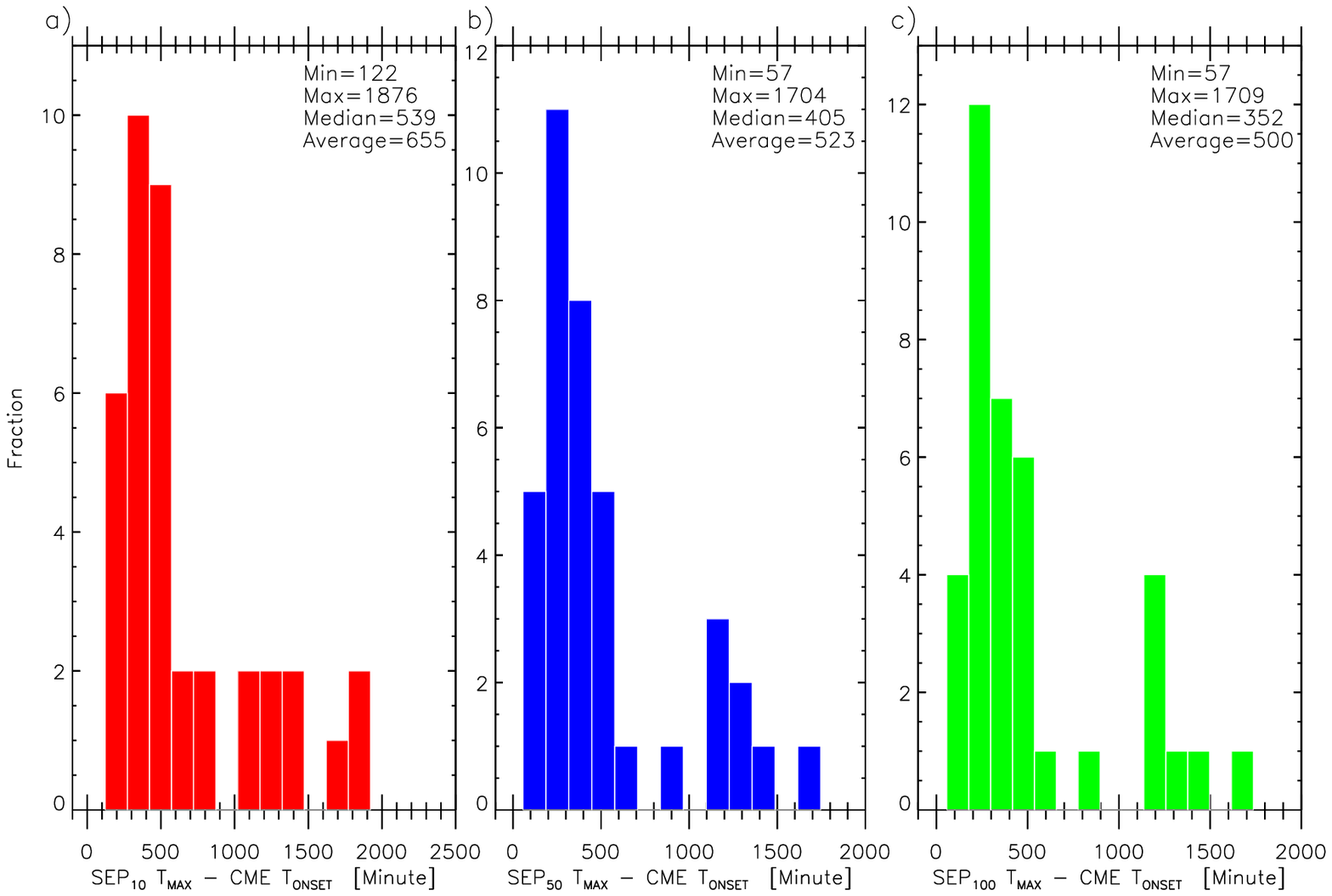}
   \includegraphics[width=9.1cm,height=6cm]{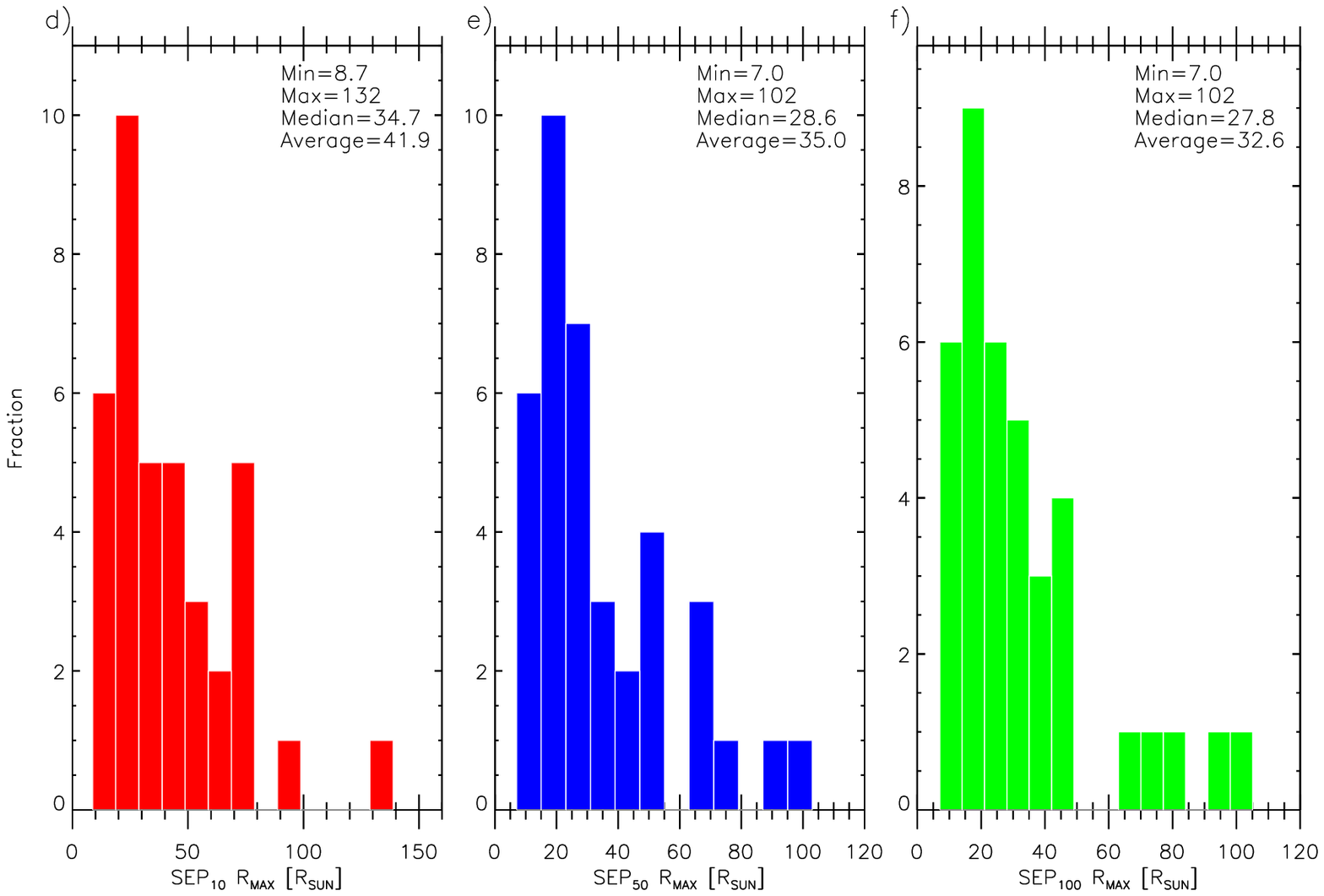}
   \caption{Distributions of time taken by SEPs to reach peak flux after the onset of the CME (panels a, b, and c) and distance at which the SEPs reach peak flux (panels c, d, and e) in three energy channels (>10 MeV [SEP$_{10}$ T$_{MAX}$, R$_{MAX}$], in red), (>50 MeV [SEP$_{50}$ T$_{MAX}$, R$_{MAX}$], in blue), and (>100 MeV [SEP$_{100}$ T$_{MAX}$, R$_{MAX}$], in green).}
              \label{FigGam}%
    \end{figure*}

 The results presented in  figures 3 and 4 are the basis of our study, and  are summarized in Table~6. In columns 2-6 we have date and time, average velocity (V$_{AVG}$), maximum velocity (V$_{MAX}$), distance (R$_{MAX}$), and time (T$_{MAX}$) at V$_{MAX}$ of a given CME taken from the STEREO/SECCHI observations. Columns 7-9 show the peak SEP fluxes in the three energy channels. The next three columns present CME speeds at peak SEP fluxes for these energy channels. Columns 13-15 give the maximum Mach number (M$_{MAX}$) and distance (MR$_{MAX}$) and time (MT$_{MAX}$) at M$_{MAX}$ of a given CME. Columns 16-18 show the Mach number at maximum SEP peak flux in the three energy channels. The location of solar flares associated with CMEs from the GOES data is shown in column 19. The start and end times of the associated type II burst are shown in the last two columns.

For all the  considered correlation coefficients in the study we tested their significance. These tests confirmed (with a significance level of p=0.05) that there is a significant linear relationship between the considered parameters of CMEs and SEPs except for events originating at eastern longitudes. We also performed statistical test determining the significance of the difference between a pair of correlation coefficients. These tests confirmed that (with a significance level of p=0.95) the respective correlation coefficients are not significantly different from each other. The statistical values are  shown in Tables 1, 2, 3, 4, and 5.

\section{Analysis and results}

Our article focuses on recognizing how different kinematic parameters of CME affect the generation of SEP events. In our study we use the STEREO instruments; they allow us to track CMEs to very long distances from the Sun, which is important because the SEPs are generated by the shocks driven by CMEs up to the orbit of the Earth and beyond. The results of the study are presented in following sections.

\subsection{Basic speeds of CMEs generating SEPs}

The CME speeds can be determined in different ways. A linear fit of the height--time measurements can be useful for determining an average CME speed, but will fail to capture the significant changes in velocity that can occur during CME expansion; therefore, in our considerations we use the speeds determined in our new approach. These speeds were described in section 2.2. In figure 4 we present histograms comparing the values of average and maximum velocities in panels a and b, respectively. The considered CMEs in our sample have an average speed range of 345 to 1277 km~s{$^{-1}$} and a maximum speed range of 524 to 2627 km~s{$^{-1}$}. On average, the maximum speeds in the STEREO field of view are 223\% larger than the  speed obtained from linear fits to all height--time points (average speeds). This clearly shows that the average speeds in the STEREO field of view are not practical for studies. Panels c and d of figure 4 show the time (T$_{MAX}$) and distance (R$_{MAX}$) when CMEs reach maximum velocity, respectively. The CMEs in our sample have a range of 18 to 187 minutes at a distance in the range 2.63--13.04 R$_{sun}$ to reach maximum velocity. Therefore, on average, they take about 68 minutes at 6.38 R$_{sun}$ distance to reach maximum velocity. Instead, the results obtained in \citealt{Anitha2020a} with 25 events show that the CMEs, on average, take about 60 minutes at 7.61 R$_{sun}$ distance to reach maximum velocity. It is worth noting that the CMEs achieve maximum velocity very close to the Sun, but SEP peak fluxes could be observed when CMEs have propagated very far away from the Sun. This discrepancy is described in the next subsection. Thus, for the purposes of SEP generation, determining speeds closer to the Sun are more important for predicting what is seen at the Earth.

%6
 \begin{figure*}[h!] 
   \centering
   \includegraphics[width=9.1cm,height=6cm]{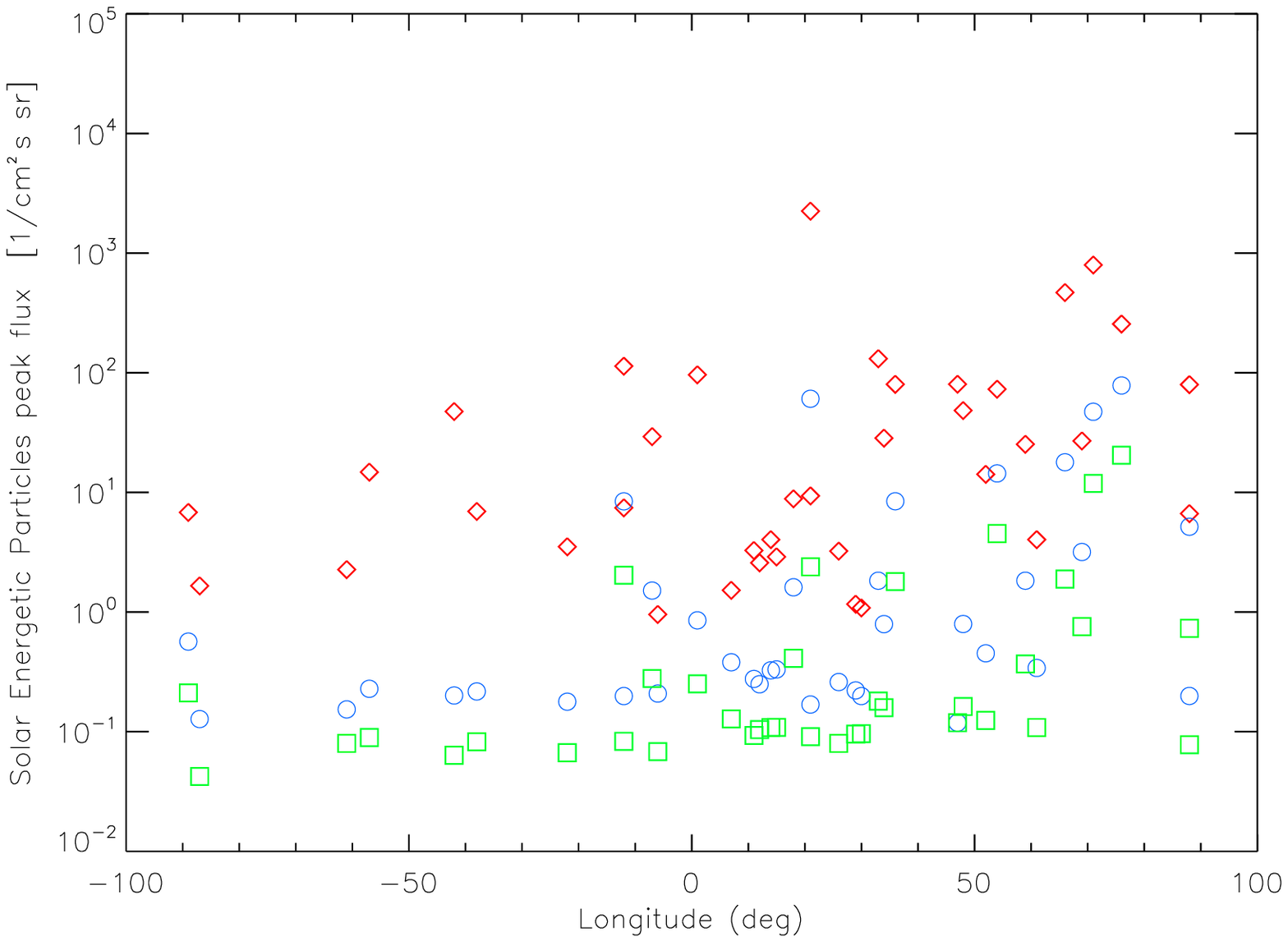}
   \includegraphics[width=9.1cm,height=6cm]{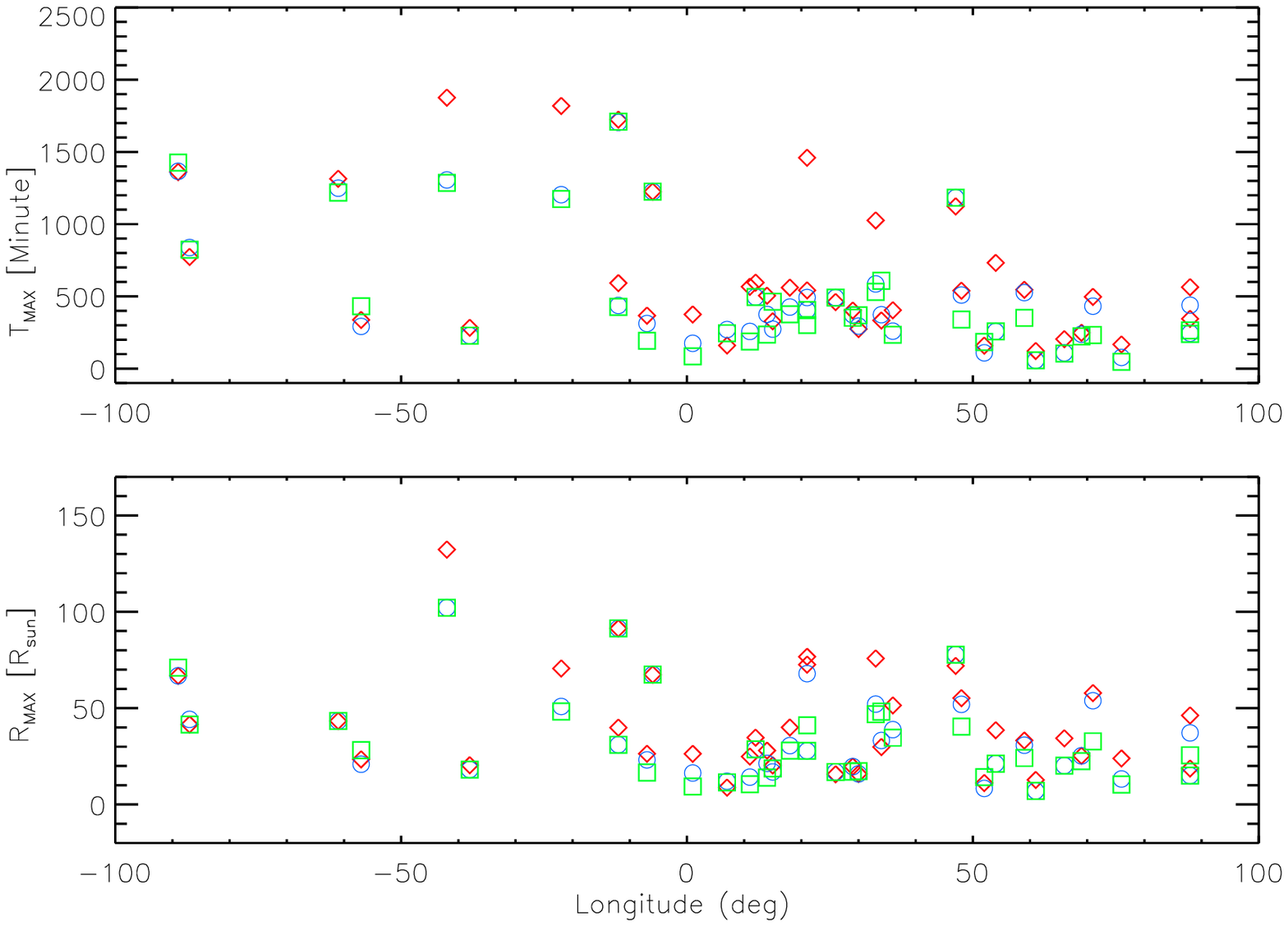}
   \caption{Scatter plot of longitude of the solar flare associated with the respective CMEs vs SEP peak flux (left panel) and the time and distance at which SEPs reach peak flux (right panels). Colors are assigned to the SEPs in the energy channels: >10 MeV (red), >50 MeV (blue), and >100 MeV (green).}
              \label{FigGam}%
    \end{figure*}

 A comparison of the errors determined from the standard deviation for average values of T$_{MAX}$ and R$_{MAX}$ for the samples considered in our papers show <T$_{MAX}$>=72.04$\pm$6.66 and <R$_{MAX}$>=6.29$\pm$0.53 for the current paper comprising 38 events, and <T$_{MAX}$>=60.50$\pm$12.71 and <R$_{MAX}$>=7.61$\pm$1.18 for \citet{Anitha2020a}  comprising 25 events. The errors in the present study are within the error range of the previous study. The errors are higher for the latter for both the considered parameters, which may be due to inaccuracies in measurements. Hence, our completely new height--time measurements for CMEs in our current sample are much more precise.

\subsection{The CME onset, SEP peak fluxes, and their source location}

The relationship between the CME onset and the SEP peak fluxes are shown in figure 5. Panels a, b, and c show the time taken by the SEPs to reach peak flux after the onset of the CME and panels d, e, and f show the distance at which the SEP peak fluxes are observed.  On average, all the events in our considered sample take 655, 523, and 500 minutes and at a distance of about 41.9, 35.0, and 32.6 R$_{sun}$ in the >10, >50, and >100 MeV energy bands, respectively, to reach peak fluxes. The delay in both time and distance is less for events located in the western limb, and these delays increase as we move towards the eastern limb (\citealt{Dalla2017a}; \citealt{Dalla2017b}). Specifically, the western events, on average, take 437, 348, and 290 minutes and at a distance of about 35, 30, and 25 R$_{sun}$ in the >10, >50, and >100 MeV energy bands, respectively, to reach peak fluxes. Disk events, on average, take 626, 485, and 470 minutes and at a distance of about 40, 32, and 30 R$_{sun}$ in the >10, >50, and >100 MeV energy bands, respectively, to reach peak fluxes. The eastern events, on average, take 1109, 926, and 941 minutes and at a distance of about 56, 49, and 50 R$_{sun}$ in the >10, >50, and >100 MeV energy bands, respectively, to reach peak fluxes. Similarly, in the left panel of Figure 6 we see the variation in SEP peak flux along its location on the heliographic longitude of the Sun. The longitude can be divided into disk center (-20$^\circ$ < longitude < 45$^\circ$), west limb (longitude > 45$^\circ$), and east limb (longitude < -20$^\circ$). A modest trend is observed in the plot where the higher intensity SEPs seem to originate in the west limb and this intensity gradually reduces as we move towards the east limb. This gradual decrease in peak intensity is seen clearly for >10 MeV protons. The trend similar to >10 MeV protons is also seen for >50 and >100 MeV protons at longitudes greater than 0$^\circ$, but at longitudes less than 0$^\circ$; their peak fluxes are  constant and lie in the range 0.1-1 cm{$^{-2}$} s{$^{-1}$} sr{$^{-1}$} (i.e., moving towards the east). These are a consequence of the magnetic connectivity of Sun and Earth (i.e., Parker spiral IMF) (\citealt{Marsh2013}). West limb events are well connected to the Earth, and this connectivity decreases as we move towards the east limb. In addition, \citealt{Dalla2017a} and \citealt{Dalla2017b} have pointed out that the SEP propagation is affected, also due to the drifts caused by the gradient and curvation of the Parker spiral IMF, with their importance increasing with the energy of the particle. The peak flux of >50 and >100 MeV particles occur much before the instance when the associated shock gets connected to the Earth. For  this reason, though we observe these fluxes, they are much reduced and the instruments missed detecting the maximum fluxes. Hence, the maximum  SEP fluxes from the west longitudes can be accurately measured in situ, while those from  east longitudes cannot.

Additionally, in figure 5 we observe the delay in both time and distance decrease with increasing energy bands of protons; in other words,  >100 MeV protons take less time and distance to reach peak flux and >10 MeV protons take more time and are observed at farther distances away from the Sun. The same delays are represented in figure 6 (right panels) varying with longitude. An explanation provided by \citealt{reames2020} tells us that the distinction between impulsive and gradual SEP events becomes unclear as the CME-driven shock waves can reaccelerate the impulsive ions pre-accelerated in the magnetic reconnection. The impulsive suprathermal seed ions are preferentially accelerated by shock waves at active regions, and can even be dominated and reaccelerated SEPs produced by weaker shocks (\citealt{Desai2003}; \citealt{Tylka2005}; \citealt{Tylka2006}). Thus, the higher energy ions we observe are most likely pre-accelerated in the magnetic reconnection close to the Sun and are further seeded into reacceleration by CME-driven shocks at farther distances from Sun. For this reason, we observe the >50 and >100 MeV protons reaching peak fluxes much earlier than >10 MeV protons.

%7
\begin{figure*}
\begin{minipage}{22cm}
\includegraphics[width=6.3cm,height=6cm]{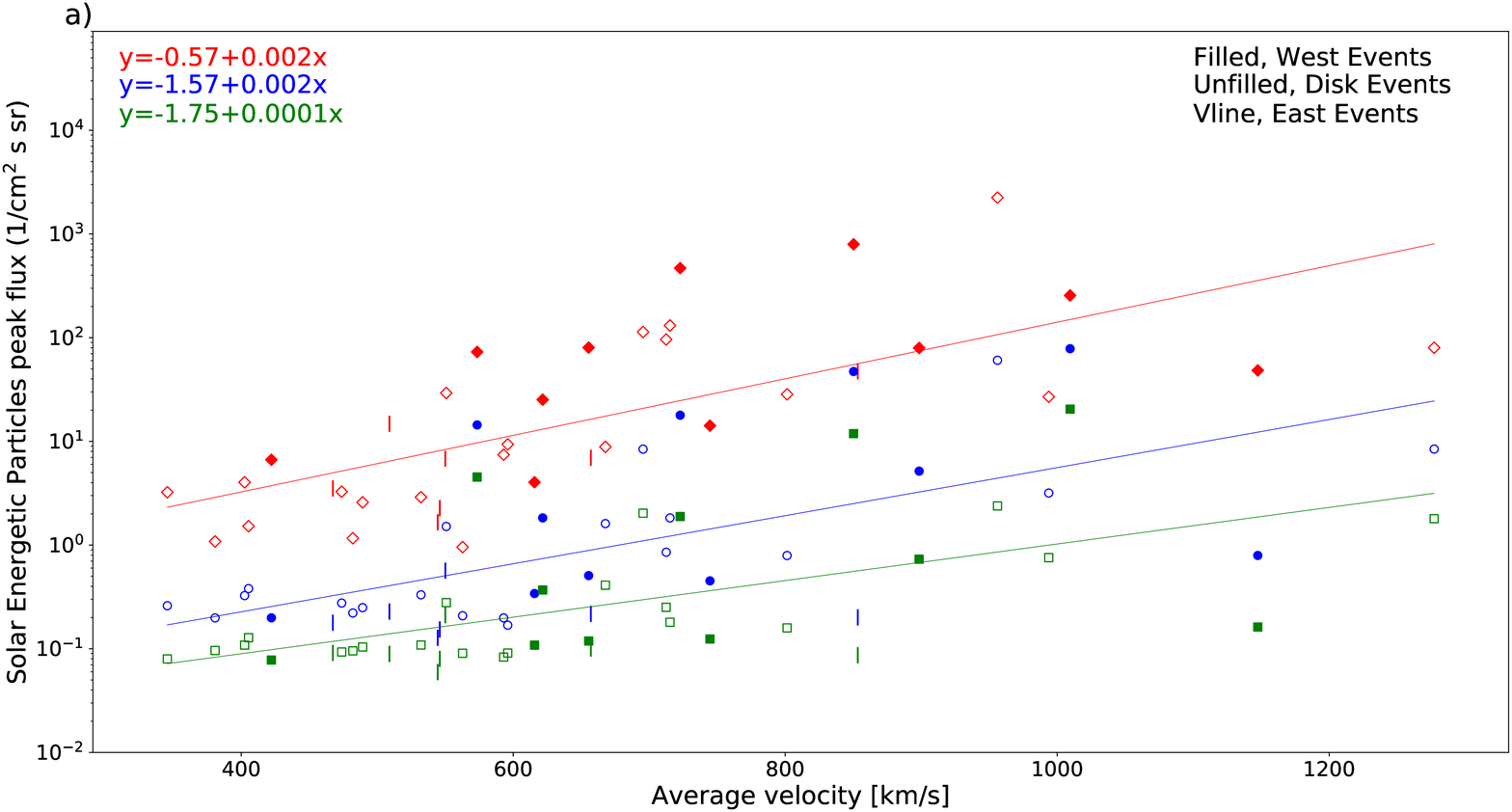}
\includegraphics[width=6.3cm,height=6cm]{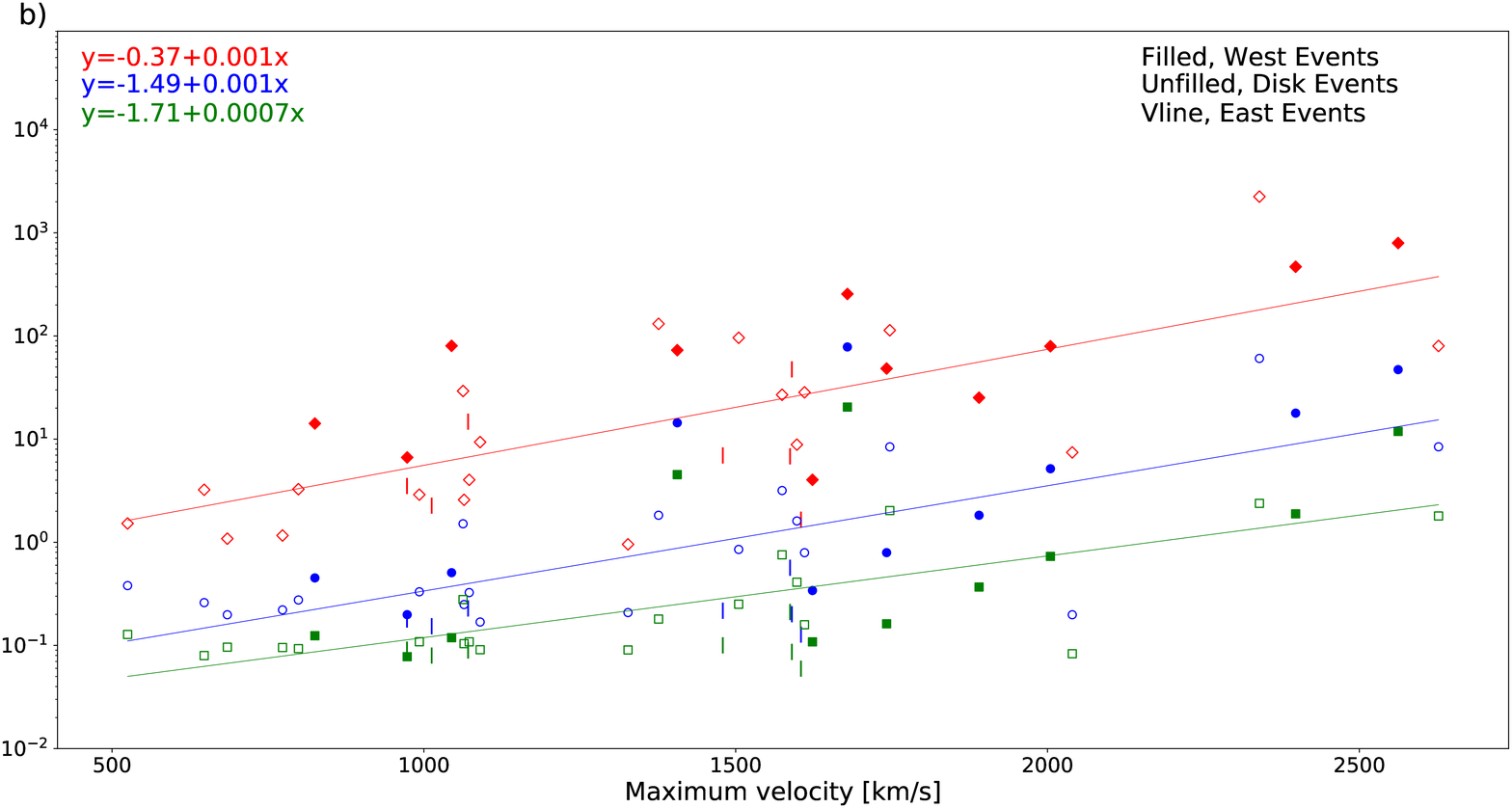}
\includegraphics[width=6.3cm,height=6cm]{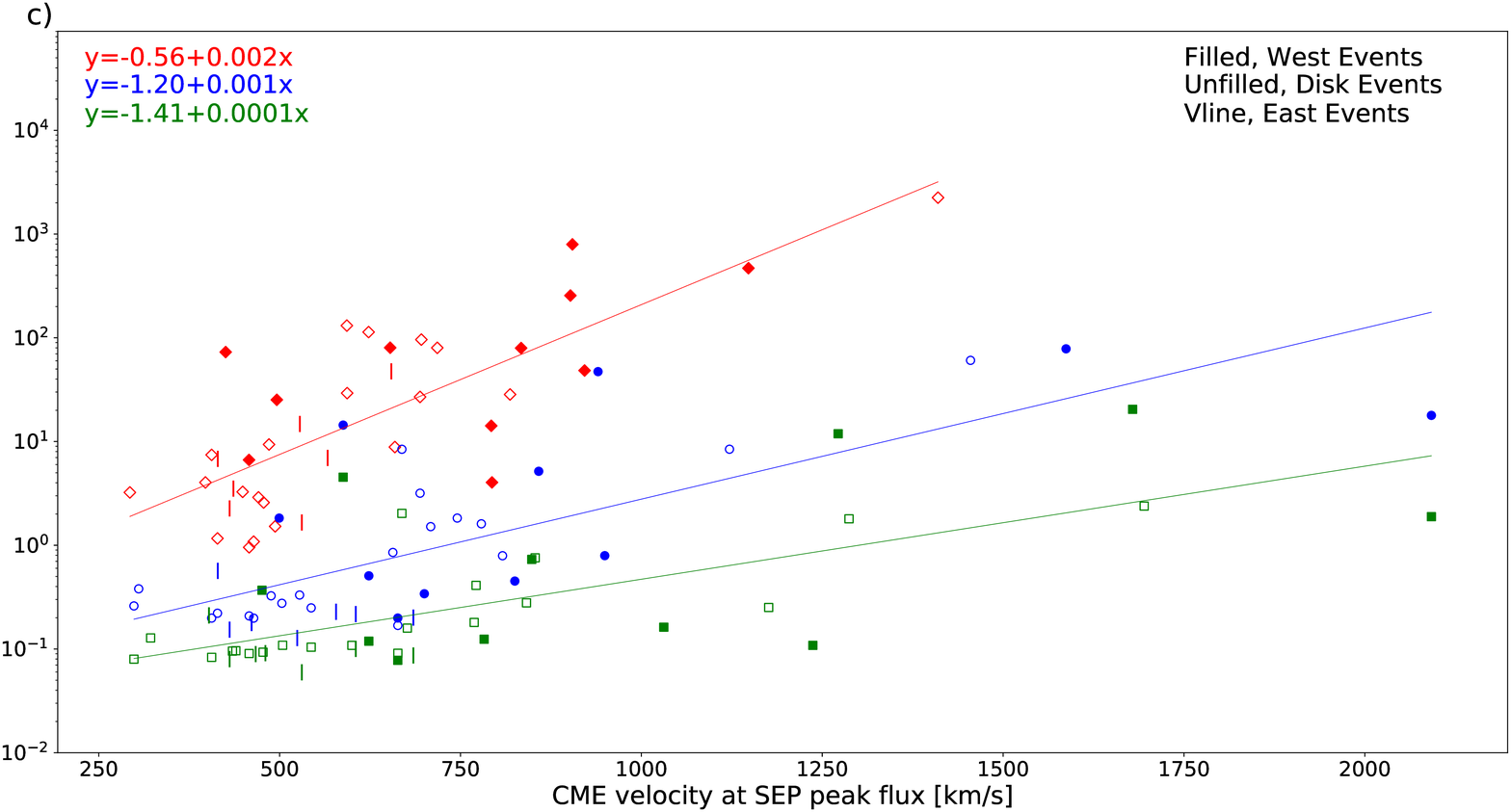}
\end{minipage}
\caption{Scatter plots of the average velocity (panel a), maximum velocity (panel b), and the CME velocity at SEP peak flux (panel c) vs SEP peak flux in the >10 MeV (red), >50 MeV (blue), and >100 MeV (green) energy channels. The open symbols represent disk events (longitude -20 < L < 45) and the filled symbols represent west events (longitude > 45) and vertical lines represent east events (longitude < -20).}
\end{figure*}

%t1
 \begin{table*}[!h]
\caption{Correlation coefficients of CME velocities vs SEP peak flux and their probabilities (significance at p-value<0.05), as shown in Figure 7. The number of events in each subdivision of the longitude is given in parentheses.}\label{YSOtable}
\centering
\tiny
\begingroup
\setlength{\tabcolsep}{1pt} % Default value: 6pt
\renewcommand{\arraystretch}{1.4} % Default value: 1

\begin{tabular}{|c |c| c| c| c| c| c| c| c| c| c| c|}
\hline
\hline
\multirow{3}{*}{\begin{tabular}[c]{@{}l@{}}\\\\a) Average velocity\end{tabular}}  & \multicolumn{1}{c|}{Energy channel} & \begin{tabular}[c]{@{}l@{}}All events \\      (38)\end{tabular} & \multicolumn{1}{c|}{ p-value } & \begin{tabular}[c]{@{}l@{}}Disk+West \\      (31)\end{tabular} & \multicolumn{1}{c|}{ p-value } & \begin{tabular}[c]{@{}l@{}}Disk+East \\      (26)\end{tabular} &\multicolumn{1}{c|}{ p-value } & \begin{tabular}[c]{@{}l@{}}Disk events \\      (19)\end{tabular}& \multicolumn{1}{c|}{ p-value } & \begin{tabular}[c]{@{}l@{}}East events \\      (7)\end{tabular} & \multicolumn{1}{c|}{ p-value }  \\

\hline
  & >10 MeV  &  0.69 & .00001  & 0.68 & .000026 & 0.74 & .000016 & 0.74 & .000292 & 0.71 & .073861\\
  & >50 MeV  &  0.61 & .000048 & 0.63 & .000146 & 0.67 & .000181 & 0.76 & .000159 & 0.0014 & .997623\\
 &  >100 MeV &  0.57 & .000187 & 0.59 & .000477 & 0.66 & .000244 & 0.76 & .000159 & 0.14 & .764651\\
  
\hline
\multirow{3}{*}{b) Maximum velocity} & \multicolumn{1}{c|}{} & \multicolumn{1}{c|}{} & \multicolumn{1}{c|}{} & \multicolumn{1}{c|}{} & \multicolumn{1}{c|}{} & \multicolumn{1}{c|}{}  & \multicolumn{1}{c|}{} & \multicolumn{1}{c|}{} & \multicolumn{1}{c|}{} & \multicolumn{1}{c|}{} & \multicolumn{1}{c|}{}  \\

  & >10 MeV  &  0.70 & .00001  & 0.74 & .00001 & 0.68 & .000133 & 0.73 & .000388 & 0.24 & .604195\\
  & >50 MeV  &  0.68 & .00001  & 0.72 & .00001 & 0.64 & .00043  & 0.72 & .000509 & 0.29 & .528119\\
 &  >100 MeV &  0.63 & .000023 & 0.67 & .000037 & 0.63 & .000562 & 0.73 & .000388 & 0.06 & .898324\\
 
 \hline
 \multirow{3}{*}{\begin{tabular}[c]{@{}l@{}}c) CME velocity\\at SEP peak flux\end{tabular}} & \multicolumn{1}{c|}{} & \multicolumn{1}{c|}{} & \multicolumn{1}{c|}{} & \multicolumn{1}{c|}{} & \multicolumn{1}{c|}{} & \multicolumn{1}{c|}{}  & \multicolumn{1}{c|}{} & \multicolumn{1}{c|}{} & \multicolumn{1}{c|}{} & \multicolumn{1}{c|}{} & \multicolumn{1}{c|}{}  \\

  & >10 MeV  &  0.77 & .00001 & 0.77  & .00001 & 0.81 & .00001 & 0.82 & .00001 & 0.66 & .106682\\
  & >50 MeV  &  0.73 & .00001 & 0.72  & .00001 & 0.83 & .00001 & 0.87 & .00001 & 0.24 & .604195\\
 &  >100 MeV &  0.71 & .00001 & 0.70  & .00001 & 0.75 & .00001 & 0.79 &  .000057 & 0.52 & .231562\\
\hline\hline

\end{tabular}
\endgroup
\end{table*}

\subsection{CME velocities versus SEP peak flux}

With the above introduction to the fundamental properties of the CMEs and the associated SEPs, we proceed to the main aim of the article, which is  to investigate the best velocity parameter to study the acceleration of SEPs to their peak intensities. We  chose the average, the maximum, and the CME velocity at SEP peak flux to analyze their correlation with SEP peak fluxes, as shown respectively in panels a, b, and c of figure 7. Linear fits are suitable for all the scatter plots, and their formulae are shown in the left corner of the figures. In addition to dividing the sample into disk center, west, and east limb events, we also classify the sample into disk+west, disk+east, and  disk-only events to study the variation of their correlations. The correlation coefficients and their probability values (significance at p-value<0.05) are given in Table 1. Panel a displays a modest correlation (0.69, 0.61, and 0.57 for the >10, >50, and >100 MeV energy channels, respectively, for all events) and the least correlation  compared to panels b and c. As explained in detail in the first paragraph of section 3.1, the average velocity may not be the best parameter to consider for correlation studies (\citealt{ravishankar}, \citealt{Anitha2020a}). Therefore, this leads to utilizing  instantaneous velocities for accurate correlation studies and also to studying SEP peak fluxes that are attained at farther distances away from the Sun.

Figure 7, panel b, shows a good correlation (0.70, 0.68, and 0.63 for the >10, >50, and >100 MeV energy channels, respectively, for all events) between CME maximum velocity and SEP peak intensities. This parameter is good, but not the best to use while studying a sample comprising events originating at all locations on the Sun. The west and disk events can be studied well with maximum velocity as they have good magnetic connectivity to the Earth, but the CMEs originating in the eastern longitudes are poorly connected to Earth, which leads to a delay in SEP peak flux with respect to CME maximum speed (as shown in figure 2). Therefore, a better approach to studying all CMEs, irrespective of their location, is to use CME velocity at the SEP peak flux. The fluxes of energetic particles are produced during the entire CME passage to the Earth, so it is also important to determine their velocities during the same distance, if possible. Investigation of this parameter has improved the correlation as shown in panel c (0.77, 0.73, and 0.71 for the >10, >50, and >100 MeV energy channels, respectively, for all events).

     %8
\begin{figure*}[h!] 
   \centering
   \includegraphics[width=9.1cm,height=6cm]{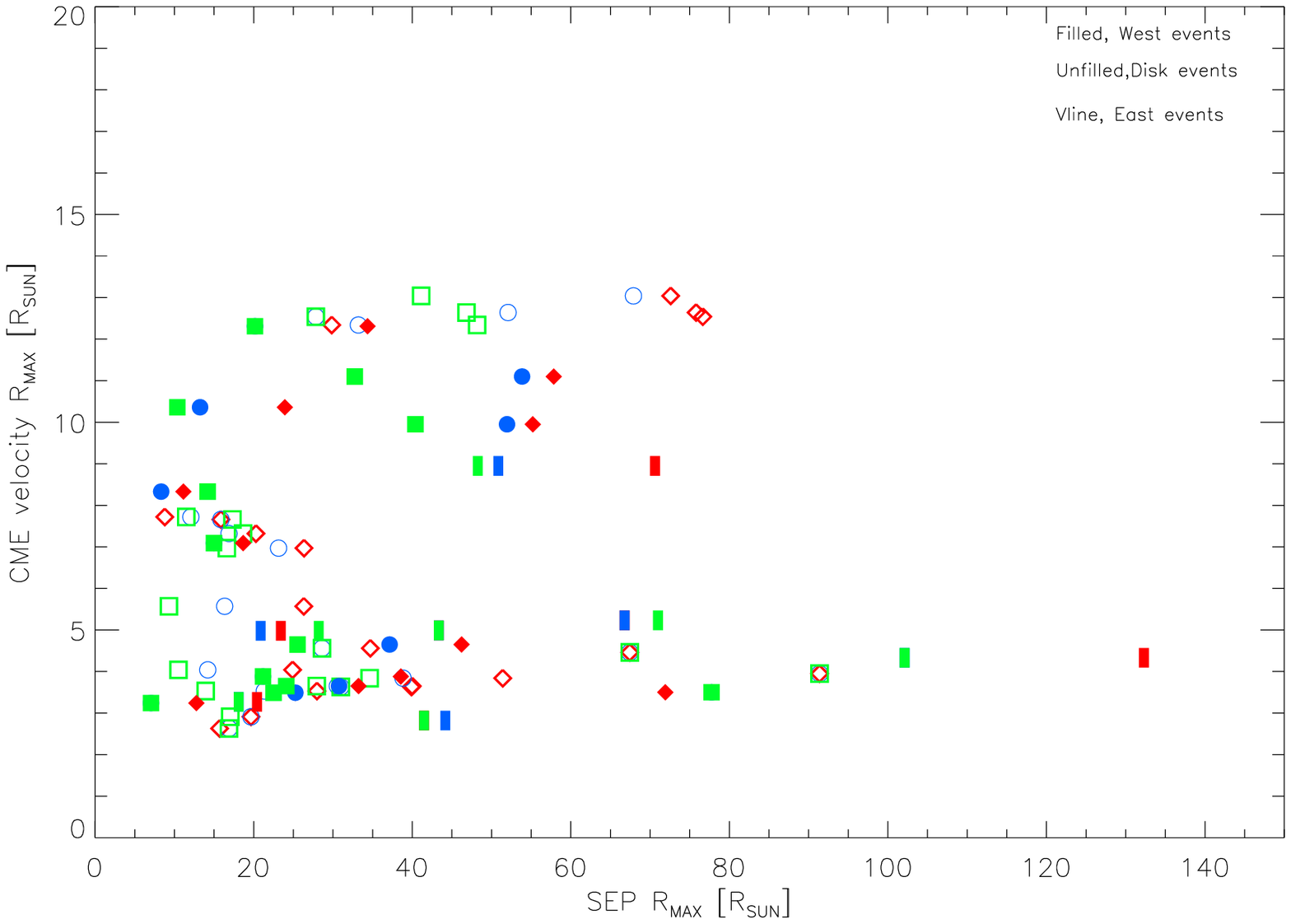}
   \includegraphics[width=9.1cm,height=6cm]{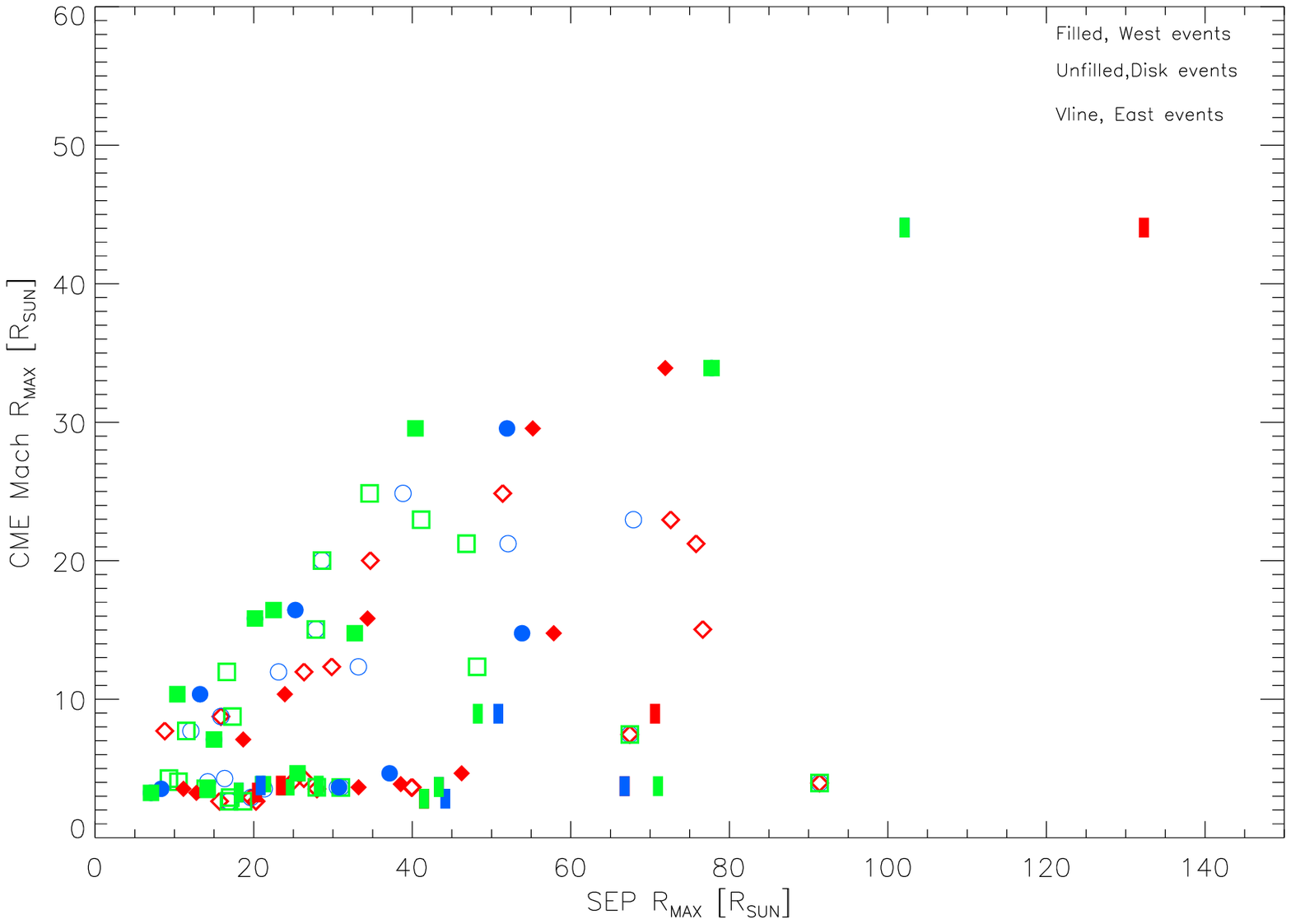}
   \caption{Scatter plots showing distance at CME maximum velocity (left panel) and distance at CME maximum Mach number (right panel) vs distance at SEP peak flux in the >10 MeV (red), >50 MeV (blue), and >100 MeV (green) energy channels.}
              \label{FigGam}%
    \end{figure*}

%t2
 \begin{table*}[!h]
\caption{Correlation coefficients of parameters and their probabilities (significance at p-value<0.05), as shown in Figure 8.  The number of events in each subdivision of the longitude are given in parentheses.}\label{YSOtable}
\centering
\tiny
\begingroup
\setlength{\tabcolsep}{1pt} % Default value: 6pt
\renewcommand{\arraystretch}{1.4} % Default value: 1

\begin{tabular}{|c |c| c| c| c| c| c| c| c| c| c| c|}
\hline
\hline
\multirow{3}{*}{\begin{tabular}[c]{@{}l@{}}\\\\SEP R$_{MAX}$ vs\\ CME velocity R$_{MAX}$\end{tabular}}  & \multicolumn{1}{c|}{Energy channel} & \begin{tabular}[c]{@{}l@{}}All events \\      (38)\end{tabular} & \multicolumn{1}{c|}{ p-value } & \begin{tabular}[c]{@{}l@{}}Disk+West \\      (31)\end{tabular} & \multicolumn{1}{c|}{ p-value } & \begin{tabular}[c]{@{}l@{}}Disk+East \\      (26)\end{tabular} &\multicolumn{1}{c|}{ p-value } & \begin{tabular}[c]{@{}l@{}}Disk events \\      (19)\end{tabular}& \multicolumn{1}{c|}{ p-value } & \begin{tabular}[c]{@{}l@{}}East events \\      (7)\end{tabular} & \multicolumn{1}{c|}{ p-value }  \\

\hline

  & >10 MeV  &  0.15 & .36872 & 0.24  & .193445 & 0.21 & .303165 & 0.34 & .154372 & 0.22 & .635489\\
  & >50 MeV  &  0.02 & .90513 & 0.09  & .630171 & 0.07 & .734011 & 0.17 & .486556 & 0.11 & .814386\\
 &  >100 MeV & -0.03 & .858099 & 0.03 & .87272 & 0.03 & .884332 & 0.13 & .595802 & 0.11 & .814386\\
  
\hline

\multirow{3}{*}{\begin{tabular}[c]{@{}l@{}}SEP R$_{MAX}$\\ vs CME Mach R$_{MAX}$\end{tabular}} & \multicolumn{1}{c|}{} & \multicolumn{1}{c|}{} & \multicolumn{1}{c|}{} & \multicolumn{1}{c|}{ }&\multicolumn{1}{c|}{ }&\multicolumn{1}{c|}{}&\multicolumn{1}{c|}{}&\multicolumn{1}{c|}{}&\multicolumn{1}{c|}{} &\multicolumn{1}{c|}{} &\multicolumn{1}{c|}{}\\

  & >10 MeV  &  0.62 & .000033 & 0.50  & .004181 & 0.65 & .000325 & 0.44 & .059404 & 0.22 & .635489\\
  & >50 MeV  &  0.56 & .000256 & 0.50  & .004181 & 0.51 & .007775 & 0.31 & .196488 & 0.11 & .814386\\
  & >100 MeV &  0.50 & .001391 & 0.43  & .015761 & 0.45 & .021073 & 0.20 & .411681 & 0.10 & .831082\\
 
\hline\hline

\end{tabular}
\endgroup
\end{table*}

A comparison of the correlation coefficients for the subdivided events according to their location, presented in Table 1, shows the best correlation for events located at disk center, the next best for events located at disk+east, and last for disk+west events. These subdivisions give us an even more clear understanding of the magnetic connectivity of the Sun and Earth at different longitudes. In addition we observe that the correlation coefficient decreases with increasing energy band, meaning that  >10 MeV protons are best correlated and >100 MeV protons are least correlated for all the considered velocity parameters. For all considered subsamples, the correlation coefficient are highly significant (probability>0.95). Only for the seven eastern events does the test give high p-values (p>0.05), which means that at the significance level 0.05 we should reject the hypothesis that the parameters are linearly correlated.

Since this is a continuation of the results obtained in our work presented in \citet{Anitha2020a}, we wanted to perform comparative studies of the obtained linear fits and correlation coefficients shown in the figures. \citet{Anitha2020a} comprised a sample of 25 events with 1 eastern event, whereas the current work comprises 38 events with 7 eastern events. The number of eastern events are particularly highlighted here because significant delays in observations of the peak fluxes are seen for these events due to their poor magnetic connectivity. For this reason there are notable changes to the overall correlation and the fits applied. The average velocity versus SEP peak flux display the previous paper results (slope=0.001, 0.001, 0.0008 and y-intercept=0.05, -0.74, -0.92 for >10, >50, >100 MeV, respectively) and the current paper results (slope=0.002, 0.002, 0.0001 and y-intercept=-0.57, -1.57, -1.75 for >10, >50, >100 MeV, respectively). The maximum velocity versus SEP peak flux display the previous paper results (the second order quadratic equation coefficients a=-7.205e-07 and b=3.825e-03 and y-intercept=-2.846 for >10 MeV and the slope=0.001, 0.0009 and y-intercept=1.51, -1.72 for  >50, >100 MeV, respectively) and the current paper results (slope=0.001, 0.001, 0.0007 and y-intercept=-0.37, -1.49, -1.71 for >10, >50, >100 MeV, respectively). Lastly, the CME velocity at SEP peak flux versus SEP peak flux show  the previous paper results (slope=0.001, 0.001, 0.001 and y-intercept=-0.34, -1.14, -1.35 for >10, >50, >100 MeV, respectively) and the current paper results (slope=0.002, 0.001, 0.0001 and y-intercept=-0.56, -1.20, -1.41 for >10, >50, >100 MeV, respectively). All three comparisons show that the slopes do not exhibit much variation, but the values at which the fit intercepts the y-axis are far  lower in the current paper.

\subsection{Mach number}

%9
\begin{figure*}[h!] 
   \centering
   \includegraphics[width=18.5cm,height=9cm]{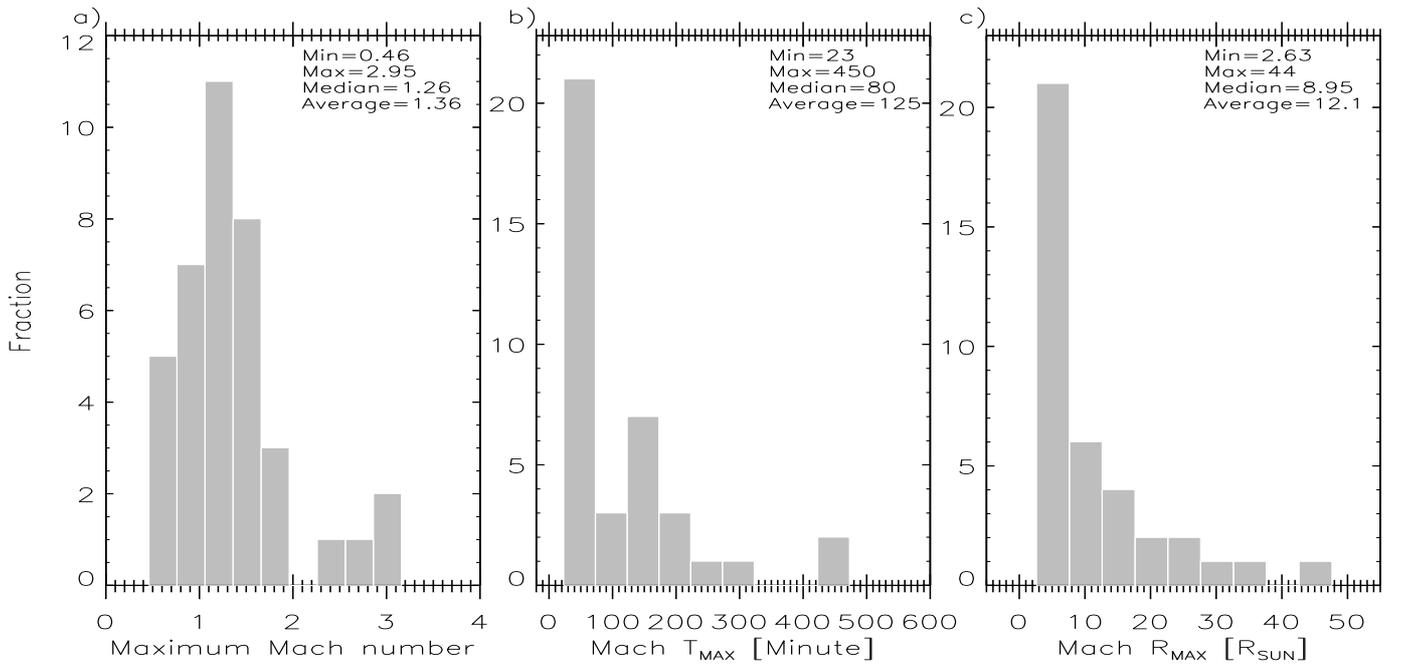}
   \caption{Distributions of maximum Mach number (panel a), time (Mach T$_{MAX}$ , panel b), and distance (Mach R$_{MAX}$ , panel c) at maximum Mach number of CMEs.}
              \label{FigGam}%
    \end{figure*}

Mach number (M$_A$) is one of the most significant parameters determining the efficiency of acceleration of particles in the shock vicinity (\citealt{Li2012a}, \citealt{Li2012b}). As explained by \citet{gopalswamy2010} and referenced  in section 1, the formation of shock occurs when the velocity of the CME (V$_{CME}$) exceeds the sum of Alfvén (V$_{A}$) and solar wind (V$_{SW}$) speed (i.e., V$_A$ + V$_{SW}$) in  interplanetary space. The threshold value for the onset of SEP is, at M$_A$, equal to 1, where the V$_{CME}$ and V$_A$ + V$_{SW}$ are equal. From this point of view, investigating the Mach number parameter with the associated SEP intensities must provide better results than the velocities of CMEs. These parameters are represented in figure 2. The sum V$_A$ + V$_{SW}$ is represented as the dashed orange line and the instant at which the M$_A$=1 is shown as the dotted orange line. The continuous orange line representing the Mach number was calculated using the formula M$_{A}$ = V$_{CME}$/(V$_{A}$ + V$_{SW}$), and mainly varies with the distance parameter (r) away from the Sun; more specifically, the parameters decrease with distance from the Sun. Near the Sun the corona holds dense streamers and tenuous regions that vary the magnetic field significantly; therefore, M$_A$ may vary significantly  (\citealt{gopalswamy08a})  compared to a much farther distance in the interplanetary medium away from the Sun. The described models for estimating the V$_A$ are not perfect, hence we must consider the obtained M$_A$ only as an approximate value.

%10    
\begin{figure*}[h!] 
   \centering
   \includegraphics[width=9.1cm,height=6cm]{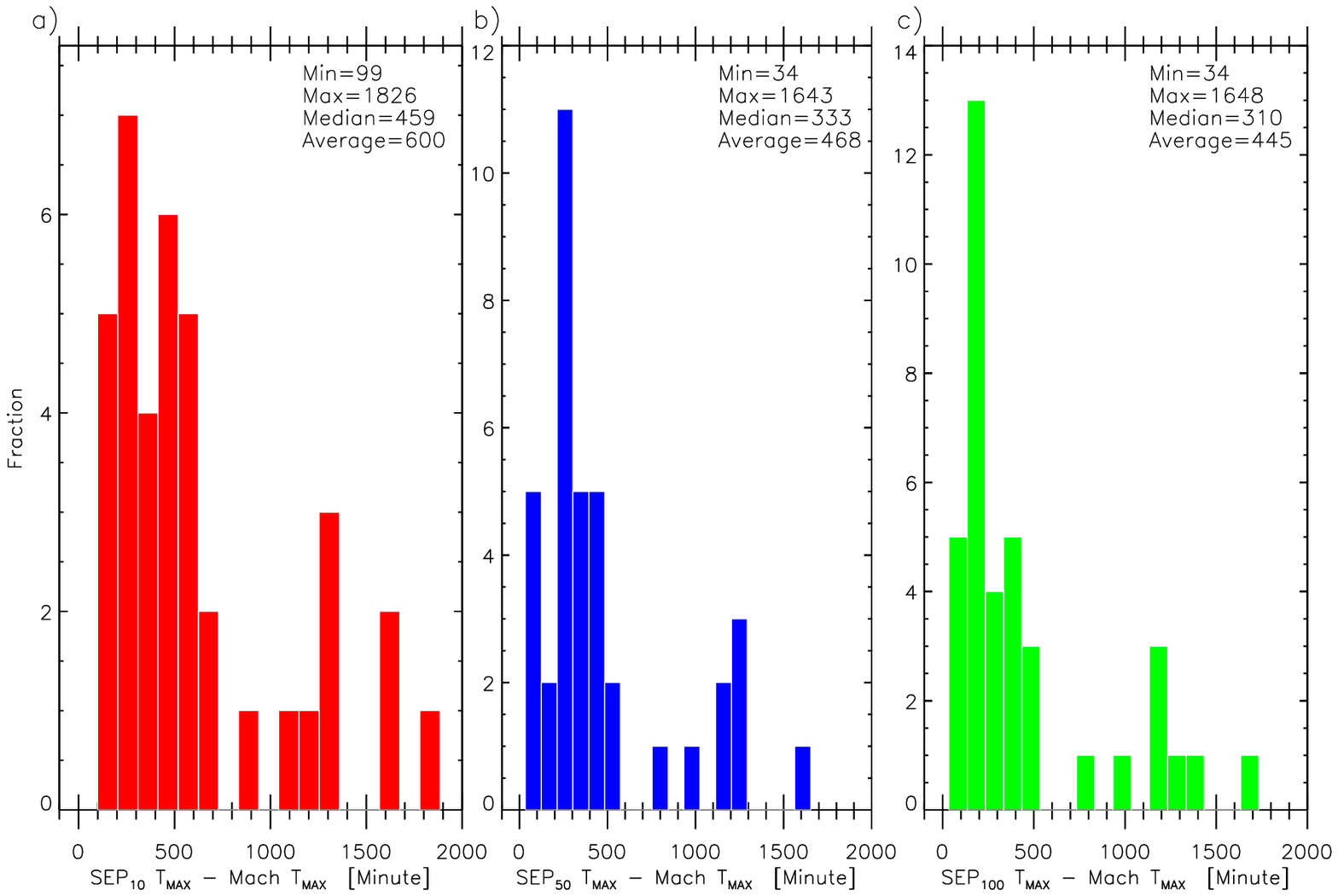}
   \includegraphics[width=9.1cm,height=6cm]{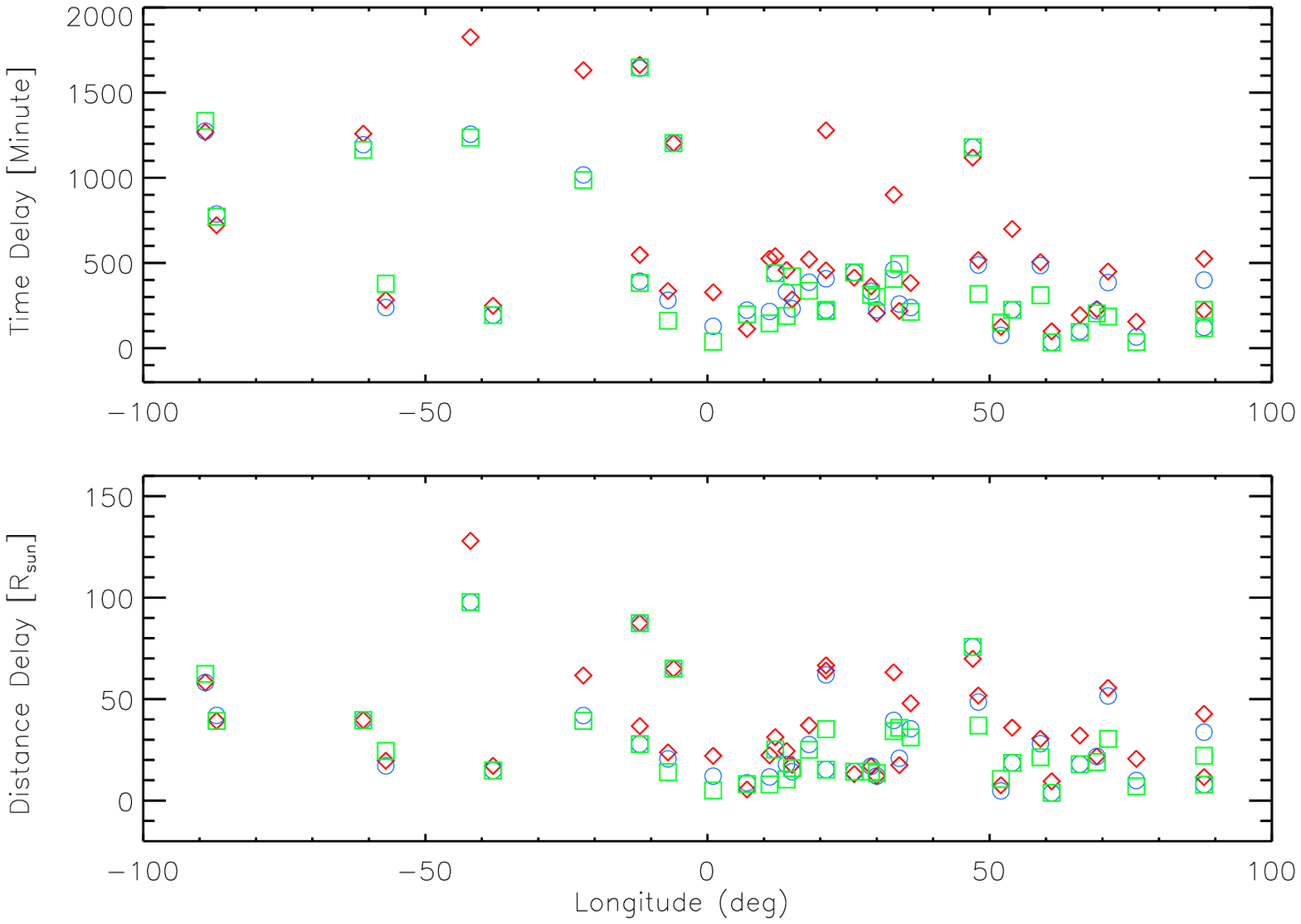}
   \caption{Distribution showing the time difference between SEP peak flux and maximum Mach number. Left: panel a for >10 MeV (red), panel b for >50 MeV (blue), and panel c for >100 MeV (green); Right:  Variation of the time and distance delay with longitude. }
              \label{FigGam}%
    \end{figure*}

%11
\begin{figure*}[h!] 
   \centering
   \includegraphics[width=9.1cm,height=6cm]{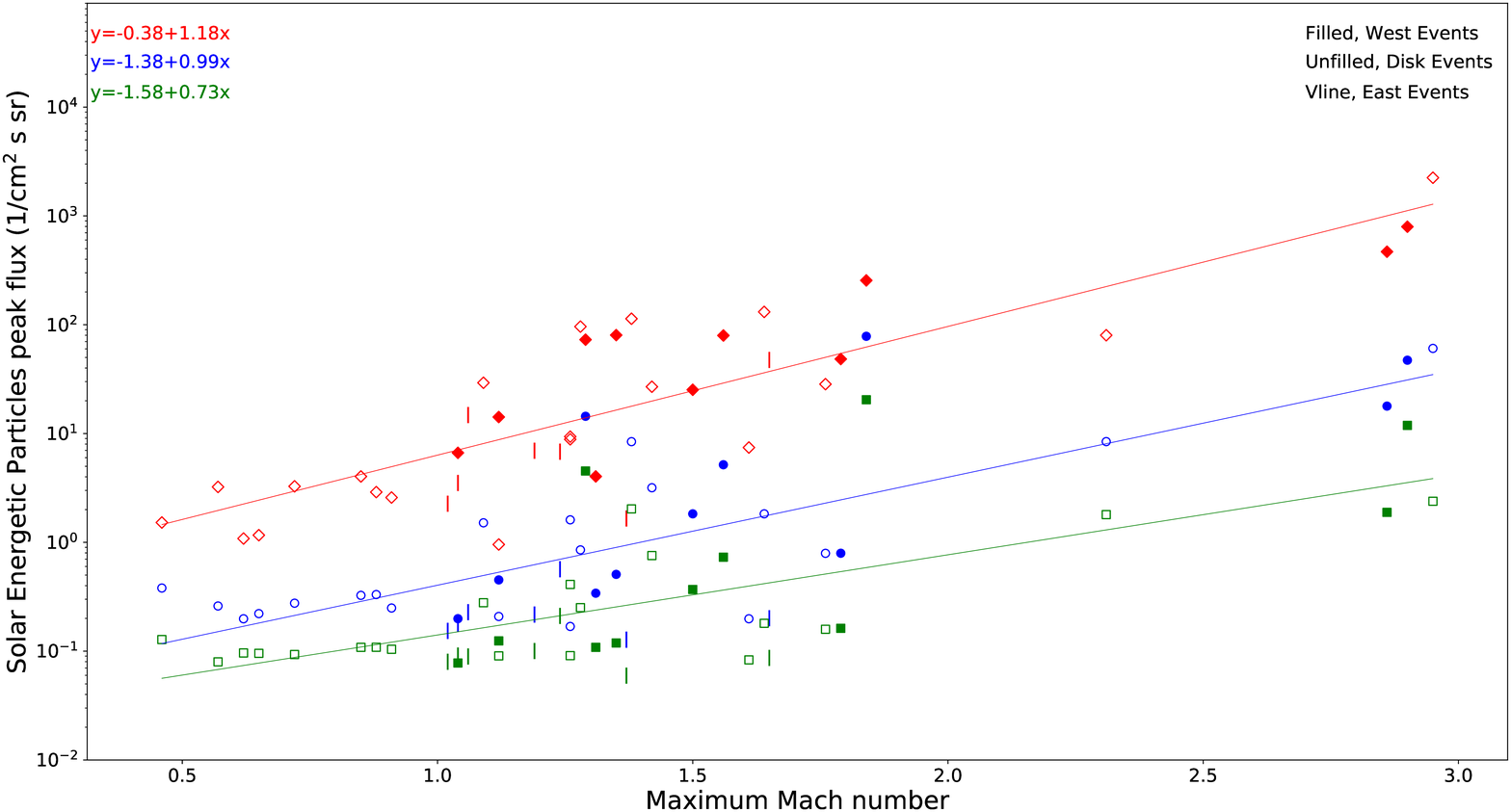}
   \includegraphics[width=9.1cm,height=6cm]{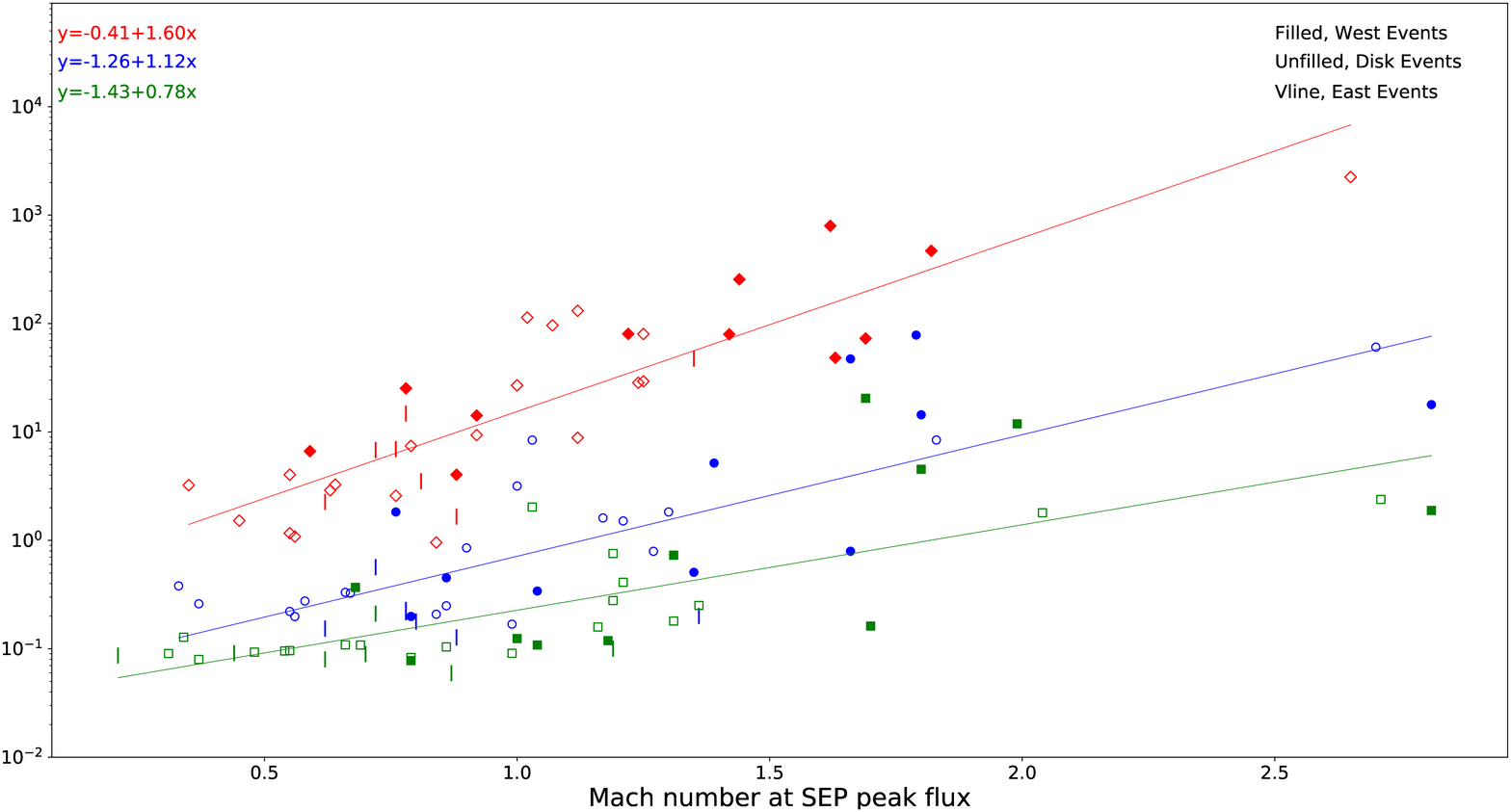}
   \caption{Scatter plot of maximum Mach number (left panel) and Mach number at SEP peak flux (right panel) vs SEP peak flux in the >10 MeV (red), >50 MeV (blue), and >100 MeV (green) energy channels. The open symbols represent disk events (longitude -20 < L < 45), the filled symbols   west events (longitude > 45), and the vertical lines  east events (longitude < -20).}
              \label{FigGam}%
    \end{figure*}

In order to choose the Mach number parameter suitable for the study, we first compared maximum velocity and maximum Mach number, and investigated the outcome. In figure 8 we see the relationship between distance at maximum velocity (left panel) and distance at maximum Mach number (right panel) versus distance at SEP peak flux in the considered energy channels. The correlation coefficients and their probability values (significance at p-value<0.05) are given in Table 2. For magnetically well-connected events the distance at which we observe the R$_{MAX}$ for  the three considered parameters should be approximately same and must provide good correlation. But as we see in both panels, the correlations are insignificant due to the contribution by disk and east events which show delay at the distance at which the SEP attains peak flux. In the left panel we observe the best correlation for disk events and least good for events located at longitudes less than 0, comprising a few disk+east events. A significant improvement is seen in the right panel. This is  evidence that maximum Mach number offers a better correlation for all the considered events to study the SEP peak fluxes.

%t3
\begin{table*}[!h]
\caption{Correlation coefficients of parameters and their probability (significance at p-value<0.05) as shown in Figure 11. The number of events in each subdivision of the longitude are given in the brackets.}\label{YSOtable}
\centering
\tiny
\begingroup
\setlength{\tabcolsep}{1pt} % Default value: 6pt
\renewcommand{\arraystretch}{1.4} % Default value: 1

\begin{tabular}{|c |c| c| c| c| c| c| c| c| c| c| c|}
\hline
\hline
\multirow{3}{*}{\begin{tabular}[c]{@{}l@{}}\\\\Maximum Mach number\end{tabular}}  & \multicolumn{1}{c|}{Energy channel} & \begin{tabular}[c]{@{}l@{}}All events \\      (38)\end{tabular} & \multicolumn{1}{c|}{ p-value } & \begin{tabular}[c]{@{}l@{}}Disk+West \\      (31)\end{tabular} & \multicolumn{1}{c|}{ p-value } & \begin{tabular}[c]{@{}l@{}}Disk+East \\      (26)\end{tabular} &\multicolumn{1}{c|}{ p-value } & \begin{tabular}[c]{@{}l@{}}Disk events \\      (19)\end{tabular}& \multicolumn{1}{c|}{ p-value } & \begin{tabular}[c]{@{}l@{}}East events \\      (7)\end{tabular} & \multicolumn{1}{c|}{ p-value }  \\

\hline

  & >10 MeV  &  0.83 & 0.00001 & 0.85  & 0.00001 & 0.82 & 0.00001 & 0.85 & 0.00001 & 0.53 & 0.221096\\
  & >50 MeV  &  0.72 & 0.00001 & 0.75  & 0.00001 & 0.72 & 0.000034 & 0.80 & 0.000039 & 0.004 & 0.993209\\
 &  >100 MeV &  0.65 & 0.00001 & 0.67  & 0.000076 & 0.64 & 0.00043 & 0.72 & 0.000509 & 0.24 & 0.604195\\
  
\hline

\multirow{3}{*}{Mach number at SEP peak flux} & \multicolumn{1}{c|}{} & \multicolumn{1}{c|}{} & \multicolumn{1}{c|}{} & \multicolumn{1}{c|}{} & \multicolumn{1}{c|}{} & \multicolumn{1}{c|}{}  & \multicolumn{1}{c|}{} & \multicolumn{1}{c|}{} & \multicolumn{1}{c|}{} & \multicolumn{1}{c|}{} & \multicolumn{1}{c|}{}  \\

  & >10 MeV  &  0.85 & 0.00001 & 0.86  & 0.00001 & 0.83 & 0.00001 & 0.85 & 0.000034 & 0.71 & 0.073861\\
  & >50 MeV  &  0.77 & 0.00001 & 0.78  & 0.00001 & 0.79 & 0.00001 & 0.86 & 0.00001 & 0.08 & 0.864622\\
  & >100 MeV &  0.71 & 0.00001 & 0.72  & 0.00001 & 0.72 & 0.000034 & 0.79 & 0.000057 & 0.34 & 0.455574\\
 
\hline\hline

\end{tabular}
\endgroup
\end{table*}

 Figure 9 shows the distribution of maximum Mach number (panel a), and the time (panel b) and distance (panel c) at which they reach maximum Mach number. The average value of maximum Mach number of all the events in our sample is about 1.36 and, on average, they take about 125 minutes at 12.1 R$_{sun}$ to reach the maximum. On comparing these results with figure 4, we observe that CMEs, on average, take about 68 minutes at 6.38 R$_{sun}$ distance to reach maximum velocity. Hence, Mach number takes a longer time and farther distance to reach maximum after the CME eruption. Here, since the estimation of the Mach number depends on the V$_A$ and V$_{SW}$ models along with V$_{CME}$, such differences are seen. The CME speed and Mach number are independent of magnetic connectivity. Therefore, it is obvious that we do not see any parity in these parameters with longitude.  We can compare these parameters with similar indicator describing SEP peak fluxes. Panels a, b, and c in the left panel of figure 10 show the time taken by the SEPs to reach peak fluxes after the maximum Mach number is attained. We observe that it takes, on average, 600, 468, and 445 minutes at a distance of about 41, 35, and 32 R$_{sun}$ to reach peak intensities in the >10, >50, and >100 MeV energy bands, respectively. Panel c shows that the >100 MeV protons take less time to reach peak fluxes   compared to the >10 MeV protons, as shown in panel a. The same trend is observed for delay in distance. Similar conclusions to those shown in figure 5 can be drawn from these results. Based on their magnetic connectivity, as explained in detail in section 3.1, we observe the events located at the eastern limb exhibit more delay in time and distance, and this decreases as we move towards the western limb. This means that maximum Mach number is not   related very significantly with SEP peak flux, especially for magnetically poorly connected events. Therefore, it is reasonable to analyze the instantaneous Mach number at SEP peak fluxes. These parameters should be best correlated with SEP peak fluxes and should not depend on the source location of CMEs.

%t4
 \begin{table*}[!h]
\caption{ Probability values (significance at p>0.05) of the difference between two correlation coefficients presented in Tables 1 and 3. A probability value of more than 0.05 indicates that the two correlation coefficients are significantly the same.}\label{YSOtable}
\centering
\tiny
\begingroup
\setlength{\tabcolsep}{1pt} % Default value: 6pt
\renewcommand{\arraystretch}{1.4} % Default value: 1

\begin{tabular}{|c |c| c| c| c| c| c|}
\hline
\hline
\multirow{3}{*}{\begin{tabular}[c]{@{}l@{}}\\a) Average velocity\\ vs\\Mach number at SEP peak flux\end{tabular}} & \multicolumn{1}{c|}{Energy channel} & \multicolumn{1}{c|}{ All events (38) } & \multicolumn{1}{c|}{ Disk+West (31) } & \multicolumn{1}{c|}{ Disk+East (26) }&\multicolumn{1}{c|}{Disk events (19) }&\multicolumn{1}{c|}{East events (7)} \\

\hline

  & >10 MeV  &  0.08 & 0.08  & 0.83 & 0.38 & 1.00\\
  & >50 MeV  &  0.19 & 0.25  & 0.69 & 0.40 & 0.91\\
 &  >100 MeV &  0.31 & 0.38  & 0.82 & 0.83 & 0.76\\
 
 \hline

\multirow{3}{*}{\begin{tabular}[c]{@{}l@{}}b) Maximum velocity\\ vs\\Mach number at SEP peak flux\end{tabular}} & \multicolumn{1}{c|}{} & \multicolumn{1}{c|}{} & \multicolumn{1}{c|}{} & \multicolumn{1}{c|}{} & \multicolumn{1}{c|}{} & \multicolumn{1}{c|}{}   \\

  & >10 MeV  &  0.03  & 0.19   & 0.22  & 0.35  & 0.36 \\
  & >50 MeV  &  0.42  & 0.60   & 0.28  & 0.27  & 0.75 \\
  & >100 MeV &  0.54  & 0.71   & 0.57  & 0.68  & 0.67 \\ 
  
  \hline

\multirow{3}{*}{\begin{tabular}[c]{@{}l@{}}c) Velocity at SEP peak flux\\ vs\\Mach number at SEP peak flux\end{tabular}} & \multicolumn{1}{c|}{} & \multicolumn{1}{c|}{} & \multicolumn{1}{c|}{} & \multicolumn{1}{c|}{} & \multicolumn{1}{c|}{} & \multicolumn{1}{c|}{}   \\

  & >10 MeV  &  0.32 & 0.30  & 0.83 & 0.77 & 0.89\\
  & >50 MeV  &  0.70 & 0.60  & 0.69 & 0.91 & 0.81\\
 &  >100 MeV &  1.00 & 0.88  & 0.82 & 1.00 & 0.75\\
 
\hline

\multirow{3}{*}{\begin{tabular}[c]{@{}l@{}}d) Maximum Mach number\\ vs\\Mach number at SEP peak flux\end{tabular}} & \multicolumn{1}{c|}{} & \multicolumn{1}{c|}{} & \multicolumn{1}{c|}{} & \multicolumn{1}{c|}{} & \multicolumn{1}{c|}{} & \multicolumn{1}{c|}{}   \\

  & >10 MeV  &  0.77  & 0.88   & 0.91  & 1.00  & 0.67 \\
  & >50 MeV  &  0.63  & 0.78   & 0.57  & 0.58  & 0.91 \\
  & >100 MeV &  0.63  & 0.71   & 0.61  & 0.64  & 0.87 \\

\hline\hline

\end{tabular}
\endgroup
\end{table*}

We restrict our analysis to using instantaneous parameters of Mach number to study their correlation with SEP intensities. As  preliminary evidence seen in \citet{Anitha2020a}, the properties of the Mach number   shows better correlation with SEP peak intensities. We further investigate in detail this instantaneous parameter with a larger number of events in our sample. We observe reduced correlation for maximum Mach number versus SEP peak flux (left panel)  compared to the Mach number at SEP peak flux versus SEP peak flux (right panel) of figure 11 as shown in Table 3. The reduction in correlation coefficient and its significance is much more prominent while considering east events. This is an important result as the comparison shows that in order to study the correlation between CMEs and the associated SEP peak flux, the best parameter to consider is the Mach number at SEP peak flux. The maximum Mach number is suitable for events magnetically well connected to the Earth as the delay between the peaks of Mach number and SEP flux are not higher or rather are appropriate according to their propagation. Hence the SEP peak flux that we observe is accurate. But for events originating in the eastern longitudes the instances at which we observe the peaks are not accurate as they are poorly connected. As a consequence, we observe significant delays between peaks of Mach number and SEP flux. For such cases the best parameter to consider is the Mach number at SEP peak flux, and we observe high correlations in the right panel of figure 11. The optimum correlation is observed for events located at disk center (i.e., -20<longitude<45) that are well connected to the Earth. Thus, the Mach number parameter does better than CME velocities for eastern SEP events with energies >10 MeV. The coefficients of the linear fits do not display notable differences, but the fits are   steeper in the right panel  compared to the left panel of figure 11. The y-intercepts do not show any significant change.

While comparing  the use of the instantaneous parameters, CME speed and Mach number at SEP peak flux, the latter proves to be a better parameter as it shows higher correlation coefficient. This is evident as Mach number takes into account the major parameters (V$_{CME}$, V$_A$, and V$_{SW}$) involved in the onset and influence of particle acceleration, whereas CME speed alone lacks the necessary information for a detailed study. In addition, we observe in the left panel a Mach number of less than 1 for eight disk events, which can also be seen in the histogram presented in figure 9. These events are among the slowest events having maximum velocity of about 800 km~s{$^{-1}$}, maximum Mach number of about 0.7, and the associated SEP peak flux of about 5 cm{$^{-2}$} s{$^{-1}$} sr{$^{-1}$} in the >10 MeV band. Hence, the low speed may have contributed to the low Mach number. All these events were measured using the data from STEREO-A. With respect to the relative position of STEREO-A and the longitude of the event, the projection effect may have played a role in the decrease in   velocity as the  STEREO quadrature configuartion is not perfect for disk events. An additional cause may be due to the the model used to calculate the Alfv\'{e}n speed.

Table 4 shows the comparison of probabilty values of the difference between two correlation coefficients of velocity parameters (Table 1) and maximum Mach number (table 3) versus Mach number at SEP peak flux. Inspecting table 4 we have to reject, at significance level p=0.05, that the considered pairs of correlation coeeficients are significantly different. However, we can consider 1-p, which is the probablility that the respective correlation coefficients are different. In a few examples (for >10 MeV particles for All and Disk+West events in figure 7 panel c) this probability could be very high (1-p>0.8; a value of 0.8 means there is an 80\% probability that the difference is significant). The results presented in this paper, showing importance of Mach at SEP peak flux could be useful for future studies and space weather prediction.

%12
\begin{figure*}[h!] 
   \centering
   \includegraphics[width=18.5cm,height=9cm]{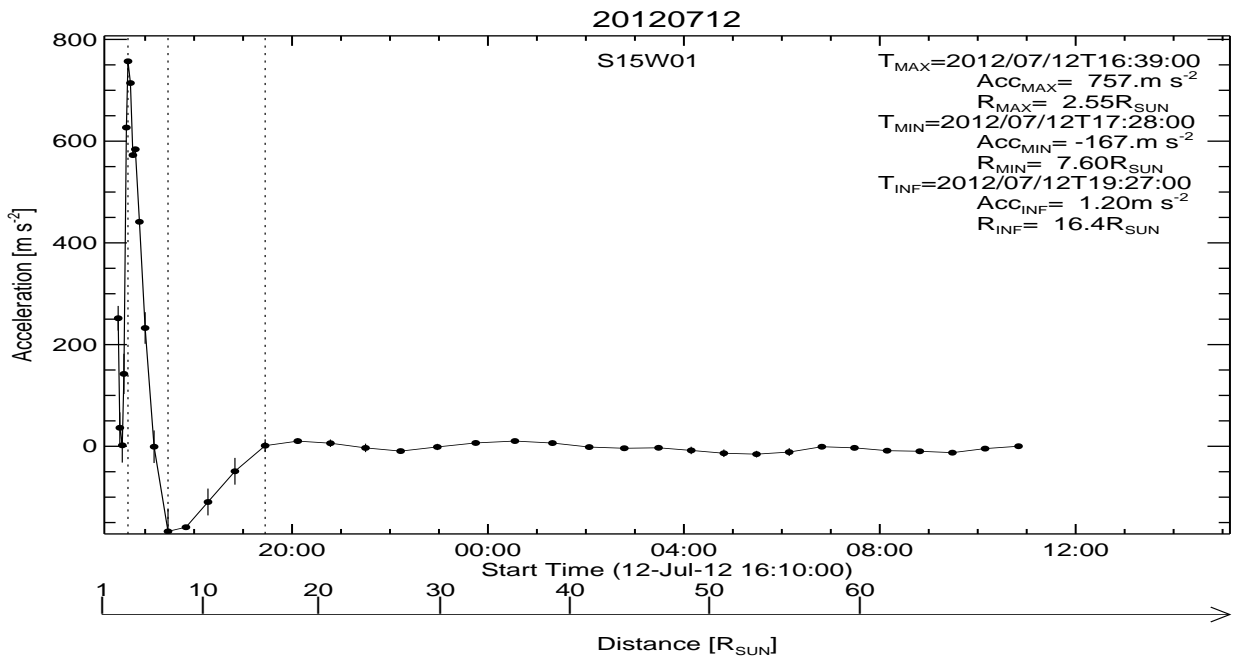}
   \caption{12 July 2012 CME acceleration profile with the active region located at disk center (longitude=01). The parameters of maximum acceleration (Acc$_{MAX}$, T$_{MAX}$, R$_{MAX}$), minimum acceleration (Acc$_{MIN}$, T$_{MIN}$, R$_{MIN}$), and acceleration at the point of inflection (Acc$_{INF}$, T$_{INF}$, R$_{INF}$) are presented in the  top right corner.}
              \label{FigGam}%
    \end{figure*}

In regards to space weather forecasting, the best CME kinematic parameter that could be used to predict the  SEPs is the Mach number at SEP peak flux. As we observe that the Mach number at SEP peak flux and SEP peak occur at the same time, and agrees with events originating at all longitudes, this could be best utilized. Fortunately, SEPs are delayed by 69 (>10 MeV), 31 (>50 MeV), and 22 (>100 MeV) minutes to reach the Earth with respect to white light measurements by coronagraph. So at least for the lower energetic particles that are comparatively slow (>10 MeV), we can determine the Mach number one hour before the SEPs reach the Earth. In figure 2 we have shifted the profiles of the Mach number and the >10 MeV SEP flux by about one hour to take into account the delay, but in reality we first observe the white light and then we measure the height--time data points to determine the Mach number. Therefore,   predictions of the arrival of lower energetic SEPs to the Earth can be made by this method.

However, accurate determination of the CME Mach number may be difficult for a few reasons. First, the CMEs in consideration must be strictly non-interacting throughout their propagation in the interplanetary space as in a CME-CME interaction  the  kinematics can vary significantly. Interacting CMEs are frequent during solar maximum, hence the application of our method could be inaccurate. Next, while the empirical models used in our analysis are satisfactory, the errors on the Mach number depend on the models of plasma density and magnetic field as they are crucial for determining the Alfv\'{e}n speed. To improve our results we considered two different models for magnetic field, but a model never reflects perfectly the real scenario, especially  for magnetic fields around active regions where most CMEs appear. Hence, a further investigation and better approach is required for these parameters. Lastly, the SEP peak fluxes are observed to be achieved at average distances of about R$_{MAX}$= 41.9, 35, and 32.6 R$_{sun}$ for >10, >50, and >100 MeV particles (figure 5). The CME associated with the production of SEPs must be sufficiently strong (or rather, not too weak) to be visible in the coronagraphs;  farther away from the Sun the CMEs are poorer, making measurements difficult and consequently affecting the determination of the Mach numbers. The errors on the measurements depend on the quality or brightness of the CMEs (\citealt{Michalek17}). Hence, one must take into consideration these limitations in real time prediction of CMEs and their associated SEPs.

An important note on SEP fluxes is that it is impossible to determine their peak flux until after the conclusion of an event at 1 AU using the in situ observations, whereas the CME maximum Mach number, on average, is attained at distance R$_{MAX}$=12.1 R$_{sun}$ (figure 9, panel c). With the help of the linear model presented in figure 11 (left panel), which shows CME maximum Mach number versus SEP peak flux, we can obtain the associated SEP peak flux and the time and/or distance of attaining peak,  which can be ultimately used for Mach number at SEP peak flux versus SEP peak flux correlation. This proves to be an important advantage in space weather forecasts, but the determined values must be considered  an underestimation due to the limitations mentioned above.

\subsection{CME acceleration and SEP intensities}

In this section we investigate whether the acceleration parameters can influence the intensities of the associated SEPs. Figure 12 shows an interesting trend in the variation of the CME acceleration with heliocentric distance and time for the event on 12 July 2012. The speed and acceleration of the CME after its eruption increases rapidly as their dynamics are dominated by the propelling Lorentz force (e.g., \citealt{Vrsnak06}; \citealt{Bein11}; \citealt{Carley12}). This expansion phase ends when the CMEs reach maximum velocities leading to a drop in their acceleration to zero. At this point the forces acting on the CMEs (i.e., the propelling Lorentz force and the drag force of the surrounding solar wind) are balanced. This first phase of CME propagation is called the initial or main acceleration. After the maximum speed is reached, the CMEs are gradually slowed down by the ambient medium until they reach the speed of the solar wind (e.g., \citealt{Zhang06}; \citealt{Subramanian07}; \citealt{Gopalswamy13}). This phase of expansion is called the residual acceleration. Using the method described by \citet{Anitha2020b}, the initial or main acceleration is obtained from the formula

 $$
Acc_{INI}={{V_{MAX}}\over{ Time_{MAX}-Time_{ONSET}}},
$$
where $V_{MAX}$ is the maximum velocity of a given CME, $Time_{MAX}$ is the time at maximum velocity, and $Time_{ONSET}$ is the onset time of a given CME on the Sun. These parameters are obtained for each event as shown in figure 2. The initial acceleration is about 3.7 m~s{$^{-2}$} for the fastest and 145 m~s{$^{-2}$} for the slowest CME, and on average the CMEs have Acc$_{INI}$ of 28 m~s{$^{-2}$} until they reach V$_{MAX}$.

The maximum acceleration (Acc$_{MAX}$) and the time and distance at Acc$_{MAX}$ (T$_{MAX}$, R$_{MAX}$), and the minimum acceleration (Acc$_{MIN}$) and the time and distance at Acc$_{MIN}$ (T$_{MIN}$, R$_{MIN}$) are represented by the first and second vertical dotted lines in figure 12, respectively. On average, Acc$_{MAX}$ is about 161 m~s{$^{-2}$} and is achieved at about 30 minutes at 4 R$_{sun}$, and Acc$_{MIN}$ is about -118 m~s{$^{-2}$} and  is achieved at about 80 minutes at 10 R$_{sun}$ after the CME onset. The point at which the CME ceases to decelerate as its kinematics is completely dominated by the interaction with the solar wind is called the point of inflection, and the acceleration at this point is represented by Acc$_{INF}$. The CME acceleration from this point onwards is 0 m~s{$^{-2}$} as it begins to travel at the same velocity as the surrounding solar wind. The Acc$_{INF}$ and the time and distance at Acc$_{INF}$ (T$_{INF}$, R$_{INF}$) is represented by the third vertical dotted line in figure 12. On average, Acc$_{INF}$ is about -10 m~s{$^{-2}$} and  is achieved at about 300 minutes at 25 R$_{sun}$.

 We compare the correlation between the acceleration parameters, Acc$_{INI}$ (left panel) and Acc$_{MAX}$ (right panel), with the associated SEP peak flux in figure 13. The correlation coefficients and their probability values are given in Table 5. Acc$_{INI}$ displays a higher correlation  compared to Acc$_{MAX}$. We can understand that Acc$_{INI}$  is in some sense the total acceleration of a CME in the first or initial phase of expansion. Therefore, it is a better indicator of SEP peak intensities  compared to the instantaneous point of Acc$_{MAX}$. Although there is no significant correlation observed in the left panel, a general trend of increasing Acc$_{INI}$ and SEP peak intensity is observed (i.e., higher initial acceleration leads to higher intensity peak fluxes of SEPs). Furthermore, the correlation shows a slight improvement for higher energy channels of SEPs. In comparison, the best correlation is observed for events originating at disk center.

%13
\begin{figure*}[h!] 
\includegraphics[width=9.1cm,height=6cm]{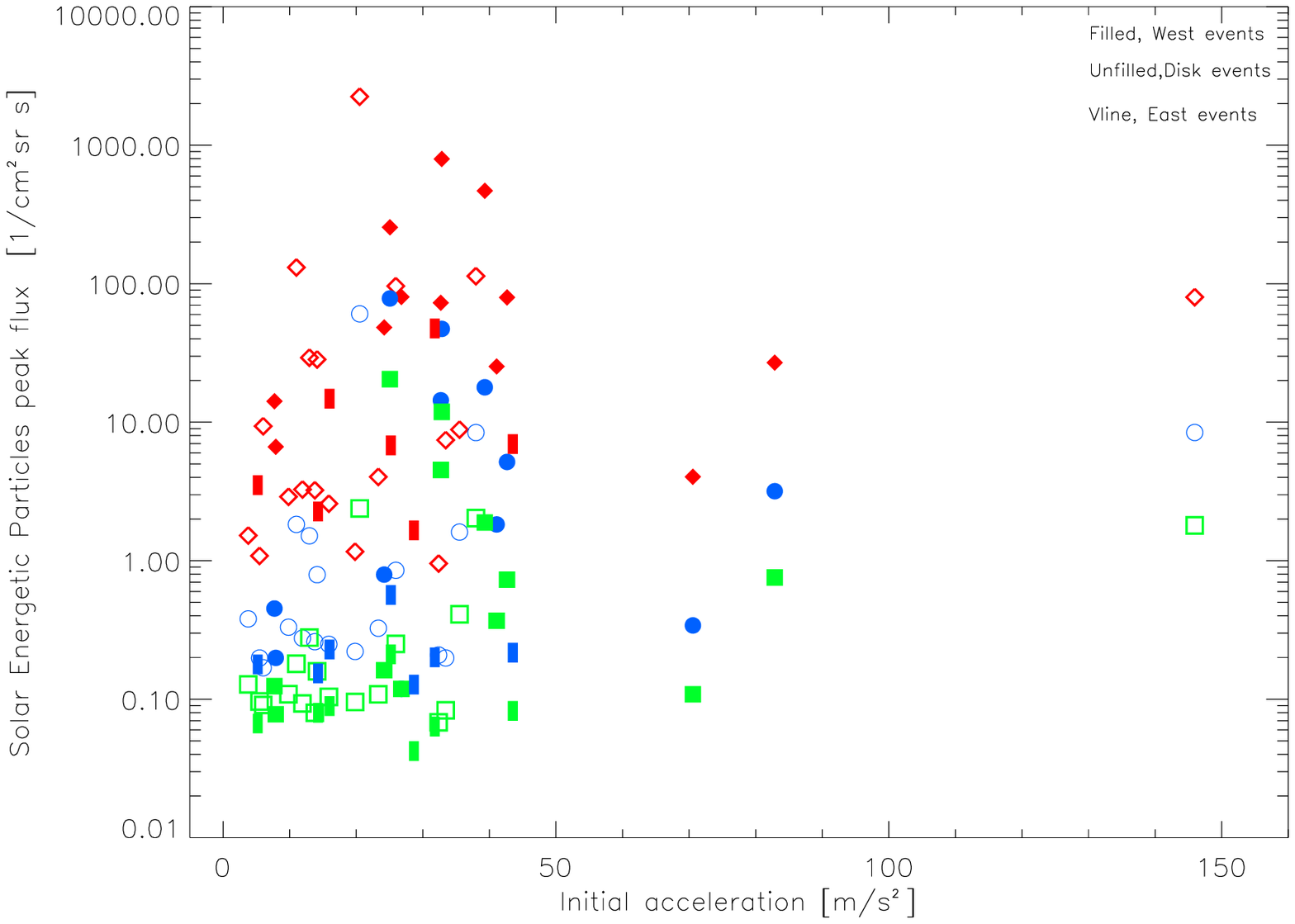}
   \includegraphics[width=9.1cm,height=6cm]{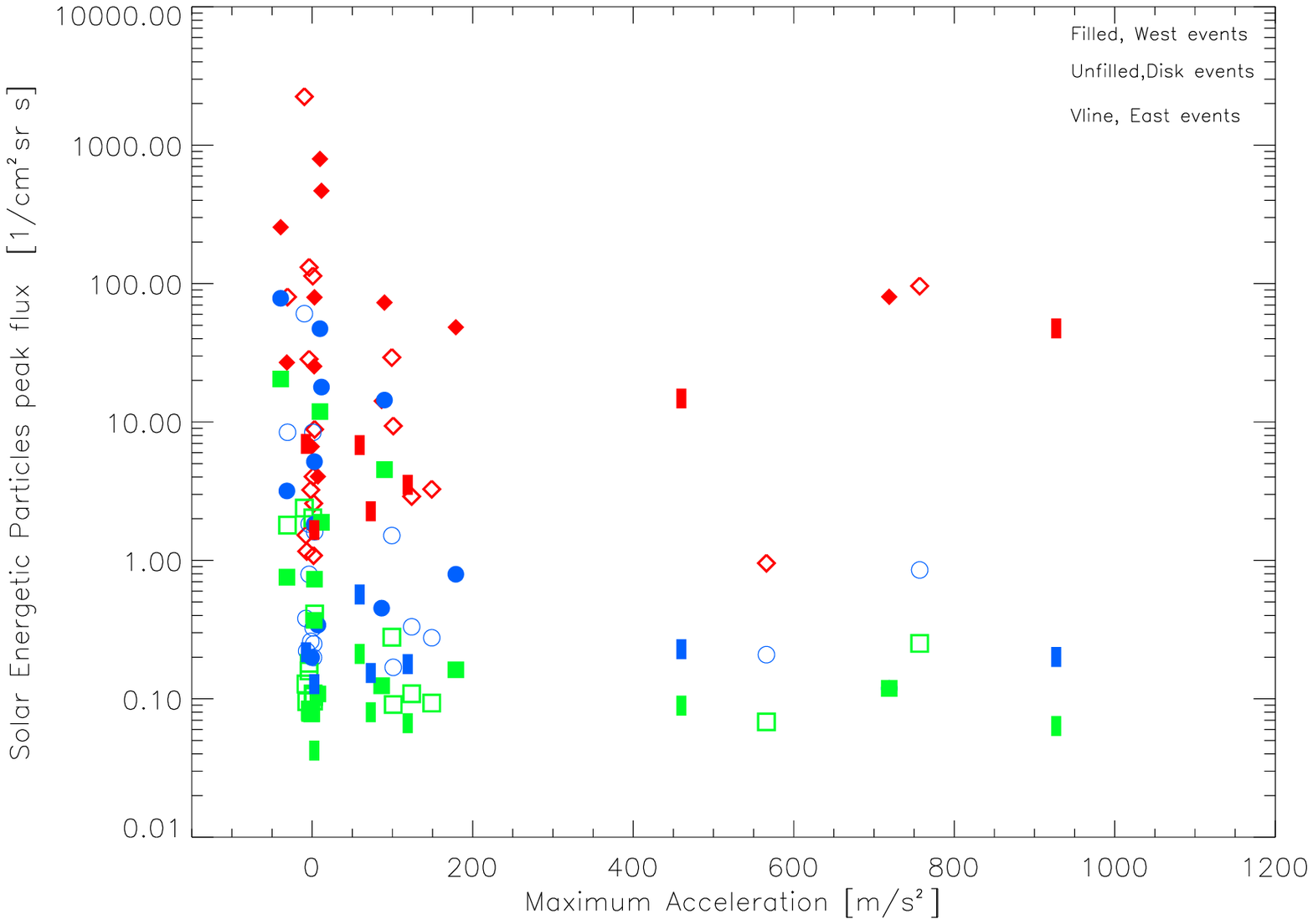}
\caption{Scatter plots of initial acceleration (left panel) and maximum acceleration (right panel) vs SEP peak flux in the >10 MeV (red), >50 MeV (blue), and >100 MeV (green) energy channels. The open symbols represent disk events (longitude -20 < L < 45),  filled symbols represent western events (longitude > 45), and the vertical lines represent eastern events (longitude < -20).}

\end{figure*}

 %t5   
\begin{table*}[!h]
\caption{Correlation coefficients and their probabilities (significance at p-value<0.05) of CME acceleration versus SEP peak flux, as shown in Figure 13. The number of events in each subdivision of the longitude are given in parentheses.}\label{YSOtable}
\centering
\tiny
\begingroup
\setlength{\tabcolsep}{1pt} % Default value: 6pt
\renewcommand{\arraystretch}{1.4} % Default value: 1

\begin{tabular}{|c |c| c| c| c| c| c| c| c| c| c| c|}
\hline
\hline
\multirow{3}{*}{\begin{tabular}[c]{@{}l@{}}\\\\Initial acceleration\end{tabular}}  & \multicolumn{1}{c|}{Energy channel} & \begin{tabular}[c]{@{}l@{}}All events \\      (38)\end{tabular} & \multicolumn{1}{c|}{ p-value } & \begin{tabular}[c]{@{}l@{}}Disk+West \\      (31)\end{tabular} & \multicolumn{1}{c|}{ p-value } & \begin{tabular}[c]{@{}l@{}}Disk+East \\      (26)\end{tabular} &\multicolumn{1}{c|}{ p-value } & \begin{tabular}[c]{@{}l@{}}Disk events \\      (19)\end{tabular}& \multicolumn{1}{c|}{ p-value } & \begin{tabular}[c]{@{}l@{}}East events \\      (7)\end{tabular} & \multicolumn{1}{c|}{ p-value }  \\

\hline

  & >10 MeV  &  0.25 & .130083 & 0.23  & .213231 & 0.28 & .165929 & 0.28 & .245625 & 0.29 & .528119\\
  & >50 MeV  &  0.34 & .036747 & 0.34  & .061285 & 0.40 & .042896 & 0.43 & .066128 & 0.12 & .797745\\
 &  >100 MeV &  0.34 & .036747 & 0.36  & .046669 & 0.49 & .011052 & 0.55 & .014698 & -0.01 & .983024\\
  
\hline

\multirow{3}{*}{Maximum acceleration} & \multicolumn{1}{c|}{} & \multicolumn{1}{c|}{} & \multicolumn{1}{c|}{} & \multicolumn{1}{c|}{ }&\multicolumn{1}{c|}{ }&\multicolumn{1}{c|}{} &\multicolumn{1}{c|}{} &\multicolumn{1}{c|}{} &\multicolumn{1}{c|}{} &\multicolumn{1}{c|}{} &\multicolumn{1}{c|}{} \\

  & >10 MeV  &  -0.006 & .971481 & -0.07  & .708265 & 0.05  & .808343 & -0.05 & .83892 & 0.87 & .010899\\
  & >50 MeV  &  -0.31  & .058213 & -0.32  & .079269 & -0.24 & .237628 & -0.26 & .282375 & -0.009 & .984722\\
  & >100 MeV &  -0.27  & .10113 & -0.27  & .141844 & -0.23 & .258336 & -0.25 & .301953 & -0.13 & .781165\\
 
\hline\hline

\end{tabular}
\endgroup
\end{table*}

\subsection{Shocks and type II bursts}

CME-driven shocks accelerate not just protons, but also the electrons in the solar corona (\citealt{Holman1983}; \citealt{Schlickeiser1984}; \citealt{Kirk1994}; \citealt{Mann1995}; \citealt{Mann2001}; \citealt{Mann2005}). These accelerated electron beams can be observed as type II bursts in the solar radio radiation in the metric wave range (\citealt{Wild1950}; \citealt{Uchida1960}). Type II bursts require electrons escaping from the shock front, and the lack of these bursts implicates the absence of accelerated electrons as type II bursts occur when 0.2-10 KeV electrons are accelerated in the shock front (see, e.g., \citealt{Bale1999}; \citealt{Knock2001}; \citealt{Mann2005}). Energetic electrons are unstable to Langmuir waves, thus they are converted into radio emission at the local plasma frequency and its harmonic (see \citealt{Nelson1985}). Therefore, type II radio bursts hold crucial information of both the shock and the surrounding ambient medium in which the CME-driven shock propagates (\citealt{gopalswamy08a}). Although almost every large SEP event is accompanied by a type II radio burst (\citealt{gopalswamy2003}; \citealt{Cliver2004}) that indicates CME-driven particle acceleration (\citealt{gosling1993}; \citealt{reames1999}), we have ten events in our sample that lack a type II burst: 14 August 2010, 03 August 2011, 04 March 2012, 26 May 2012, 27 May 2012, 14 June 2012, 08 September 2012, 14 December 2012, 21 April 2013, and 06 November 2013. Of  the ten events, two  originate in the west, three in the east, and five at the disk center. On average, these events have a maximum velocity of about 1000 km~s{$^{-1}$} and maximum Mach number of about 0.94. The protons accelerated by these events have peak fluxes of about 23.2 cm{$^{-2}$} s{$^{-1}$} sr{$^{-1}$} in the >10 MeV band, making these events the slowest and weakest SEPs in the sample. As the propagation of radio bursts does not depend on the magnetic connectivity, a possible explanation for their absence could be that the path of the radio burst did not coincide with the instrument on board the satellite or that the detection of the waves was below the range of the radio instrument, hence missing the signature. Detailed investigation is required to understand the absence of DH type II radio bursts in these events.

%14    
\begin{figure*}[h!] 
   \centering
   \includegraphics[width=18.5cm,height=9cm]{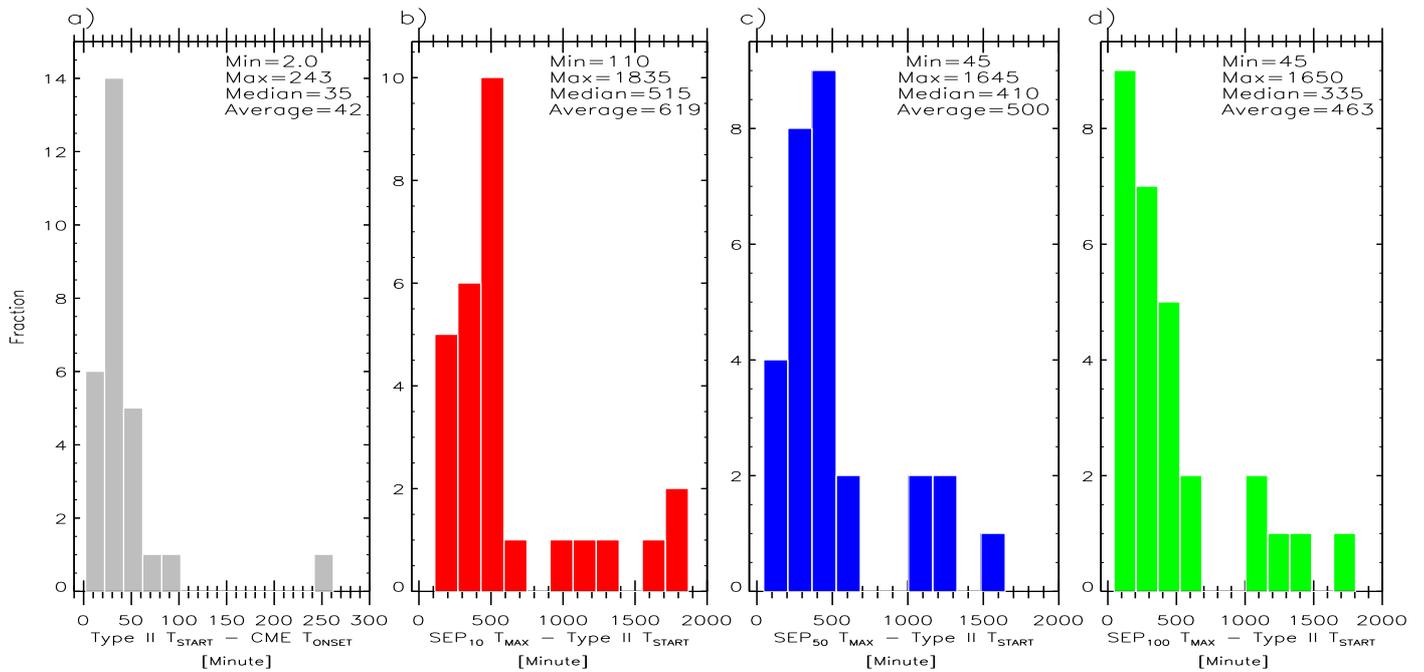}
   \caption{Distribution showing the time difference between shock and CME onset (panel a), and time at SEP peak flux and shock onset (panels b, c, and d) in the energy channels: >10 MeV (red), >50 MeV (blue), and >100 MeV (green), respectively.}
              \label{FigGam}%
    \end{figure*}

%15
\begin{figure*}
\begin{minipage}{22cm}
\includegraphics[width=6.3cm,height=6cm]{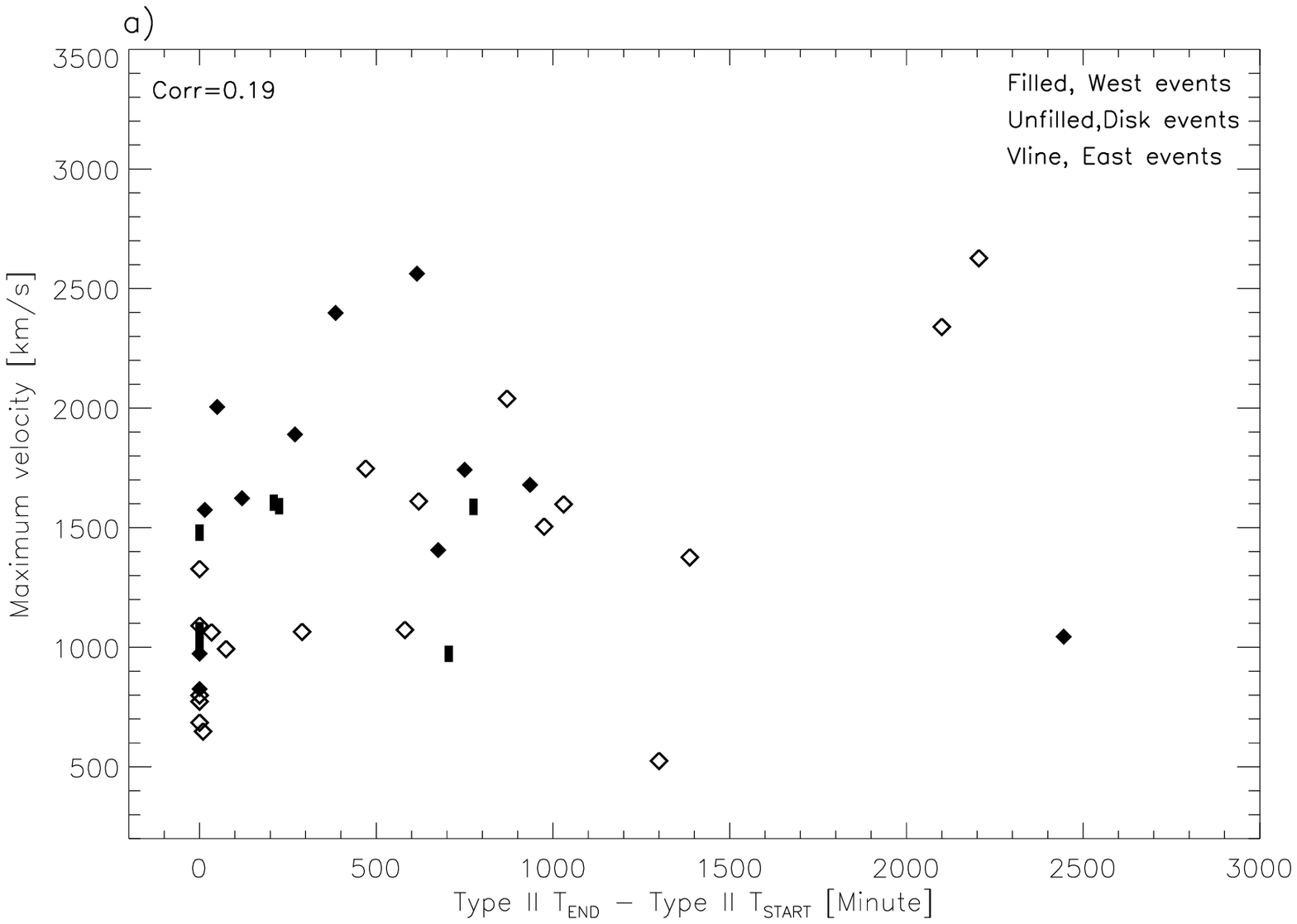}
\includegraphics[width=6.3cm,height=6cm]{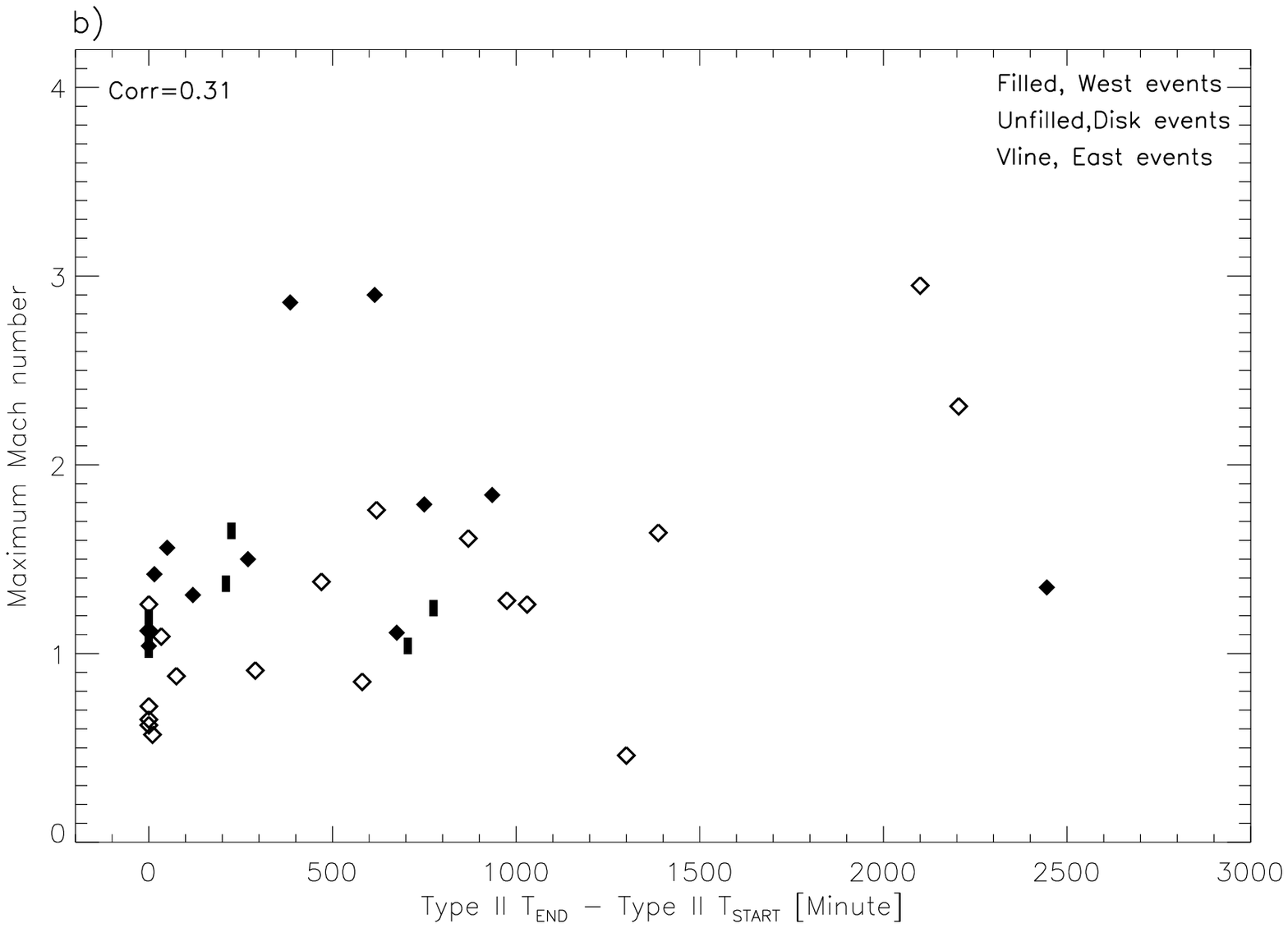}
\includegraphics[width=6.3cm,height=6cm]{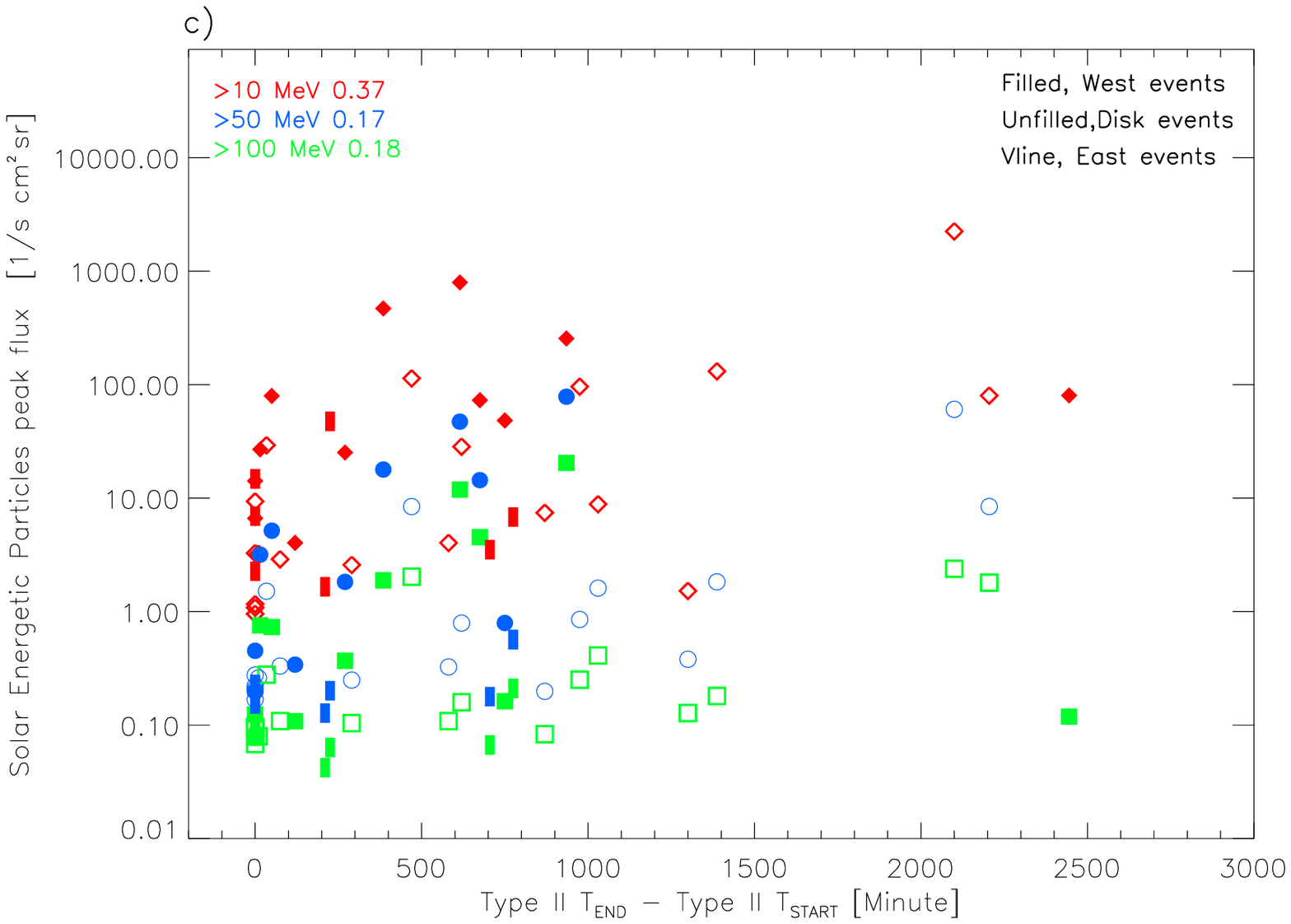}
\end{minipage}
\caption{Scatter plots showing maximum velocity (panel a), maximum Mach number (panel b), and SEP peak flux (panel c) vs duration of the shock. The SEPs are shown in the >10 MeV (red), >50 MeV (blue), and >100 MeV (green) energy channels. The open symbols represent disk events (longitude -20 < L < 45),  filled symbols represent western events (longitude > 45), and vertical lines  represent eastern events (longitude < -20). The presented correlation coefficient in the  top left corner is for all events.}
\end{figure*}

In order to investigate the time taken to observe the start of type II bursts after the onset of the associated CME, figure 14 panel a, clearly shows their distribution. On average, shocks arise 42 minutes after the onset of the CME. The time taken for the shock to accelerate the SEPs to peak fluxes is shown in panel b, c, and d for the >10 MeV (red), >50 MeV (blue), and >100 MeV (green) energy channels, respectively. Again, the >100 MeV protons take less time  to reach peak flux  compared to the >10 and >50 MeV protons. Comparison with the results presented in figure 2 for all events show that the Mach number is equal to 1 at the same point when we observe the start of a type II burst, so our profiles seem to be correct. Similary, the burst disappears when again the Mach number decreases to approximately  1. Although we still observe the SEP flux, the type II burst disappears due to the threshold of the instruments. The instruments measure the radio signal, but particles are observed in situ; therefore, the threshold is much lower for SEP detection in comparison with the radio signal coming from very far away.

We also explore the relationship between the shock duration, i.e., type II Time$_{END}$ - type II Time$_{START}$ with the CME and SEP parameters. The scatter plots are presented in figure 15 showing CME maximum velocity (panel a) maximum Mach number (panel b) and SEP peak flux (panel c) versus shock duration. Although there are no significant correlations observed in these parameters, a general trend of increasing Mach number and SEP peak intensity in the >10 MeV energy channel leads to the longer duration of shocks. An interesting conclusion that can be drawn from panel c is that the duration of the shock is closely correlated with >10 MeV protons,  and less with  >50 and >100 MeV protons. The strength of the shock decreases with distance from the Sun, and the reaccelerated suprathermal SEPs may not reach the spacecraft and the maximum fluxes may not be detected. In addition, although shocks accelerate impulsive seed ions when they are available, they can only result in a small fraction of the SEPs observed (\citealt{Mason1999}). Due to this, the >10 MeV protons predominantly accelerated by CME-driven shocks exhibits the best correlation  compared to the higher energy protons.

\section{Conclusions}

To determine the best instantaneous kinematic parameter of a CME to conduct correlation studies with the intensities of energetic particles, we conducted a statistical study of 38 non-interacting CMEs and their associated SEPs during the ascending phase of solar cycle 24 (i.e., 2009-2013). On further investigation the events were classified as halo and partial-halo events. This particular period was chosen as the STEREO twin spacecraft were near quadrature configuration with respect to the Earth. This position offered a big advantage in the accurate determination of the plane-of-sky speed, which is close to the true radial speed of the halo CMEs. It is worth noting that the presented sample of events is the complete list of non-interacting halo or partial-halo CMEs that generate SEPs with flux values $\geq$1 pfu in the >10 MeV, $\geq$0.1 pfu in >50 MeV, and $\geq$0.05 pfu in >100 MeV bands during the above-mentioned period. This work is a continuation of our previous paper \citet{Anitha2020a}, where the main limitation was the small population of the considered CMEs (25 events). The comparative studies presented in \citet{Anitha2020a} have also shown that STEREO/SECCHI offers a wider range of observation (1.5 R$_{sun}$ – 318 R$_{sun}$) in contrast to the  SOHO/LASCO C2/C3 field of view (1.5 R$_{sun}$ – 32 R$_{sun}$). Therefore, in this paper we completely dedicated the kinematic study of CMEs with STEREO/SECCHI data as we were able to study the CME evolution at large distances from Sun during peak SEP intensities in the heliosphere. Manual measurements of height--time data points were employed to determine instantaneous velocities. Using the empirical models by \citet{dulk}, \citet{leblanc}, \citet{Mann1999}, \citet{gopalswamy01}, \citet{eselevich}, and  \citet{sheeley}, for Alfv\'{e}n and solar wind speed, we derived the instantaneous Mach number parameters for the CMEs. The obtained Mach number was verified with the start and end time of type II radio bursts, which are signatures of CME-driven shock in the interplanetary medium. The start and end times of type II radio bursts should be consistent with instances when the speeds of CMEs reach Mach Number=1. GOES-13 and Wind/WAVES data were used to study the SEPs (in the  >10, >50, and >100 MeV energy channels) and shock profiles, respectively. Their properties are summarized in Table~6.

Electromagnetic waves such as X-rays and radio bursts travel directly from the Sun to the Earth, but energetic ions and electrons propagate along the interplanetary magnetic field lines. Kinematic studies of SEPs are influenced by two uncertainties: 1) the magnetic field may significantly vary for each event from the Parker spiral and  2) pitch angle scattering may occur due to   interplanetary turbulence,  thus distorting the propagation profile. Therefore, studies of these energetic protons in varied energy ranges is required for to determine the path length and propagation times as accurately as possible (\citealt{Tylka2003}; \citealt{Li2012c}). The key results of this article are summarized here.

 Of the 38 events in the sample, 19 events (50\%) originate at the disk center (-20$^\circ$ < longitude < 45$^\circ$), 12 events (31\%) originate at the western limb (longitude > 45$^\circ$), and 7 events (18\%) originate at the eastern limb (longitude < -20$^\circ$). The location of the event is the most important parameter that affects the propagation of the particles. Western limb events have the best connectivity, and SEPs reach peak fluxes quickly after the CME onset, but the connectivity becomes poor as we move towards the east and to the farther western limb (longitude > 60$^\circ$), causing delay as the CMEs must expand wide enough so that their shock fronts are well connected magnetically to the Earth. Due to this variation in connectivity along the solar disk we observe delays in the time at which the SEPs reach maximum intensity, as shown in figures 2 and 5. Additionally, we observe in figure 6, left panel, a gradual decrease in peak intensity for >10 MeV protons, but >50 and >100 MeV protons exhibit this trend only at longitudes greater than 0$^\circ$. At longitudes less than 0$^\circ$ (i.e., moving towards the east), their peak fluxes are  constant and lie in the range 0.1-1 cm{$^{-2}$} s{$^{-1}$} sr{$^{-1}$}. The peak flux of >50 and >100 MeV particles occurs long before the instance when the associated shock gets connected to the Earth. Thus, though we observe these fluxes, they are very reduced and the instruments miss detecting the maximum fluxes. Hence, the maximum SEP fluxes from the western longitudes can be more accurately measured   in situ  compared to eastern longitudes. A consequence of this can also be seen in figure 6, right panel, showing the delays in protons attaining peak flux after the CME onset. The >10 MeV protons take longer    to reach peak fluxes and are observed at farther
distances from the Sun  compared to the >100 MeV protons which are observed close to the Sun (in the range 0-30 R$_{sun}$). As pointed out by \citealt{Dalla2017a} and \citealt{Dalla2017b},  the SEP propagation is also affected   by the drifts caused by the gradient and curvation of the Parker spiral IMF, with their importance increasing with energy of the particle. \\

  We also observe the delay in both time and distance decreases with increasing energy bands of protons, i.e., >100 MeV protons take less time and distance to reach peak flux, and >10 MeV protons take more time and are observed at greater distances  from the Sun. CME-driven shock waves can reaccelerate the impulsive ions pre-accelerated in the magnetic reconnection close to the Sun, and are further seeded into reacceleration by CME-driven shocks at farther distances from the Sun. For this reason we observe the >50 and >100 MeV protons reaching peak fluxes much earlier than >10 MeV protons (\citealt{Desai2003}; \citealt{Tylka2005}; \citealt{Tylka2006}; \citealt{reames2020}).\\

  The main objective of the paper was to determine the ideal kinematic parameter of the CME that offers the best correlation with the associated SEP peak fluxes, irrespective of the location of their origin. First, in figure 3 we compare instantaneous CME speed and Mach number versus SEP fluxes for the western and eastern events; we observed a high correlation for western events and an anticorrelation for eastern events. The anticorrelation is observed as the SEP peak intensties are achieved farther away from the Sun for poorly connected events. Of the two parameters, the Mach number offers higher correlation. Next, the comparative studies shown in figures 7 and 11 show that the CME velocity and Mach number at SEP peak flux offer high correlation of about 0.77 and 0.85, respectively, for the >10 MeV protons. For all of the considered subsamples, correlation coefficients are highly significant (probability>0.95). Only for the seven eastern events did the test give high p-values (p>0.05), which means that at the significance level 0.05 we should reject the hypothesis that parameters are linearly correlated. Comparatively, the Mach number is recommended as it takes into account the major parameters (V$_{CME}$, V$_A$, and V$_{SW}$) involved in the onset and influence of particle acceleration, whereas CME speed alone lacks the necessary information for a detailed study. This good correlation is observed also for eastern limb events where the magnetic connectivity of the Sun and the Earth are poor. The results support the works of \citealt{Liou2011} on their study of correlating fast-forward shock Mach numbers with the intensity of SEPs. Inspecting Table 4, which shows the difference between two correlation coefficients presented in Table 1 and 3, we have to reject (at significance level p=0.05) that the considered pairs of correlation coefficients are significantly different. However, we can consider 1-p, which is the probablility that the respective correlation coefficients are different. In a few examples (for >10 MeV particles for All and Disk+West events in figure 7 panel c) this probability could be very high (1-p>0.8;  a value of 0.8 means there is an 80\% probability that the difference is significant). And the results presented in this paper, showing the importance
of Mach at SEP peak flux, could be useful for future studies and space weather prediction. \\

 For space weather forecasting the best CME kinematic parameter that could be used to predict SEPs is the Mach number at SEP peak flux. As we observed that the  Mach number at SEP peak flux and SEP peak occur at the same time, and that it agrees with events originating at all longitudes, this could be best utilized. Fortunately, for the >10 MeV particles that are comparatively slow, we can determine the Mach number one hour before the SEPs reach the Earth with respect to white light measurements by coronagraph. In figure 2 we have shifted the profiles of Mach number and >10 MeV SEP flux by about one hour to take into account the delay, but in reality we first observe the white light and then we measure the height--time data points to determine the Mach number. Therefore, a prediction of the arrival of the lower energetic SEPs to the Earth can be made by this method. \\

  A thorough analysis of the active regions of five backside events in our sample allowed us to determine the source location that is responsible for the production of energetic particles. Two of them (28 January 2011 and 23 July 2012) were located just over the edge of the solar disk, but the other three (21 March 2011, 04 June 2011, and 08 November 2012) were located as far as $\approx$30$^\circ$ behind the western limb of the Sun. This means that SEPs can be produced from sources located not only on the visible part of the Sun's disk, but even very far ($\approx$30$^\circ$) beyond the eastern (\citealt{gopalswamy2020}) and western (our study) limb of the disk. These results demonstrate the ability of backsided events to cause space weather effects at Earth, and  accurate predictions of SEPs will therefore need to include such events. \\

 Ten events in our sample do not show the associated type II burst. Additionally, for all events we observed that the Mach number is equal to 1 at the same point when we observe the start of the type II burst and that the burst disapears when again the Mach number decreases to approximately   1. Even though we still observe SEP flux, the type II burst disappears due to threshold of instruments. The instrument measures radio signal, but particles are observed in situ; therefore, the threshold is much lower for SEP detection in comparison with radio signals coming from a very great distance.
Although   no significant correlation is observed in these parameters, a general trend of increasing Mach number and SEP peak intensity in the >10 MeV energy channel lead to the longer duration of shocks. The duration of the shock is well correlated with >10 MeV protons  compared to >50 and >100 MeV protons. The shock strength  decreases with distance from the Sun, and the reaccelerated suprathermal SEPs may not reach the spacecraft and the maximum fluxes may not be detected. In addition, although shocks accelerate impulsive seed ions when they are available, they can only result in a small fraction of the SEPs observed (\citealt{Mason1999}). Consequently, the >10 MeV protons predominantly accelerated by CME-driven shocks exhibit the best correlation  compared to the higher energy protons.\\

%t6
\begin{sidewaystable*}[!h]
\caption{Observational parameters of 38 CMEs and the associated SEPs and type II bursts during the period 2009-2013.}\label{YSOtable}
\centering
\tiny
\begingroup
\setlength{\tabcolsep}{1pt} % Default value: 6pt
\renewcommand{\arraystretch}{1.4} % Default value: 1

\begin{tabular}{ |l| c c c c c| c c c| c c c| c c c| c c c| c | c c | }
\hline
\hline
\multicolumn{1}{|c|}{} & \multicolumn{5}{|c|}{CME observations} & \multicolumn{3}{c|}{SEP peak flux }&\multicolumn{3}{c|}{CME speed at SEP peak flux }&\multicolumn{3}{c|}{Maximum Mach number} & \multicolumn{3}{c|}{Mach number at SEP peak flux}&\multicolumn{1}{c|}{Solar Flare} &\multicolumn{2}{c|}{Type II burst}\\

\hline
\# & Date \& Time & V$_{AVG}$ & V$_{MAX}$ & R$_{MAX}$ & T$_{MAX}$ & >10 MeV & >50 MeV & >100 MeV & >10 MeV & >50 MeV & >100 MeV & M$_{MAX}$ & MR$_{MAX}$ & MT$_{MAX}$ &  >10 MeV & >50 MeV & >100 MeV  & Location & Start time & End time \\

   &              & [km/s]    &  [km/s]   & [R$_{SUN}$] & [Minute]&   [pfu] &  [pfu]  &   [pfu]  &   [km/s]& [km/s]  &   [km/s] &           & [R$_{SUN}$]& [Minute]   &          &         &             &     &   &  \\
\hline

  1 &  20100814 10:06 & 744  & 825  & 8.33 & 107 & 14.17 & 0.45  & 0.12 & 792  & 825  & 782  & 1.12 & 3.53  & 34  & 0.92 & 0.86 & 1.00 & N17W52 &         --       &            --    \\
  2 &  20110215 01:20 & 489  & 1064 & 4.56 & 67  & 2.58  & 0.24  & 0.10 & 477  & 543  & 543  & 0.91 & 20.01 & 292 & 0.76 & 0.86 & 0.86 & S20W12 & 20110215 02:10 &  20110215 07:00\\
  3 &  20110307 19:56 & 1147 & 1742 & 9.95 & 72  & 48.36 & 0.79  & 0.16 & 921  & 949  & 1031 & 1.79 & 29.55 & 245 & 1.63 & 1.66 & 1.70 & N30W48 & 20110307 20:00 &  20110308 08:30\\
  4 &  20110607 06:08 & 573  & 1406 & 3.88 & 43  & 72.86 & 14.40 & 4.52 & 425  & 587  & 587  & 1.11 & 3.87  & 43  & 1.69 & 1.80 & 1.80 & S21W54 & 20110607 06:45 &  20110607 18:00\\
  5 &  20110802 05:42 & 532  & 992  & 7.32 & 101 & 2.89  & 0.33  & 0.10 & 470  & 527  & 504  & 0.88 & 2.63  & 42  & 0.63 & 0.66 & 0.66 & N14W15 & 20110802 06:15 & 20110802 07:30 \\
  6 &  20110803 12:36 & 380  & 684  & 7.66 & 125 & 1.08  & 0.19  & 0.09 & 464  & 464  & 439  & 0.62 & 8.75  & 144 & 0.56 & 0.56 & 0.55 & N19W36 &        --       &          --      \\
  7 &  20110804 03:46 & 1277 & 2627 & 3.84 & 18  & 80.05 & 8.42  & 1.79 & 717  & 1121 & 1287 & 2.31 & 24.86 & 144 & 1.25 & 1.83 & 2.04 & N17W69 & 20110804 04:15 & 20110805 17:00 \\
  8 &  20110808 17:58 & 615  & 1623 & 3.24 & 23  & 4.03  & 0.34  & 0.10 & 793  & 700  & 700  & 1.31 & 3.24  & 23  & 0.88 & 1.04 & 1.04 & N14W07 & 20110808 18:10 & 20110808 20:10 \\
  9 &  20110809 08:02 & 993  & 1574 & 3.49 & 19  & 26.91 & 3.17  & 0.75 & 694  & 694  & 853  & 1.42 & 16.44 & 136 & 1.00 & 1.00 & 1.19 & N14W18 & 20110809 08:20 & 20110809 08:35 \\
  10 & 20110906 01:25 & 405  & 524  & 7.72 & 139 & 1.52  & 0.38  & 0.12 & 493  & 305  & 321  & 0.46 & 7.71  & 139 & 0.45 & 0.33 & 0.34 & N11W47 & 20110906 02:00 & 20110906 23:40 \\
  11 & 20110906 21:54 & 667  & 1598 & 3.65 & 45  & 8.84  & 1.61  & 0.41 & 659  & 778  & 771  & 1.26 & 3.64  & 45  & 1.12 & 1.17 & 1.21 & S22W26 & 20110906 22:30 & 20110907 15:40 \\
  12 & 20110922 10:19 & 550  & 1587 & 5.23 & 63  & 6.80  & 0.56  & 0.21 & 414  & 414  & 402  & 1.24 & 3.72  & 51  & 0.76 & 0.76 & 0.75 & N27W71 & 20110922 11:05 & 20110922 24:00 \\
  13 & 20111126 06:42 & 655  & 1044 & 3.50 & 39  & 80.26 & 0.11  & 0.11 & 652  & 623  & 623  & 1.35 & 33.91 & 447 & 1.22 & 1.18 & 1.18 & N17W66 & 20111126 07:15 & 20111127 24:00 \\
  14 & 20111225 17:54 & 345  & 647  & 2.63 & 47  & 3.23  & 0.26  & 0.07 & 293  & 298  & 298  & 0.57 & 2.63  & 47  & 0.35 & 0.37 & 0.37 & N11W76 & 20111225 18:45 & 20111225 18:55 \\
  15 & 20120119 13:31 & 467  & 972  & 8.95 & 187 & 3.51  & 0.17  & 0.06 & 436  & 461  & 480  & 0.91 & 8.95  & 187 & 0.81 & 0.80 & 0.82 & S17E06 & 20120119 15:00 & 20120120 02:45 \\
  16 & 20120123 02:48 & 956  & 2339 & 13.0 & 114 & 2243  & 60.59 & 2.38 & 1410 & 1455 & 1695 & 2.95 & 22.96 & 168 & 2.65 & 2.70 & 2.71 & S13W59 & 20120123 04:00 & 20120124 15:00 \\
  17 & 20120127 17:48 & 850  & 2562 & 11.1 & 78  & 795.5 & 47.19 & 11.8 & 904  & 940  & 1272 & 2.90 & 14.76 & 96  & 1.62 & 1.66 & 1.99 & S15W01 & 20120127 18:30 & 20120128 04:45 \\
  18 & 20120304 10:46 & 546  & 1012 & 4.99 & 71  & 2.26  & 0.15  & 0.07 & 430  & 430  & 430  & 0.92 & 3.66  & 54  & 0.62 & 0.62 & 0.62 & S13W88 &         --       &          --      \\
  19 & 20120313 17:21 & 722  & 2397 & 12.3 & 61  & 468.7 & 17.86 & 1.88 & 1148 & 2092 & 2092 & 2.86 & 15.83 & 80  & 1.82 & 2.81 & 2.81 & S19E42 & 20120313 17:35 & 20120313 24:00 \\
  20 & 20120517 01:43 & 1009 & 1679 & 10.3 & 67  & 255.4 & 78.29 & 20.4 & 901  & 1587 & 1679 & 1.84 & 10.36 & 42  & 1.44 & 1.79 & 1.69 & S06W34 & 20120517 01:45 & 20120517 17:20 \\
  21 & 20120526 20:32 & 657  & 1479 & 3.27 & 34  & 6.94  & 0.21  & 0.08 & 566  & 605  & 605  & 1.19 & 3.26  & 34  & 0.76 & 0.78 & 0.78 & N07E12 &         --       &          --      \\
  22 & 20120527 05:08 & 509  & 1070 & 4.98 & 67  & 14.77 & 0.22  & 0.08 & 528  & 578  & 466  & 0.96 & 3.77  & 53  & 0.78 & 0.78 & 0.70 & N14W11 &         --       &          --      \\
  23 & 20120614 13:00 & 562  & 1327 & 4.46 & 41  & 0.95  & 0.20  & 0.06 & 457  & 457  & 457  & 1.12 & 7.46  & 68  & 0.84 & 0.84 & 0.84 & N10W33 &         --       &          --      \\
  24 & 20120706 22:39 & 621  & 1890 & 3.65 & 46  & 25.24 & 1.82  & 0.36 & 496  & 499  & 475  & 1.50 & 3.64  & 46  & 0.78 & 0.76 & 0.68 & S11W88 & 20120706 23:10 & 20120707 03:40 \\
  25 & 20120712 16:10 & 712  & 1504 & 5.57 & 58  & 96.08 & 0.85  & 0.25 & 695  & 656  & 1176 & 1.28 & 4.27  & 47  & 1.07 & 0.90 & 1.36 & S18E07 & 20120712 16:45 & 20120713 09:00 \\
  26 & 20120719 05:01 & 898  & 2004 & 4.65 & 47  & 79.60 & 5.16  & 0.73 & 833  & 858  & 848  & 1.56 & 4.65  & 47  & 1.42 & 1.39 & 1.31 & S13W59 & 20120719 05:30 & 20120719 06:20 \\
  27 & 20120831 19:19 & 853  & 1590 & 4.33 & 50  & 47.44 & 0.20  & 0.06 & 654  & 684  & 684  & 1.65 & 44.05 & 450 & 1.35 & 1.36 & 1.36 & S15W01 & 20120831 20:00 & 20120831 23:45 \\
  28 & 20120908 09:48 & 482  & 773  & 2.91 & 39  & 1.16  & 0.22  & 0.09 & 414  & 414  & 434  & 0.65 & 2.90  & 39  & 0.55 & 0.55 & 0.54 & S13W88 &         --       &          --      \\
  29 & 20120927 23:12 & 801  & 1610 & 12.3 & 114 & 28.43 & 0.79  & 0.15 & 818  & 808  & 676  & 1.76 & 12.34 & 114 & 1.24 & 1.27 & 1.16 & S19E42 & 20120927 23:55 & 20120928 10:15 \\
  30 & 20121214 01:35 & 596  & 1090 & 12.5 & 181 & 9.36  & 0.16  & 0.09 & 485  & 663  & 663  & 1.26 & 15.04 & 211 & 0.92 & 0.99 & 0.99 & S06W34 &         --       &         --      \\
  31 & 20130116 17:57 & 544  & 1604 & 2.82 & 56  & 1.65  & 0.12  & 0.04 & 530  & 524  & 530  & 1.37 & 2.82  & 56  & 0.88 & 0.88 & 0.88 & N07E12 & 20130116 22:00 & 20130117 01:30\\
  32 & 20130315 06:01 & 593  & 2039 & 3.95 & 61  & 7.43  & 0.19  & 0.08 & 406  & 406  & 406  & 1.61 & 3.95  & 61  & 0.79 & 0.79 & 0.79 & N14W11 & 20130315 07:00 & 20130315 21:30\\
  33 & 20130411 06:53 & 695  & 1747 & 3.63 & 46  & 113.1 & 8.41  & 2.02 & 623  & 669  & 669  & 1.38 & 3.63  & 46  & 1.02 & 1.03 & 1.03 & N10W33 & 20130411 07:10 & 20130411 15:00\\
  34 & 20130421 07:13 & 473  & 798  & 4.04 & 67  & 3.27  & 0.27  & 0.09 & 448  & 503  & 477  & 0.72 & 4.04  & 67  & 0.64 & 0.58 & 0.48 & S11W88 &         --       &          --     \\
  35 & 20130929 21:29 & 715  & 1376 & 12.6 & 125 & 131.1 & 1.82  & 0.18 & 592  & 745  & 768  & 1.64 & 21.23 & 209 & 1.12 & 1.30 & 1.31 & S18E07 & 20130929 21:53 & 20130930 21:00\\
  36 & 20131106 22:50 & 422  & 973  & 7.09 & 123 & 6.64  & 0.19  & 0.07 & 457  & 663  & 663  & 0.84 & 7.09  & 123 & 0.59 & 0.79 & 0.79 & N10W33 &        --       &          --     \\
  37 & 20131119 10:00 & 402  & 1072 & 3.53 & 46  & 4.03  & 0.32  & 0.10 & 397  & 488  & 599  & 0.85 & 3.53  & 46  & 0.55 & 0.67 & 0.69 & S11W88 & 20131119 10:39 & 20131119 20:20\\
  38 & 20131228 17:08 & 550  & 1062 & 6.97 & 82  & 29.27 & 1.51  & 0.27 & 593  & 708  & 841  & 1.09 & 11.97 & 138 & 0.87 & 1.00 & 1.00 & S18E07 & 20131228 17:31 & 20131228 18:05\\

\hline\hline

\end{tabular}
\endgroup
\end{sidewaystable*}

\begin{acknowledgements}
     Anitha Ravishankar and Grzegorz Micha$\l$ek were supported by NCN through the grant UMO-2017/25/B/ST9/00536 and DSC grant N17/MNS/000038. This work was also supported by NASA LWS project led by Dr. N. Gopalswamy. The authors thank the referee for the useful comments and suggestions that have greatly improved the quality of the manuscript. We thank all the members of the STEREO/SECCHI and GOES consortium who built the instruments and provided the data used in this study.\\
     
\end{acknowledgements}


\begin{thebibliography}{}

\bibitem[Aschwanden(2012)]{Aschwanden12} Aschwanden, M.~J.\ 2012, \ssr, 171, 3.

\bibitem[Bale et al.(1999)]{Bale1999} Bale, S.~D., Reiner, M.~J., Bougeret, J.-L., et al.\ 1999, \grl, 26, 1573.

\bibitem[Bein et al.(2011)]{Bein11} Bein, B.~M., Berkebile-Stoiser, S., Veronig, A.~M., et al.\ 2011, \apj, 738, 191.

\bibitem[Bougeret et al., (1995)]{bougeret} Bougeret, J.-L., Kaiser, M.~L., Kellogg, P.~J., et al.\ 1995, \ssr, 71, 231.

\bibitem[Bronarska \& Michalek(2018)]{Bronarska2018} Bronarska, K. \& Michalek, G.\ 2018, Advances in Space Research, 62, 408.

\bibitem[Brueckner et al.(1995)]{Brueckner95} Brueckner, G.~E., Howard, R.~A., Koomen, M.~J., et al.\ 1995, \solphys, 162, 357.

\bibitem[Cane et al., (1986)] {cane1986} Cane, H.~V., McGuire, R.~E., \& von Rosenvinge, T.~T.\ 1986, \apj, 301, 448.

\bibitem[Cane et al., (1987)] {cane1987} Cane, H.~V., Sheeley, N.~R., \& Howard, R.~A.\ 1987, \jgr, 92, 9869.

\bibitem[Cane et al.(2002)]{Cane2002} Cane, H.~V., Erickson, W.~C., \& Prestage, N.~P.\ 2002, Journal of Geophysical Research (Space Physics), 107, 1315.

\bibitem[Carley et al.(2012)]{Carley12} Carley, E.~P., McAteer, R.~T.~J., \& Gallagher, P.~T.\ 2012, \apj, 752, 36.

\bibitem[Cho et al.(2008)]{cho2008} Cho, K.-S., Bong, S.-C., Kim, Y.-H., et al.\ 2008, \aap, 491, 873.

\bibitem[Cliver et al., (1999)] {cliver} Cliver, E.~W., Webb, D.~F., \& Howard, R.~A.\ 1999, \solphys, 187, 89.

\bibitem[Cliver et al.(2004)]{Cliver2004} Cliver, E.~W., Kahler, S.~W., \& Reames, D.~V.\ 2004, \apj, 605, 902.

\bibitem[Dalla et al.(2017a)]{Dalla2017a} Dalla, S., Marsh, M.~S., \& Battarbee, M.\ 2017, \apj, 834, 167.

\bibitem[Dalla et al.(2017b)]{Dalla2017b} Dalla, S., Marsh, M.~S., Zelina, P., et al.\ 2017, \aap, 598, A73.

\bibitem[Desai et al.(2003)]{Desai2003} Desai, M.~I., Mason, G.~M., Dwyer, J.~R., et al.\ 2003, \apj, 588, 1149.

\bibitem[Desai \& Giacalone(2016)]{DesaiGiacalone2016} Desai, M., \& Giacalone, J.\ 2016, Living Reviews in Solar Physics, 13, 3.

\bibitem[Dulk and McLean (1978)] {dulk}  Dulk, G.~A., \& McLean, D.~J.\ 1978, \solphys, 57, 279.

\bibitem[Eselevich and Eselevich (2008)] {eselevich} Eselevich, M.~V., \& Eselevich, V.~G.\ 2008, \grl, 35, L22105.

\bibitem[Gleisner and Watermann(2006)]{gleisner} Gleisner, H., \& Watermann, J.\ 2006, Space Weather, 4, S06006.

\bibitem[Gopalswamy et al.(2000)]{gopalswamy2000} Gopalswamy, N., Lara, A., Lepping, R.~P., et al.\ 2000, \grl, 27, 145.

\bibitem[Gopalswamy et al., (2001)] {gopalswamy01} Gopalswamy, N., Lara, A., Kaiser, M.~L., et al.\ 2001, \jgr, 106, 25261.

\bibitem[Gopalswamy(2003)]{gopalswamy2003} Gopalswamy, N.\ 2003, \grl, 30, 8013.

\bibitem[Gopalswamy et al.(2005)]{gopalswamy2005} Gopalswamy, N., Aguilar-Rodriguez, E., Yashiro, S., et al.\ 2005, Journal of Geophysical Research (Space Physics), 110, A12S07.

\bibitem[Gopalswamy(2006)]{gopalswamy2006} Gopalswamy, N.\ 2006, Journal of Astrophysics and Astronomy, 27, 243.

\bibitem[Gopalswamy et al.(2008a)]{gopalswamy08a} Gopalswamy, N., Yashiro, S., Akiyama, S., et al.\ 2008, Annales Geophysicae, 26, 3033.

\bibitem[Gopalswamy et al.(2008b)]{gopalswamy08b} Gopalswamy, N., Yashiro, S., Xie, H., et al.\ 2008, \apj, 674, 560.

\bibitem[Gopalswamy et al., (2009)] {gopalswamy2009a} Gopalswamy, N., Yashiro, S., Michalek, G., et al.\ 2009, Earth Moon and Planets, 104, 295.

\bibitem[Gopalswamy et al.(2009)]{gopalswamy2009b} Gopalswamy, N., Thompson, W.~T., Davila, J.~M., et al.\ 2009, \solphys, 259, 227.

\bibitem[Gopalswamy et al.(2010)]{gopalswamy2010} Gopalswamy, N., Xie, H., M{\"a}kel{\"a}, P., et al.\ 2010, \apj, 710, 1111.

\bibitem[Gopalswamy(2013)]{Gopalswamy13} Gopalswamy, N.\ 2013, Astronomical Society of India Conference Series, 11.

\bibitem[Gopalswamy et al.(2020)]{gopalswamy2020} Gopalswamy, N., M{\"a}kel{\"a}, P., Yashiro, S., et al.\ 2020, \solphys, 295, 18.

\bibitem[Gosling (1993)] {gosling1993} Gosling, J.~T.\ 1993, \jgr, 98, 18937.

\bibitem[Holman \& Pesses(1983)]{Holman1983} Holman, G.~D. \& Pesses, M.~E.\ 1983, \apj, 267, 837.

\bibitem[Howard et al.(2008)]{Howard08} Howard, R.~A., Moses, J.~D., Vourlidas, A., et al.\ 2008, \ssr, 136, 67.

\bibitem[Jokipii(1982)]{Jokipii1982} Jokipii, J.~R.\ 1982, \apj, 255, 716.

\bibitem[Kahler(1982)]{Kahler1982} Kahler, S.~W.\ 1982, \apj, 261, 710.

\bibitem[Kahler et al.(2000)]{Kahler2000} Kahler, S.~W., Reames, D.~V., \& Burkepile, J.~T.\ 2000, A Role for Ambient Energetic Particle Intensities in Shock Acceleration of Solar Energetic Particles. In High energy solar physics workshop - anticipating hess! (eds R Ramaty, N Mandzhavidze). Astronomical Society of the Pacific Conference Series, Pasadena, no. 206. p. 468.

\bibitem[Kahler (2001)] {kahler2001} Kahler, S.~W.\ 2001, \jgr, 106, 20947.

\bibitem[Kirk(1994)]{Kirk1994} Kirk, J.~G.\ 1994, Saas-Fee Advanced Course 24: Plasma Astrophysics, Lecture Notes of the Swiss Society for Astronomy and Astrophysics (SSAA), held in 1994 in Les Diablerets, Switzerland. Edited by A. O. Benz and T. J.-L. Courvoisier. Publisher: Berlin, New York: Springer, 1994, p.225.

\bibitem[Knock et al.(2001)]{Knock2001} Knock, S.~A., Cairns, I.~H., Robinson, P.~A., et al.\ 2001, \jgr, 106, 25041.

\bibitem[Krogulec et al.(1994)]{Krogulec1994} Krogulec, M., Musielak, Z.~E., Suess, S.~T., et al.\ 1994, \jgr, 99, 23489.

\bibitem[LeBlanc et al.,(1998)] {leblanc} Leblanc, Y., Dulk, G.~A., \& Bougeret, J.-L.\ 1998, \solphys, 183, 165.

\bibitem[Lee(1983)]{Lee1983} Lee, M.~A.\ 1983, \jgr, 88, 6109.

\bibitem[Lee(2000)]{Lee2000} Lee, M.~A.\ 2000, Acceleration and Transport of Energetic Particles Observed in the Heliosphere, ACE 2000 Symposium, California. AIP Conference Proceedings, Volume 528, pp. 3-18, doi:10.1063/1.1324276.

\bibitem[Lee et al.(2012)]{Lee2012} Lee, M.~A., Mewaldt, R.~A., \& Giacalone, J.\ 2012, \ssr, 173, 247.

\bibitem[Li et al.(2012a)]{Li2012a} Li, G., Ao, X., Verkhoglyadova, O., et al.\ 2012, American Institute of Physics Conference Series, 178.

\bibitem[Li et al.(2012b)]{Li2012b} Li, G., Zank, G., Verkhoglyadova, O., et al.\ 2012, American Institute of Physics Conference Series, 115.

\bibitem[Li et al.(2012c)]{Li2012c} Li, G., Moore, R., Mewaldt, R.~A., et al.\ 2012, \ssr, 171, 141. 

\bibitem[Liou et al.(2011)]{Liou2011} Liou, K., Wu, C., Dryer, M., et al.\ 2011, AGU Fall Meeting Abstracts, San Francisco.

\bibitem[M{\"a}kel{\"a} et al.(2011)]{makela2011} M{\"a}kel{\"a}, P., Gopalswamy, N., Akiyama, S., et al.\ 2011, Journal of Geophysical Research (Space Physics), 116, A08101.

\bibitem[Mann et al.(1995)]{Mann1995} Mann, G., Classen, T., \& Aurass, H.\ 1995, \aap, 295, 775.

\bibitem[Mann et al.(1999)]{Mann1999} Mann, G., Aurass, H., Klassen, A., et al.\ 1999, 8th SOHO Workshop: Plasma Dynamics and Diagnostics in the Solar Transition Region and Corona. Proceedings of the Conference held 22-25 June 1999 in CAP 15, 1-13 Quai de Grenelle, 75015 Paris, France. Sponsored by ESA, NASA, C.N.R.S.-I.N.S.U., Euroconferences, Institut d'Astrophysique Spatiale, Matra Marconi Space, SCOSTEP, Université Paris XI. ESA Special Publications 446. Edited by J.-C. Vial and B. Kaldeich-Schümann., p.477.

\bibitem[Mann et al.(2001)]{Mann2001} Mann, G., Classen, H.-T., \& Motschmann, U.\ 2001, \jgr, 106, 25323.

\bibitem[Mann et al.(2003)]{Mann2003} Mann, G., Klassen, A., Aurass, H., et al.\ 2003, \aap, 400, 329.

\bibitem[Mann \& Klassen(2005)]{Mann2005} Mann, G. \& Klassen, A.\ 2005, \aap, 441, 319.

\bibitem[Marsh et al.(2013)]{Marsh2013} Marsh, M.~S., Dalla, S., Kelly, J., et al.\ 2013, \apj, 774, 4.

\bibitem[Mason et al.(1999)]{Mason1999} Mason, G.~M., Mazur, J.~E., \& Dwyer, J.~R.\ 1999, \apjl, 525, L133.

\bibitem[Michalek et al.(2017)]{Michalek17} Michalek, G., Gopalswamy, N., \& Yashiro, S.\ 2017, \solphys, 292, 114.

\bibitem[Nelson \& Melrose(1985)]{Nelson1985} Nelson, G.~J. \& Melrose, D.~B.\ 1985, Solar Radiophysics: Studies of Emission from the Sun at Metre Wavelengths, (A87-13851 03-92). Cambridge and New York, Cambridge University Press, 1985, p. 333-359.

\bibitem[Pande et al., (2018)] {Pande} Pande, B., Pande, S., Chandra, R., et al.\ 2018, Advances in Space Research, 61, 777.

\bibitem[Ravishankar and Micha$\l$ek (2019)] {ravishankar} Ravishankar, A., \& Micha{\l}ek, G.\ 2019, \solphys, 294, 125.

\bibitem[Ravishankar \& Micha{\l}ek(2020a)]{Anitha2020a} Ravishankar, A., \& Micha{\l}ek, G.\ 2020, \aap, 638, A42.

\bibitem[Ravishankar et al.(2020b)]{Anitha2020b} Ravishankar, A., Micha{\l}ek, G., \& Yashiro, S.\ 2020, \aap, 639, A68.

\bibitem[Reames(1995)]{reames1995} Reames, D.~V.\ 1995, Advances in Space Research, 15, 41.

\bibitem[Reames (1999)] {reames1999} Reames, D.~V.\ 1999, \ssr, 90, 413.

\bibitem[Reames(2020)]{reames2020} Reames, D.~V.\ 2020, \solphys, 295, 113

\bibitem[Richardson et al.(2014)]{Richardson14} Richardson, I.~G., von Rosenvinge, T.~T., Cane, H.~V., et al.\ 2014, \solphys, 289, 3059.

\bibitem[Richardson et al.(2015)]{Richardson15} Richardson, I.~G., von Rosenvinge, T.~T., \& Cane, H.~V.\ 2015, \solphys, 290, 1741.

\bibitem[Sheeley et al., (1997)] {sheeley} Sheeley, N.~R., Wang, Y.-M., Hawley, S.~H., et al.\ 1997, \apj, 484, 472.

\bibitem[Schlickeiser(1984)]{Schlickeiser1984} Schlickeiser, R.\ 1984, \aap, 136, 227.

\bibitem[Subramanian \& Vourlidas(2007)]{Subramanian07} Subramanian, P., \& Vourlidas, A.\ 2007, \aap, 467, 685.

\bibitem[Tylka et al.(2003)]{Tylka2003} Tylka, A.~J., Cohen, C.~M.~S., Dietrich, W.~F., et al.\ 2003, International Cosmic Ray Conference, 6, 3305.

\bibitem[Tylka et al.(2005)]{Tylka2005} Tylka, A.~J., Cohen, C.~M.~S., Dietrich, W.~F., et al.\ 2005, \apj, 625, 474.

\bibitem[Tylka \& Lee(2006)]{Tylka2006} Tylka, A.~J. \& Lee, M.~A.\ 2006, \apj, 646, 1319.

\bibitem[Uchida(1960)]{Uchida1960} Uchida, Y.\ 1960, \pasj, 12, 376.

\bibitem[Vinas \& Scudder(1986)]{Vinas1986} Vinas, A.~F., \& Scudder, J.~D.\ 1986, \jgr, 91, 39.

\bibitem[Vourlidas et al.(2010)]{Vourlidas10} Vourlidas, A., Howard, R.~A., Esfandiari, E., et al.\ 2010, \apj, 722, 1522.

\bibitem[Vr{\v{s}}nak(2006)]{Vrsnak06} Vr{\v{s}}nak, B.\ 2006, Advances in Space Research, 38, 431.

\bibitem[Watanabe et al., (2012)] {watanabe} Watanabe, K., Masuda, S., \& Segawa, T.\ 2012, \solphys, 279, 317.

\bibitem[Wild \& McCready(1950)]{Wild1950} Wild, J.~P. \& McCready, L.~L.\ 1950, Australian Journal of Scientific Research A Physical Sciences, 3, 387.

\bibitem[Xie et al.(2019)]{Xie19} Xie, H., St. Cyr, O.~C., M{\"a}kel{\"a}, P., et al.\ 2019, Journal of Geophysical Research (Space Physics), 124, 6384.

\bibitem[Yashiro et al., (2004)] {yashiro2004} Yashiro, S., Gopalswamy, N., Michalek, G., et al.\ 2004, Journal of Geophysical Research (Space Physics), 109, A07105.

\bibitem[Zhang \& Dere(2006)]{Zhang06} Zhang, J., \& Dere, K.~P.\ 2006, \apj, 649, 1100.





\end{thebibliography}

\end{document}